\documentclass{aa}

\usepackage{hyperref}
\hypersetup{hidelinks,colorlinks = true, citecolor = blue ,linkcolor = blue, linktocpage}

\def\apjl {ApJL}
\def\apjs {ApJS}

\def\aap {A\&A}

\DeclareRobustCommand{\ion}[2]{%
\relax\ifmmode
\ifx\testbx\f@series
{\mathbf{#1\,\mathsc{#2}}}\else
{\mathrm{#1\,\mathsc{#2}}}\fi
\else\textup{#1\,{\mdseries\textsc{#2}}}%
\fi}

\usepackage{graphicx}
\usepackage{longtable}

\usepackage{txfonts}
\begin{document}

   \title{Towards an improvement in the spectral description of central stars of planetary nebulae}

%   \subtitle{Towards an improvement in the spectral description}

   \author{W. Weidmann
          \inst{1,2},
           R. Gamen\inst{3},
           D. Mast\inst{1,2},
           C. Fari\~na\inst{4,5},
           G. Gimeno\inst{6}, 
           E. O. Schmidt\inst{7},
           R. P. Ashley\inst{4,8},
           L.~Peralta de Arriba\inst{4,9},
           P. Sowicka\inst{4,10},         
           \and
           I. Ordonez-Etxeberria\inst{4,11} 
           }
   \institute{Universidad Nacional de Córdoba, Observatorio Astron\'omico, Laprida 854, X5000BGR, C\'ordoba, Argentina. \\
              \email{walter@oac.unc.edu.ar}
\and
Consejo de Investigaciones Cient\'{i}ficas y T\'ecnicas de la Rep\'ublica Argentina. Córdoba Argentina.
         \and
Instituto de Astrof\'isica de La Plata, CONICET--UNLP, and Facultad de Ciencias Astron\'omicas y Geof\'isicas, UNLP. Paseo del Bosque s/n, 1900, La Plata, Argentina.
\and
Isaac Newton Group of Telescopes, E-38700 Santa Cruz de La Palma, La Palma, Spain.
\and
Instituto de Astrof\'{i}sica de Canarias (IAC) and Universidad de La Laguna, Dpto. Astrof\'{i}sica, Spain.
\and
Gemini South     c/o AURA, Casilla 603, La Serena, Chile.
\and
Instituto de Astronomía Teórica y Experimental (IATE), Universidad Nacional de Córdoba, CONICET,
Observatorio Astronómico de Córdoba, Laprida 854, X5000BGR, Córdoba, Argentina.
\and
Department of Physics, University of Warwick, Gibbet Hill Road, Coventry, CV4 7AL, UK.
\and
Institute of Astronomy, University of Cambridge, Madingley Road, Cambridge CB3 0HA, UK.
\and
Nicolaus Copernicus Astronomical Center, Bartycka 18, PL-00-716 Warsaw, Poland.
\and
Dpto. de F\'{i}sica Aplicada I, Escuela de Ingenier\'{i}a de Bilbao, Universidad del País Vasco, Bilbao, Spain.
%             University of Alexandria, Department of Geography, ...\\
          }

%   \date{Received September 15, 1996; accepted March 16, 1997}

% \abstract{}{}{}{}{} 
% 5 {} token are mandatory
 
  \abstract
   {There are more than 3000  known Galactic planetary nebulae (PNe),
   but only 492 central stars of Galactic planetary nebulae (CSPN) have   
   known spectral types.
   It is vital to increase this number in order to have reliable statistics, which will lead to an increase of our understanding of these amazing objects.} %shows a wide diversity of spectral types.}
   {We aim to contribute to the knowledge of central stars of planetary nebulae and stellar evolution.
   }   
   {This observational study is based on  Gemini Multi-Object Spectrographs (GMOS)
      and 
      with the Intermediate Dispersion Spectrograph (IDS) at the Isaac
Newton Telescope (INT) %long slit 
   spectra of 78 CSPN. 
   The objects were selected because they did not have any previous 
   classification, or the present classification is ambiguous.
These new high quality spectra allowed us to identify the key stellar lines 
for determining spectral classification in the Morgan-Keenan (MK) system.}
   {We have acquired optical spectra of a large sample of CSPN.
   From the observed targets, 50 are classified here for the first time while for 28 the existing classifications have been improved.
   In seven objects we have identified a P-Cygni profile at the \ion{He}{i} lines.  Six of these CSPN are late O-type.
The vast majority of the stars in the sample exhibit an absorption-type spectrum,
and in one case we have found wide emission lines typical of [WR] stars.
We give a complementary, and preliminary, classification criterion to obtain the sub-type of the O(H)-type CSPN.
Finally, we give a more realistic value of the proportion of CSPN that are rich or poor in hydrogen.
}
   {}

             
\titlerunning{Central stars of planetary nebulae}
\authorrunning{Weidmann et al.}
\maketitle
%
%-------------------------------------------------------------------

%%%%%%%%%%%%%%%%%%%%%%%%%%%%%%%%%%%%%%%%%%%%%%%%%%%%%%%%%%%%%%%%%%%%%%%%%
%%%%%%%%%%%%%%%%%%%%%%%%%%%%%%%%%%%%%%%%%%%%%%%%%%%%%%%%%%%%%%%%%%%%%%%%%
\section{Introduction}

Planetary nebulae  are the end products of the evolution of stars 
between 0.8 and 8 M$_\odot$, before they cool as white dwarfs (WD).
After they terminate the asymptotic giant branch (AGB) phase of their evolution,
the outer layers of the progenitor star are ejected
and the remnant compact object begins to ionize the gaseous envelope.

Currently, there are more than 3000 planetary nebulae known in our Galaxy \citep{2012IAUS..283....9P}, however,  
spectroscopic information  of the central star is available for only
16.4\% of them 
\citep[][hereafter Paper I]{2011A&A...526A...6W}.
Although the number of planetary nebulae candidates has been increasing quickly
\citep{2017LAstr.131b..46L,2016JPhCS.728g2012K,2015IAUGA..2251870S},
 spectral classification, or even the detection of the ionizing star, has been progressing far more slowly.
The most important cause of this imbalance is probably the 
faintness of central stars of planetary nebulae (CSPN) with respect to their
nebula.
These objects are intrinsically faint, with  
60~\%  of  the  CSPN  listed  in  the  the Strasbourg-ESO Catalogue of Galactic Planetary Nebulae \citep[SECGPN;][]{1992secg.book.....A}
having a magnitude of $V>15.5$.

In recent years, large telescopes have made it possible
to observe fainter CSPN, further improving the spectral resolution.
This is leading to
 important progress in the observational study of CSPN as it allows the identification of stellar absorption lines, especially those that are masked by the strong nebular emissions. 
In particular, in order to identify H-rich objects, \ion{He}{i} and Balmer lines need to be detected, but these features are usually masked by nebular emissions.
Perhaps because of the difficulty 
(owing to contamination from nebular emissions)
in disentangling the stellar spectrum from the nebular one, 
only a few CSPN have been analysed
using model atmosphere techniques 
\citep[e.g.][]{2014MNRAS.440.1345F}.

Central stars of planetary nebulae
 are hot evolved stars (pre-white dwarfs) and their optical spectra cover a range of different spectral types.
 Nevertheless, 
this group of stars is comprised of two distinct populations,
so-called  H-rich and H-poor \citep{1948ApJ...108..462A}.
Despite the similarity in their evolutionary stage, they show very different spectral characteristics. 
H-rich stars display Balmer and Pickering \ion{He}{ii} lines. 
On the other hand, the CSPN spectra of the H-poor population are dominated by the
Wolf-Rayet stars
(Wolf-Rayet stars of population II are denoted by [WR]), 
which displays intense and wide lines of \ion{N}{iii}, \ion{C}{iv}
 or \ion{O}{vi}, depending on whether it is a [WN], [WC], or [WO].

The ratio of number of stars from both populations is not yet clear.
\cite{1991IAUS..145..375M} observed that
about 30~\% of CSPNs are H deficient
(with 115 objects), 
but the compilation presented in Paper I resulted in about 40~\%
(with 492 objects).
This difference may be due to the low number of CSPNs with reliable spectral 
classification considered in the mentioned studies. It is necessary to increase 
the observational data to obtain more robust statistics.

According to \cite{1989IAUS..131..463S} a single star after AGB stage should retain a normal H-rich composition.
However, the proportion of CSPN with hydrogen-deficient atmospheres is not negligible, thus other phenomena should be taken into account.
In this sense, having a realistic ratio between these two types of CSPN will result in more robust models of stellar evolution.
This is only possible by increasing the number of CSPN with spectral type determined.

In order to contribute to the understanding of this final stage in stellar
evolution, we undertook this spectroscopic survey of CSPN, 
which is part of a series of works devoted to the spectroscopic study of CSPN.
In this sense, we endeavoured to
obtain high quality spectra, which would allow us to qualitatively determine 
the spectral classification of CSPN. In particular, to
determine if their photospheres are H-rich or H-deficient.
Section~\ref{observation} describes the observations, spectrograms and reduction procedures,
Section~\ref{clasifica} explains the classification criteria, 
and Section~\ref{objetos} presents the spectral descriptions and classifications.
We present and summarize our conclusions in Section~\ref{conclu}.

%%%%%%%%%%%%%%%%%%%%%%%%%%%%%%%%%%%%%%%%%%%%%%%%%%%%%%%%%%%%%%%%%%%%%%%%%
%%%%%%%%%%%%%%%%%%%%%%%%%%%%%%%%%%%%%%%%%%%%%%%%%%%%%%%%%%%%%%%%%%%%%%%%%

\section{Observations, data reduction and sample} \label{observation}

A total of 78 CSPN spectra were studied and are presented in this paper. 
Sixty-six CSPN were observed with Gemini 
Multi-Object Spectrographs \citep[GMOS,][]{2004PASP..116..425H}
at the Gemini 
Telescopes North and South\footnote{GMOS spectrograms were acquired under programmes 
GN-2015A-Q-405,
GS-2015AQ-98,
GS-2015B-Q-103,
GN-2016A-Q-97,
GS-2016A-Q-74,
GS-2016A-Q-101, and
GS-2016B-Q-65
 (PI: W. Weidmann).}
(Table~\ref{sample}), and twelve with the IDS spectrograph at the 2.5 m Isaac Newton Telescope
(Table~\ref{sample-b}).
In addition, we increased the sample by adding other GMOS spectra available in the Gemini archive.  
All data were processed in the same way,
bias frames, dome flat-fields and arc lamp exposures were taken as
part of the Gemini baseline calibrations.

The GMOS observations were carried out  
at the 8.1 m Gemini South and North telescopes 
in their long-slit spectroscopic modes. 
At Gemini South, observations taken before June 2014 used the GMOS-S E2V detector array, 
which  consists of three adjacent 2048 $\times$ 4608 pixel CCDs separated by two  2\farcs8 wide gaps 
and a pixel scale of 0.073 arcsec pix$^{-1}$ in unbinned mode and spectral response of 
approximately 0.36 to 0.93~$\mu$m.
Observations taken after June 14
used the new Hamamatsu CCDs detector array, 
with improved the red sensitivity
\citep{2016SPIE.9908E..2SG}.
The Hamamatsu array consists of three 2048 $\times$ 4176 pixel CCDs separated by two gaps 
of 4\farcs88 wide (61 pixels in unbinned mode) and the scale is 0.08 arcsec px$^{-1}$ in unbinned mode. 
The spectral response is 0.36 to 1.03~$\mu$m.
The slits used for the observations have dimensions of 330 arcsec long and 0\farcs75 or 1\farcs5 wide. 
The diffraction gratings R400 and B600 were used with the 1\farcs5 slit (R $\simeq$ 1000) and the B1200 grating was 
used with the 0\farcs75 slit (R $\simeq$ 2300). 
The details of the observations are shown in Table~\ref{sample}. 
The configurations were chosen to allow us to detect some of 
key stellar lines from spectral classification,
such as the blend of C and N ions at 4650 \AA\ (generally detectable in Of and [WR] stars),
\ion{He}{ii} at 4686 and 5412 \AA, 
\ion{C}{iii} at 5696 \AA\ 
and \ion{C}{iv} at 5806 \AA\ (present in O and [WC]), key lines of hot O-type stars 
(4058 \AA\ of \ion{N}{iii}, 4471 \AA\ of \ion{He}{i} and 4089-4116 \AA\ of \ion{Si}{iv}) 
and key lines of WN (4604-4620-4944 \AA\ of \ion{N}{v}). 
The achieved signal-to-noise ratio (S/N) was up to 30 at 4600 \AA.

The Isaac Newton Telescope (INT) observations were performed with the Intermediate Dispersion Spectrograph (IDS) 
at the Isaac Newton Telescope\footnote{The INT is operated on the island of La Palma by the Isaac Newton Group of Telescopes in the Spanish Observatorio del Roque de los Muchachos of the Instituto de Astrofísica de Canarias.}
on La Palma, Spain. 
IDS observations were carried out using the EEV10 and Red$+$2 detectors with the 
R632V grating (reciprocal dispersion 0.9\AA\ px$^{-1}$) in a central wavelength of 5100 \AA. 
This spectrograph setup provides a resolution of R$\sim$2300 and spectral range coverage 
from approximately 4000 to 6200 \AA. For each target three exposures were taken. 
The exposure times ranged from 1800 to 2100~s depending on the target's brightness.
In all cases the slit was orientated along the parallactic angle.

The spectra were reduced following standard procedures, using
tasks from {\sc iraf}\footnote{{\sc iraf} is distributed by the National Optical Astronomical
Observatories, operated by the Association of Universities for Research
in Astronomy, Inc., under contract to the National Science Foundation
of the USA.} 
included in the GEMINI package (gemini.gmos {\sc iraf} package).
The raw images were processed as follows:
overscan and combined bias subtraction, flat-field correction, cosmic ray removal ({\sc iraf} task gscrrej or cosmicray), and
one-dimensional spectra extraction (performing sky subtraction).
The wavelength calibration was performed by fitting 
a four-order Legendre polynomial to the arcs.
Finally, we rectified the continuum, in the science spectra, 
averaged the two or three spectra of each target,
and shifted the combined spectra in wavelength to zero radial velocity.

The process for the subtraction of the nebular contribution to the 
CSPN spectra deserves a special mention here as it a delicate task.
This process, indispensable for detecting
 possible stellar absorption lines,
depends strongly on the angular size and spatial line profile of the PN.
In Table~\ref{iones}, we indicate if nebular subtraction was performed.
In general this was possible for objects with angular size larger than 10 arcsec.
The nebula spectrum was subtracted from the stellar one 
by interpolating the intensity of the nebular regions on opposite sides of the CSPN (normal to the dispersion). 
In general, we used a constant, but in more complicated cases
we required a  higher-order polynomial.
However, in some objects this procedure did not work well (e.g. when 
the intensity of emission lines was not constant spatially or when the
subtraction worked well for only some of the emission lines)
and we decided not to subtract the nebular contribution.

For the sample presented in this paper the only selection criteria applied was to observe the brightest CSPN that had, up until these observations, lacked a spectral classification. 
In addition, we obtained spectra of some CSPN that in Paper~I
are classified as weak emission-line star (WELS), O-type, H-rich or continuous, with the aim of 
improving that classification.

%%%%%%%%%%%%%%%%%%%%%%%%%%%%%%%%%%%%%%%%%%%%%%%%%%%%%%%%%%%%%%%%%%%%%%%%%
%%%%%%%%%%%%%%%%%%%%%%%%%%%%%%%%%%%%%%%%%%%%%%%%%%%%%%%%%%%%%%%%%%%%%%%%%
\section{Spectral classification}\label{clasifica}

\subsection{Main criteria}\label{mk}

We used the updated spectral-classification system for early OB stars, proposed by \citet{2011ApJS..193...24S,2014ApJS..211...10S}.
When possible, we determined the spectral qualifiers proposed by \citet[Table~3]{2011ApJS..193...24S}. 
It is appropriate to point out that the current classification system is qualitative, that is,
comparing the
unknown spectrum with the grid provided by \citet{2011ApJS..193...24S}.
In the case of O-type stars, we also added a label when the object was observed to be H-rich, following the notation of \cite{1991IAUS..145..375M}.

Some caution is required with the identification of the \ion{He}{i} features. 
A non-detection
should be interpreted as either \ion{He}{i} not being present, or present but 
masked by nebular emission. This leads to an uncertainty in the sub-type classification.

When we could not determine a specific spectral type, we adopted the
following classification:
\begin{itemize}
\item {O(H)}: Balmer and \ion{He}{ii} absorption lines detected (but no \ion{He}{i} nor \ion{N}{}
to determine sub-type).
\item {H-rich}: Balmer lines detected but neither \ion{He}{i} or \ion{He}{ii}.
\item {O}: Only \ion{He}{ii} absorptions lines detected 
(take into account that this classification includes the sub-type O(H), O(He) and O(C)).
\item {cont.}: A high S/N spectrum in which no stellar features and \ion{He}{ii} emission lines were detected.
\item {Emission line}:
Spectrum where no absorption lines are present but the observed emission lines have a stellar origin (for example \ion{C}{iv} at 5806 \AA).
\end{itemize}
%

%%%%%%%%%%%%%%%%%%%%%%%%%%%%%%%%%%%%%%%%%%%%%%%%%%%%%%%%%%%%%%%%%%%%%%%%%
%%%%%%%%%%%%%%%%%%%%%%%%%%%%%%%%%%%%%%%%%%%%%%%%%%%%%%%%%%%%%%%%%%%%%%%%%

\subsection{Other criteria: Lines in the yellow range}\label{red-arm}

As mentioned in Section~\ref{mk}, it is difficult to classify objects as O(H)-type CSPN because 
the absorption lines are often filled with nebular emission.
That is why it would be helpful to have supplementary criteria.
Thus, we explored this possibility by 
analysing the portion of stellar spectrum that extends to wavelengths longer than 5000 \AA, which is usually not taken into account for spectral classification in stars of population I.

In this spectral range, the CSPN classified as O(H)-type
show some important features that seem to be
uncontaminated by nebular emissions.
Between 5000~\AA\ and 6000~\AA, some of these lines are:
\ion{O}{v} at 5114~\AA,
\ion{N}{iv} at 5200-5203~\AA,
\ion{O}{vi} at 5290~\AA,
\ion{He}{ii} at 5412~\AA,
\ion{O}{iii} at 5592~\AA\ and
\ion{C}{iv} at 5801--12~\AA.
Of particular interest is the \ion{C}{iv} line since it is, in general, of stellar origin \citep{2014A&A...570A..26G},
 usually wider than the nebular lines and is isolated, all factors that facilitate its identification and measurement.

With the objective of studying the possible correlation between the 
presence of these lines and the spectral sub-type,
we selected the stars in which the spectral classification includes the sub-type (i.e. from O3 to B1).
This sample consists of 37 objects from this study  
and eight taken from \cite[][hereafter Paper III]{2015A&A...579A..86W}. 
We itemize the following selected remarks:
\begin{itemize}

\item {\ion{O}{v} $\lambda$5114}: Is present in absorption for objects earlier than O(H)3-5,
not detectable in O(H)6-9, and is in emission for early B-type stars.

\item {\ion{N}{iv} $\lambda\lambda$5200--03}: 
These are subtle lines. We have always detected them in absorption in early O-type CSPN, O(H)3--4.

%en algunos objetos O3 no se ve

\item {\ion{O}{vi} $\lambda$5290}:
This line is always present in emission with similar intensity, so is it not useful as sub-type indicator.

\item {\ion{He}{ii} $\lambda$5412}:
Is present in most spectra. Although it is possible that the nebula shows this 
emission, it is narrower than absorption. A surprising case is the CSPN of He 2-162, classified as B0 (see Fig.~\ref{f11}), 
where this line seems to be in emission.
We measured the equivalent width ($EW$) of this line in the whole sample
and obtained (for each range of sub-types)
$EW$(O3-4) = 1.10 $\pm$ 0.25, 
$EW$(O6-8) = 0.84 $\pm$ 0.24 and 
$EW$(B0-1) $<$ 0.2.

\item {\ion{O}{iii} $\lambda$5592}:
Is present in emission in stars earlier than O(H)4 and in absorption 
for stars later than O(H)7.

\item {\ion{C}{iv} $\lambda\lambda$5801--12}: %this line was study in \cite{2015A&A...579A..86W}
These lines are present in emission for stars earlier than O(H)4 and in absorption for stars later than O(H)7.
\end{itemize}

We searched for the following apparent correlations among high-resolution spectra obtained within the OWN Survey 
\citep[see e.g.][for details]{2010RMxAC..38...30B}.
We measured these lines in the spectra of the
standard of classification (between O4--B0) defined by \citet{2011ApJS..193...24S} for Population I stars.
We noted the same
behaviour in the $EW$ of He {\sc ii} $\lambda5412$;
the non-detection of O {\sc v-vi} lines;
the presence of the N {\sc iv} $\lambda\lambda5200-03$
absorption line between the O3 and O5 spectra;
the absorption line of O {\sc iii} being present in the
whole sample but more intense between O5 and
O8.5;
the C {\sc iv} absorption line very marginal or not detectable in the O3 and B0 sub-types, present in the
other sub-types, showing a maximum in its $EW$ at O5--O7.

The $EW$ of \ion{He}{ii} $\lambda$5412 together with the presence of the other lines of 
\ion{O}{iii}, \ion{N}{iv} and \ion{C}{iv} appear as promising features to discriminate, at least, 
between early- and late-O of population II stars, when it is not possible from the historical MK's wavelength range.
The presence of the feature identified as N {\sc iv} $\lambda\lambda$5200--03 in population I stars was 
preliminarily studied by
\citet{2002RMxAC..14...16G}.
Further analysis is necessary to assess if useful criteria can be established using these lines.

%%%%%%%%%%%%%%%%%%%%%%%%%%%%%%%%%%%%%%%%%%%%%%%%%%%%%%%%%
%%%%%%%%%%%%%%%%%%%%%%%%%%%%%%%%%%%%%%%%%%%%%%%%%%%%%%%%%

\subsection{The qualifier "z" in Pop II stars}
\label{ovz}

According to \citet{2016AJ....152...31A},
O-type stars that satisfy the following relations
between the $EW$ of \ion{He}{ii} and \ion{He}{i} absorption lines, 
$z_{4542} = EW(4686) / EW(4542) \geq$ $1.1$ or 
$z_{4471} = EW(4686) / EW(4471) \geq$ $1.1$,
should be tagged with an additional parameter "z".
It is not yet very clear what phenomenon produces 
the O~Vz spectral characteristic although it seems to be 
related with the stellar youth and low wind-strength.

We explored the determination of this qualifier in our
CSPN spectra. We find that it is applicable to nine objects. 
All of them are earlier than O4
(A~51, AMU~1, BMP~J1800$-$3408, HbDs~1,
Lo~8, Lo~16, NeVe~3$-$2, PHR~J1756$-$3342 and Sa~4$-$1).
Moreover, we found extreme values that are even higher 
than those reported by \cite{2016AJ....152...31A}.
This is the case for HbDs~1 and Lo~16 with 
$z_{4542} = 2.2$ and $z_{4542} = 2.4$ respectively. 
In addition, we found high $z$-values
in two objects that we could not determine their
sub-types - H~2$-$29 and Pe~1$-$9.
In this sense, perhaps both objects deserve early-O sub-type classification.

On the other hand, we were able to
identify P-Cygni profiles in seven spectra.
Six of them are late O-type CSPN
(H~1$-$65, IPHAS~190438, He~1$-$2, He~2$-$47, He~2$-$151 and He~2$-$162). 
The exception is the CSPN of He~2$-$107, despite being classified as O(H)4,
it displays P-Cygni profiles.
In all these objects the P-Cygni profiles were identified in the \ion{He}{i} lines.

According to the evolutionary sequence of H-rich CSPN,
they evolve from late O-type to early O-type.
In this sense, it is more likely that an object
experiences mass loss in the earliest stages of its evolution, that is, late O-type.
Our observational findings are in good agreement with
these ideas, that is, late-O stars show P-Cyg profiles and
as mass-loss is decreasing (and stars heat up) the
spectra tend to show Vz characteristics, which in this 
case is more related to mass-loss strength.

%%%%%%%%%%%%%%%%%%%%%%%%%%%%%%%%%%%%%%%%%%%%%%%%%%%%%%%%%%%%%%%%%%%%%%%%%
%%%%%%%%%%%%%%%%%%%%%%%%%%%%%%%%%%%%%%%%%%%%%%%%%%%%%%%%%%%%%%%%%%%%%%%%%

\section{Stellar descriptions and classifications}

Spectral classifications are indicated in Table~\ref{iones} along with other
parameters of interest. There, we indicate the detection, in absorption
(A) or emission (E), of 
\ion{He}{ii} $\lambda$4542--4686--5411, 
\ion{He}{i} $\lambda$4471,
\ion{N}{v} $\lambda\lambda$4603--19,
\ion{C}{iv} $\lambda\lambda$5801--12, and
H$\beta$. We also label whether the nebula spectrum was subtracted (N.S.).
The spectra are shown in Figs.~\ref{f01} to \ref{f18},
arranged  according  to  the spectral type.

Notes on some individual CSPN spectra are included in the following section.

%%%%%%%%%%%%%%%%%%%%%%%%%%%%%%%%%%%%%%%%%%%%%%%%%%%%%%%%%%%
%%%%%%%%%%%%%%%%%%%%%%%%%%%%%%%%%%%%%%%%%%%%%%%%%%%%%%%%%%%

\subsection{Notes on individual stars}\label{objetos}

\textit{PN~G000.2$-$01.9} (M~2$-$19), Fig.~\ref{f05}: 
Subtle Balmer absorption.  
Possible wide emission at 4650 \AA \ and noticeable absorption of \ion{C}{iii}
at 4647 \AA. This is a feature present in intermediate/late O-type stars.
There are absorption features of unknown origin at 4226, 4382, 
4666 and 5170 \AA\ (artifact?).
We classified this CSPN as O(H)6-8. However, the lines 4226 and 4666 are 
observed in $\tau$~Sco, a B0-type star.
In this sense, the CSPN of this object could be a O(H)8.

\textit{PN G001.7$-$04.6} (H~1$-$56), Fig.~\ref{f16B}:
\cite{2009A&A...500.1089G} report strong emission of \ion{C}{iv}
at 5806 \AA, perhaps coming from stellar atmosphere.

\textit{PN G003.3$-$04.6} (Ap~1$-$12), Fig.~\ref{f11}:
\cite{2009A&A...500.1089G} classified this CSPN as [WCL]. 
The authors reported emission of \ion{C}{iii} at 5696 \AA, we also observe this line. 
Nevertheless, it is not possible to be certain that the emission is stellar.
We disagree with this classification, as our spectrum clearly displays absorption 
lines of \ion{He}{i}, \ion{He}{ii} and \ion{C}{iii} typical of an early B-type star.
Although we could not identify the Balmer series, nebular lines are very strong.
In addition, we could identify all lines reported by
\cite[][hereafther Paper II]{2011A&A...531A.172W}.
We note that this CSPN displays absorption of \ion{C}{iv}
at 5801-11 \AA. This feature is typical of an O late star.

\textit{PN~G003.5$-$02.4} (IC~4673), Fig.~\ref{f09A}:
It is not easy to fit the nebular profile, however the presence of the Balmer series 
can be distinguished.
There is \ion{He}{i} in absorption at 4471, 4921 and 4387 \AA.
\ion{He}{ii} is not visible. Is it not clear if there is a stellar feature at 4686 \AA.
Unknown absorption at 4362 \AA.
This line is observed in $\tau$~Sco, a B0-type star.
Which is compatible with our classification of the CSPN of IC~4673 as O8.

\textit{PN~G005.0$+$03.0} (Pe~1$-$9), Fig.~\ref{f14}:
The clear lines of \ion{He}{ii} indicate that it is
an O(H) star.  
However, according to Section \ref{ovz},
this CSPN could be an early O-type star.

\textit{PN~G007.8$-$04.4} (H~1$-$65), Fig.~\ref{f09A}:
This object is very strange, with a very low excitation class (EC),
and no published spectra. 
In addition, our spectra display many emission lines.
This object may not be a planetary nebula.
The central star of this nebula was classified as [WC11] \citep{2009A&A...500.1089G}
and WELS \citep{2003A&A...403..659A}.
However, we were able to identify key lines of a supergiant late O-type star.
Emission at 4686 \AA\ has to be of stellar origin, because the EC of the nebulae is very low.
Moreover, it is likely that the \ion{He}{ii} line at 4026 \AA\ displays a P-Cygni profile.
This profile is visible in the \ion{He}{i} lines at 4471 and 5016 \AA.
Absorption features to the left of H$\varepsilon$, H$\zeta$ and H$\eta$, are probably the Balmer series of the CSPN.

\textit{PN~G011.3$-$09.4} (H~2$-$48), Fig.~\ref{f14A}:
For this object we have only detected \ion{O}{iii} in absorption at 5592 \AA.

\textit{PN~G013.1$+$05.0} (Sa~3$-$96), Fig.~\ref{f16A}:
This is a poorly studied object with a very low EC.
Our two-dimensional spectra show a bright star that is not symmetrical with
the nebular profile emission (therefore it is probably not the CSPN).
The nebular subtraction is regular, however we can infer absorption
at H$\beta$.
In addition, the spectrum shows absorption lines of unknown origin at
5167, 5179 (\ion{N}{ii}?), and 5264 \AA.
The line at 5179 \AA\ is observed in $\tau$~Sco, a B0-type star.
In this sense, the CSPN of this object could be an late O-type star.

\textit{PN~G014.8$-$08.4} (SB~20), Fig.~\ref{f03}:
Clear absorption of \ion{He}{ii} at 4200 and 4542 \AA.
Moreover, 4471\AA\ is not detected.
On the other hand, the ion \ion{O}{v} at 5114 \AA \ is in absorption, typical of hot stars.
There is absorption of unknown origin at 5125 and 5144 \AA.

\begin{figure*}
   \centering
   \includegraphics[width=0.95\textwidth]{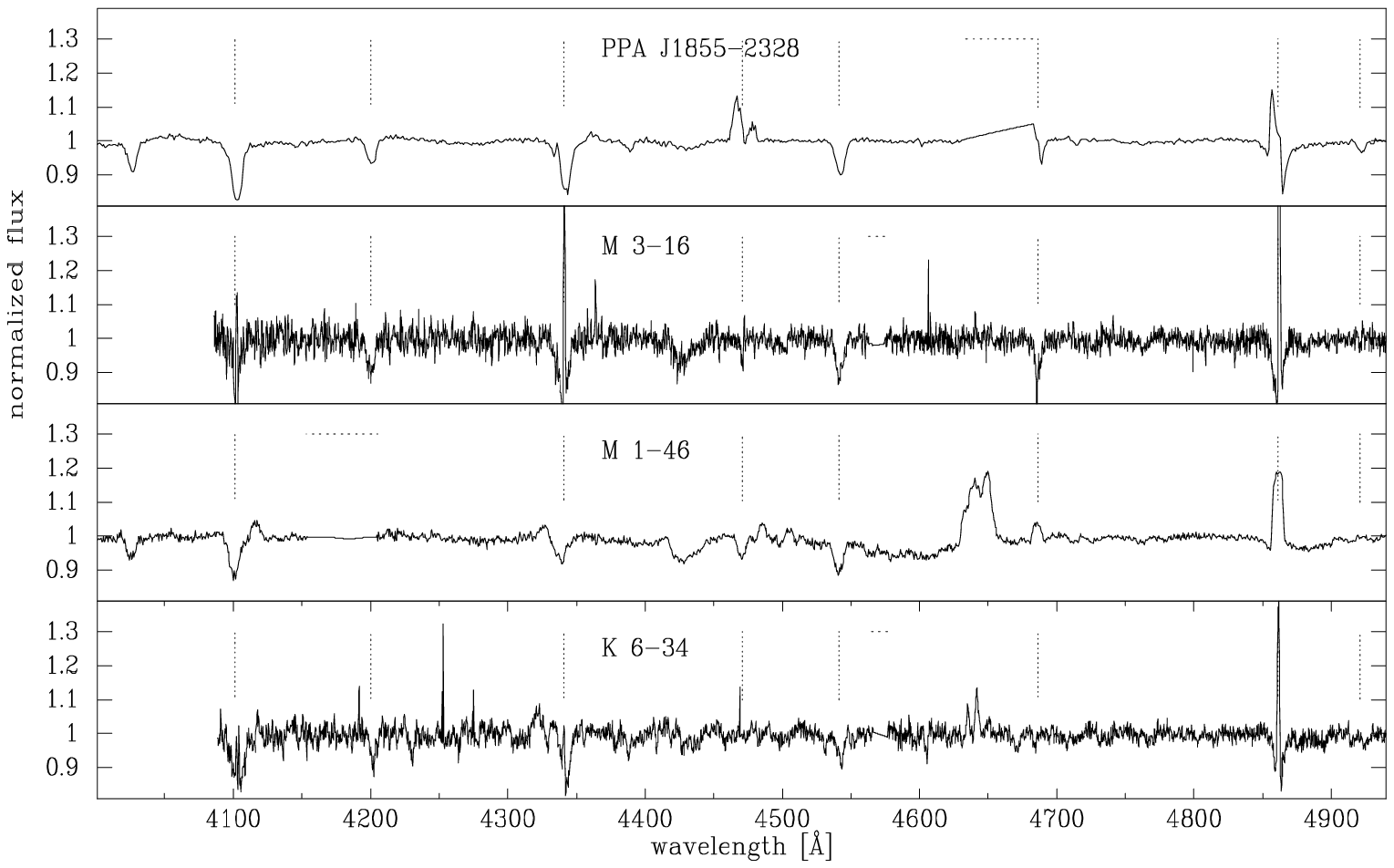}
   \includegraphics[width=0.95\textwidth]{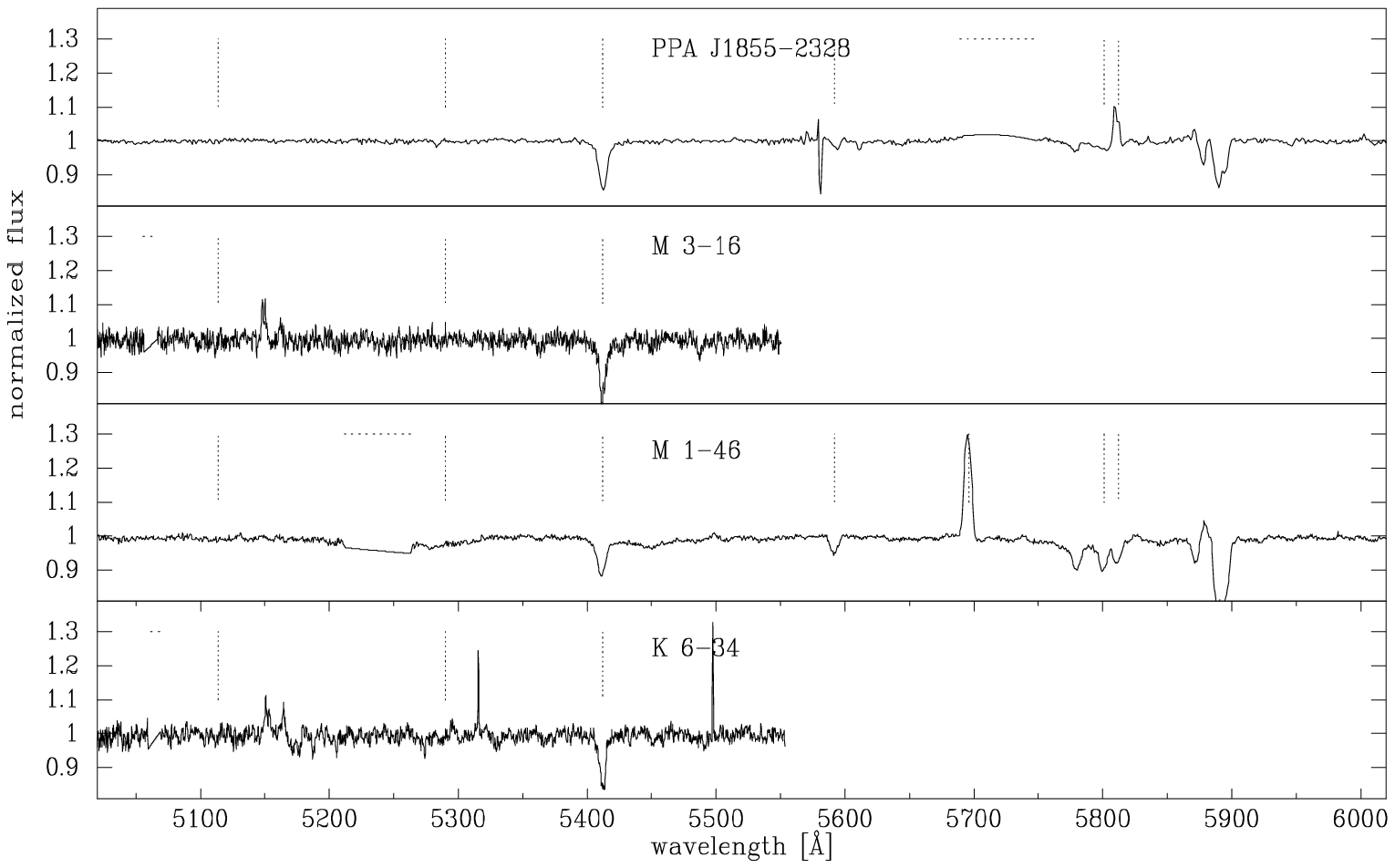}
      \caption[]{Normalized spectra of intermediate O-type CSPN 
                           (see Table~\ref{iones}).
                 The interstellar absorption bands at $\lambda$4428, the  complex at 5780 and 5890-6 are not indicated.
                 The most important spectral features (absorption and emission) identified are: H$\beta$, H$\gamma$, H$\delta$,
   \ion{He}{i} $\lambda$4471 and $\lambda$4921,
   \ion{He}{ii} $\lambda$4200, 4542, and 4686, for the blue spectral range (top panel), and \ion{O}{v}   $\lambda$5114,
 \ion{O}{vi}  $\lambda$5290,
 \ion{He}{ii} $\lambda$5412,
 \ion{O}{iii} $\lambda$5592,
 \ion{C}{iii} $\lambda$5696, and
 \ion{C}{iv}  $\lambda$5801-12, for the red spectral range (bottom panel).
}
         \label{f07}
   \end{figure*}

\textit{PN~G016.4$-$01.9} (M~1$-$46), Fig.~\ref{f07}:
The spectrum displays the key lines of 
\ion{Si}{iv} at 4116 \AA\ and \ion{C}{iii} at 4696 \AA, together with \ion{O}{iii} in absorption at 5592 \AA.
These features are typical of an intermediate O-type star.
In addition, the spectrum displays clear emission lines of \ion{N}{iii}
and \ion{C}{iii} at 4634 and 4647\AA\ respectively.
Our classification agrees with that reported by
\cite{2003IAUS..209..237H}.

\textit{PN~G017.6$-$10.2} (A~51), Fig.~\ref{f05}:
There are no traces of \ion{He}{i}, despite the high S/N.
We note the clear \ion{N}{v} in emission at 4603-19 \AA. 
The intense absorption at 4686 \AA\ is compatible with a luminosity class V.
Absorption lines are wider than the nebula, indicating an evolved object.
\cite{1964ApJ...140.1601G} classified this object as sd O8k. We disagree with 
this classification, as our spectra display clear lines of Nv, typical of an early O-type star.

\textit{PN~G034.3$+$06.2} (K~3$-$5), Fig.~\ref{f18}:
Although the object has an appreciable angular size, it is not possible
to perform a good nebular subtraction.
The two-dimensional spectrum shows clear emission lines in the central region of the object.
Although most of them originate from the nebula, there is a fraction for 
which a stellar origin can not be discarded 
(4340, 4471, 4686, 4713, 4861, 5311, 5412, 5801-12, and 5876 \AA).
This object is probably a double-envelope PN.

\textit{PN~G036.4$-$01.9} (IPHASX~J190438.6$+$021424), Fig.~\ref{f09}:
Object classified as probably PN by \cite{2014MNRAS.443.3388S}.
The object is very affected by extinction (self-extinction?).
There are no [\ion{O}{iii}] emissions.
The star displays all \ion{Fe}{ii} lines typical of a B[e] star.

\textit{PN~G038.4$-$03.3} (K~4$-$19), Fig.~\ref{f11}:
The Balmer series is visible together with \ion{He}{i}
at 4471, 4921 and 5876 (with P-Cygni profile).
In addition, the spectrum displays clear absorption of
\ion{C}{iii} at 4650 \AA\ and 
\ion{N}{iii} at 4641 \AA.

\textit{PN~G042.0$+$05.4} (K~3$-$14), Fig.~\ref{f009}:
Subtle absorption of \ion{He}{ii} at 4686 \AA\, is detected in addition to a 
narrow absorption of \ion{O}{iii} at 5592 \AA. Also, the \ion{He}{i} at 4144 \AA\, is clear.

\textit{PN~G044.3$-$05.6} (K~3$-$36), Fig.~\ref{f16B}:
The spectra of this object do not show clear lines of a O(H)-type star.
However, present in emission are  
\ion{O}{iii} at 5592 \AA\ and \ion{C}{iv} at $\lambda\lambda5801-12$.
In addition, absorption of \ion{O}{v} at 5114 \AA\ is visible.
These features, according Section~\ref{red-arm},
are characteristic of a O(H)3-4.
Hence, the CSPN of K~3-36 is probably an early O(H)-type star.

\textit{PN~G047.1$+$04.1} (K~3$-$21), Fig.~\ref{f17}:
The two-dimensional stellar continuum is not symmetrical with the nebular emission profile.
Probable absorption of \ion{He}{i} at 4385 and 4921 \AA.

\textit{PN~G052.2$+$07.6} (K~4$-$10), Fig.~\ref{f04}:
This is a poorly studied object. We only detect
\ion{O}{v} in absorption at 5114 \AA.

\textit{PN~G055.6$+$02.1} (He~1$-$2), Fig.~\ref{f009}:
Clear absorption of \ion{He}{i} at 4471 and 4921 \AA,
\ion{C}{iii} at 4650 \AA,
\ion{N}{iii} at 4097 and 4511-15 \AA,
\ion{Si}{iv} at 4089 and 4116 \AA.
Finally, absorption of \ion{O}{iii} at 5592 \AA\ is evident.
This is a strange feature for a late star. On the other hand, 
\ion{C}{iii} emission at 5696 \AA\ is evident (stellar origin?).
The spectrum displays a P-Cygni profile at 5876 \AA.

\textit{PN~G060.0$-$04.3} (A~68), Fig.~\ref{f16}:
Good nebular subtraction, but poor S/N.
There is a clear absorption at H$\alpha$.

\textit{PN~G068.7$+$14.8} (Sp~4$-$1), Fig.~\ref{f16B}:
The spectra have a very high S/N. There are absorptions of \ion{He}{ii} at 5412 \AA\ (doubtful) and
\ion{O}{v} at 5114 \AA.
On the other hand, strong emission at 4650-58 \AA, \ion{C}{iii} at 5696,  \ion{C}{iv} 5801-12 \AA.
The FWHM of this emission line is double that of the nebular ones.
In addition, \ion{O}{iii} at 5592 \AA\ is present in emission.
These features, according Section~\ref{red-arm},
are characteristic of  a O(H)3-4.
Hence the CSPN of Sp~4-1 is probably an early O(H)-type star.

\begin{figure*}[h]
   \centering
   \includegraphics[width=0.95\textwidth]{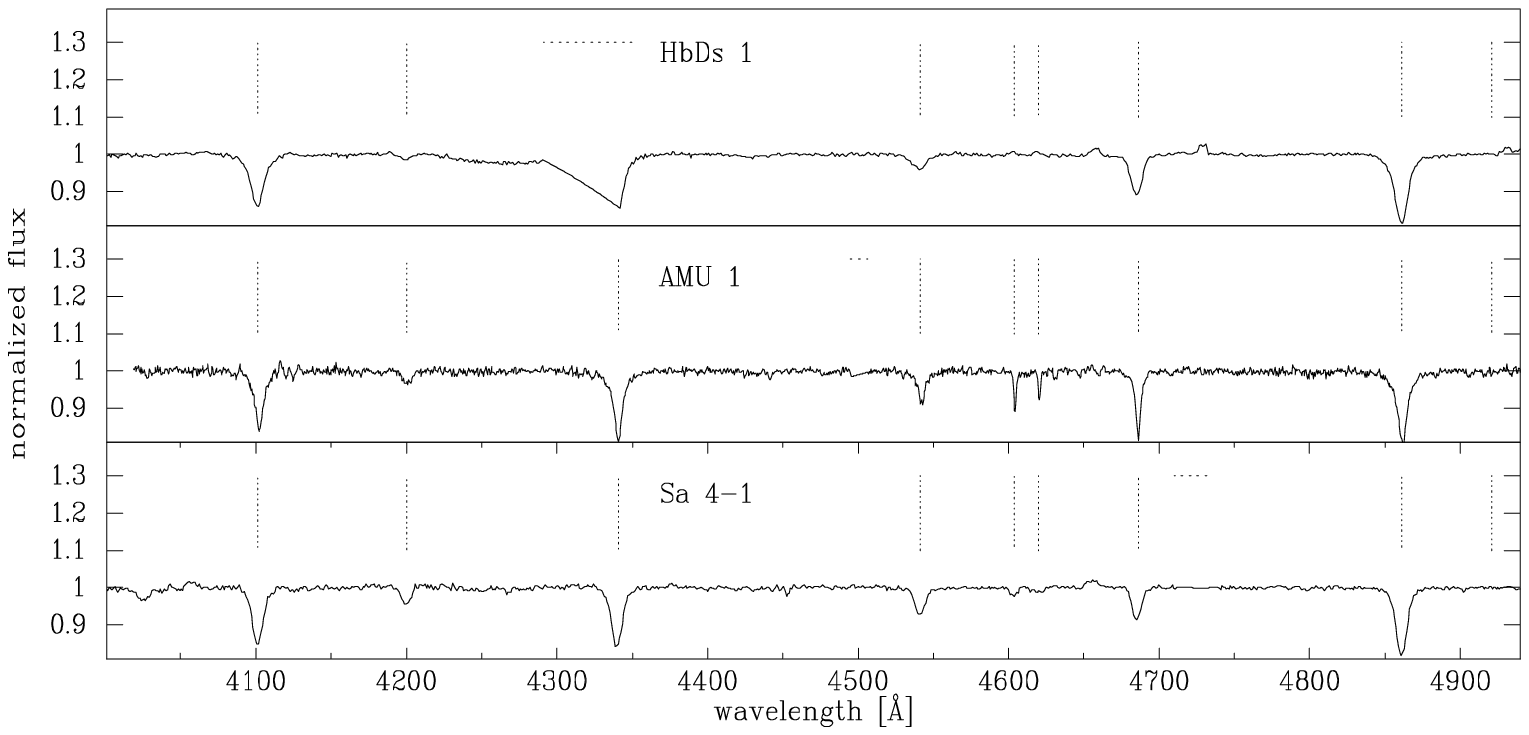}
   \includegraphics[width=0.95\textwidth]{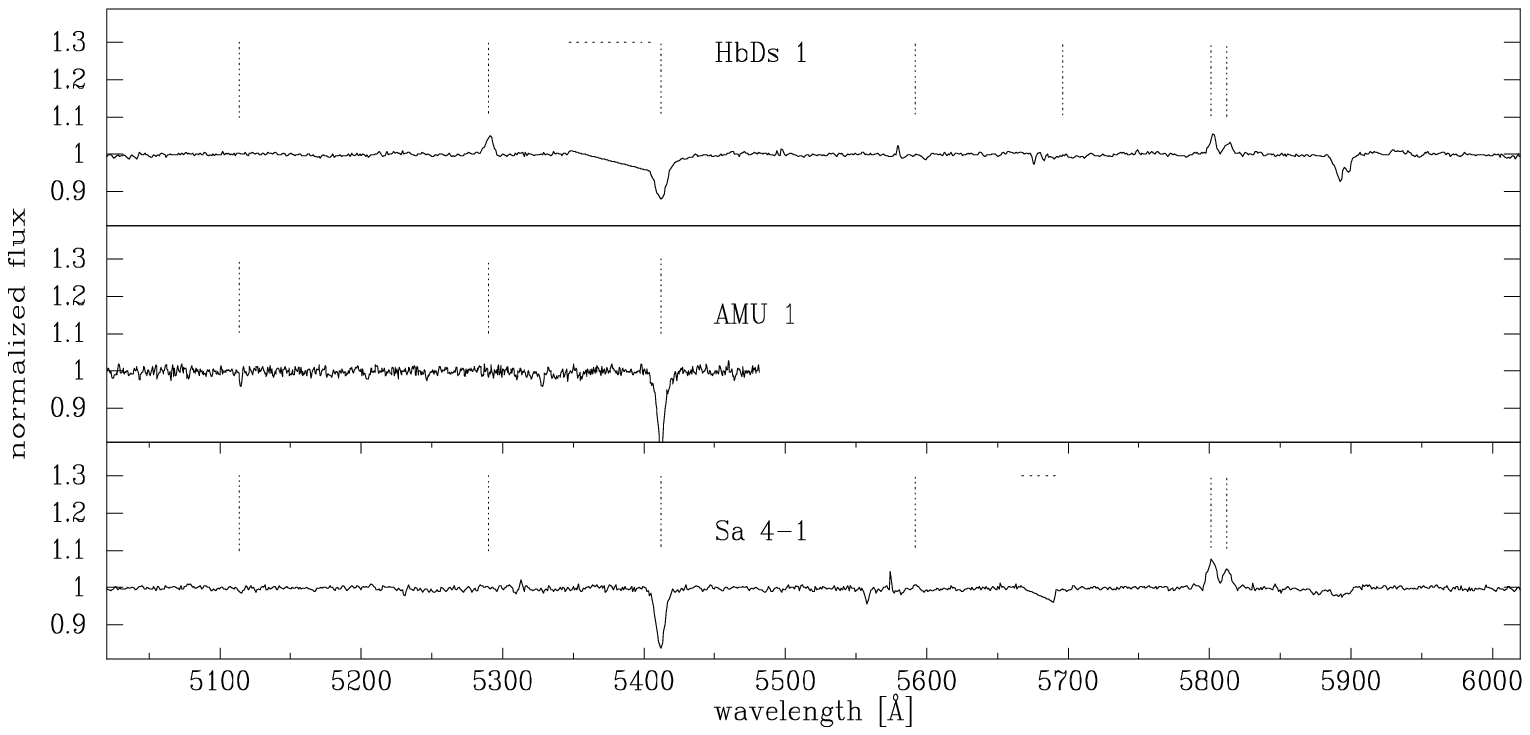}
      \caption[]{Normalized spectra of early O-type CSPN 
                           (see Table~\ref{iones}).
                 The interstellar absorption bands at $\lambda$4428, the  complex at 5780 and 5890-6 are not indicated.
                 The most important spectral features (absorption and emission) identified are: H$\beta$, H$\gamma$, H$\delta$,
   \ion{He}{ii} $\lambda$4200, 4542, and 4686,
   \ion{N}{v} $\lambda$4604-19, for the blue spectral range (top panel), and \ion{O}{v}   $\lambda$5114,
 \ion{O}{vi}  $\lambda$5290,
 \ion{He}{ii} $\lambda$5412,
 \ion{O}{iii} $\lambda$5592,
 \ion{C}{iii} $\lambda$5696,
 \ion{C}{iv}  $\lambda$5801-12, for the red spectral range (bottom panel).
}
         \label{f01}
   \end{figure*}

\begin{figure*}
   \centering
   \includegraphics[width=0.95\textwidth]{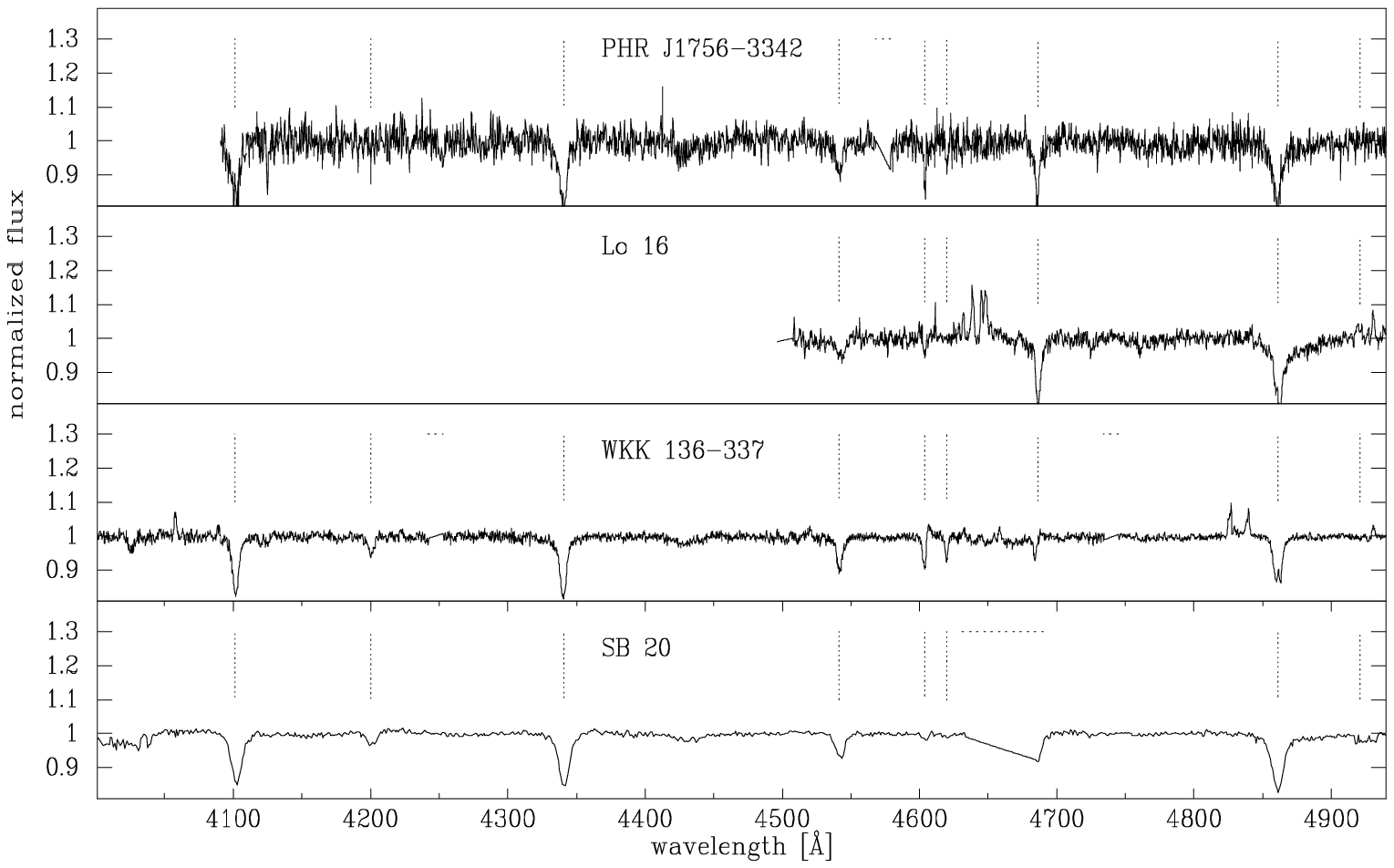}
   \includegraphics[width=0.95\textwidth]{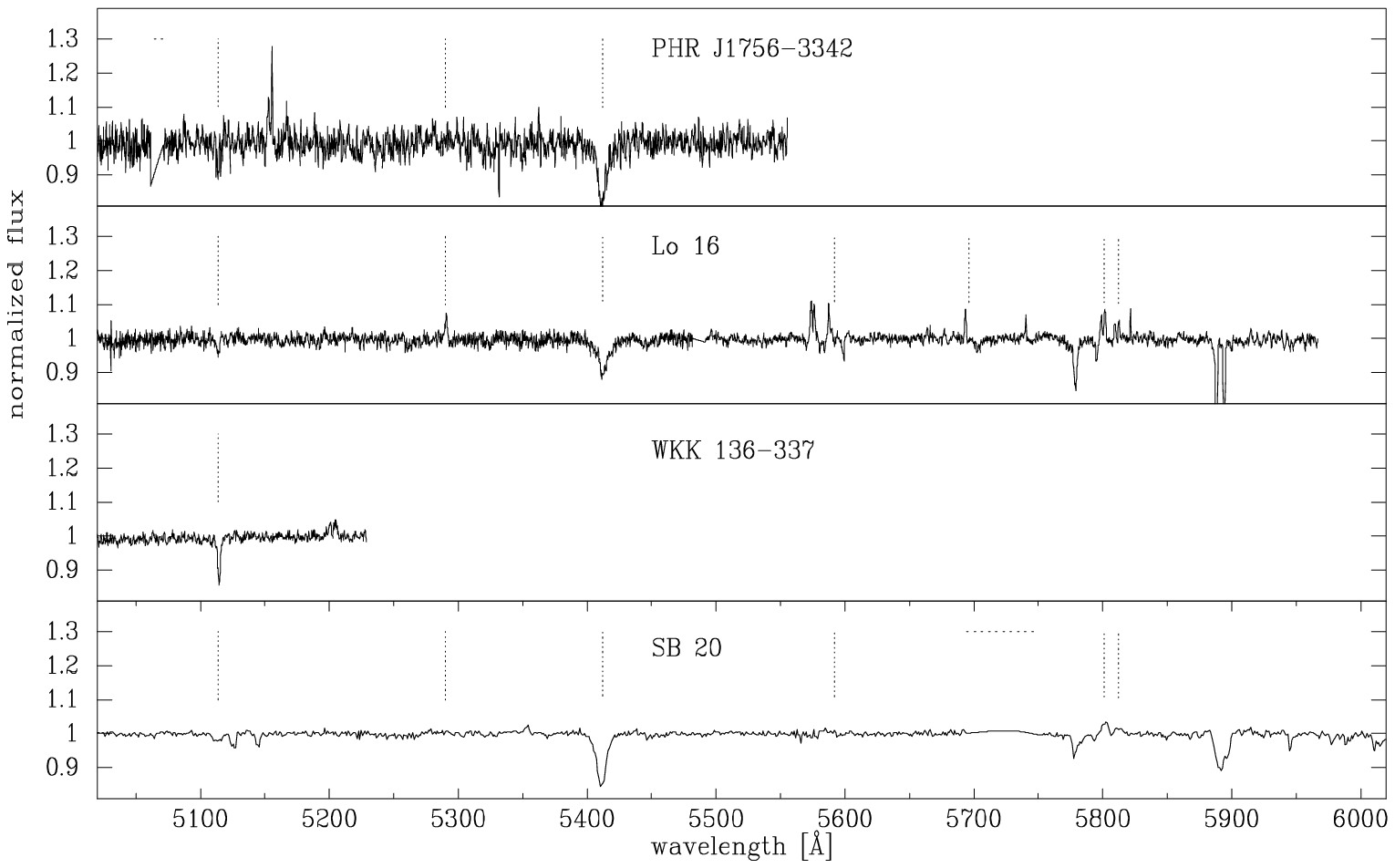}
     \caption[]{Same as Fig.~\ref{f01}. 
  }
         \label{f03}
   \end{figure*}

\begin{figure*}
   \centering
   \includegraphics[width=0.95\textwidth]{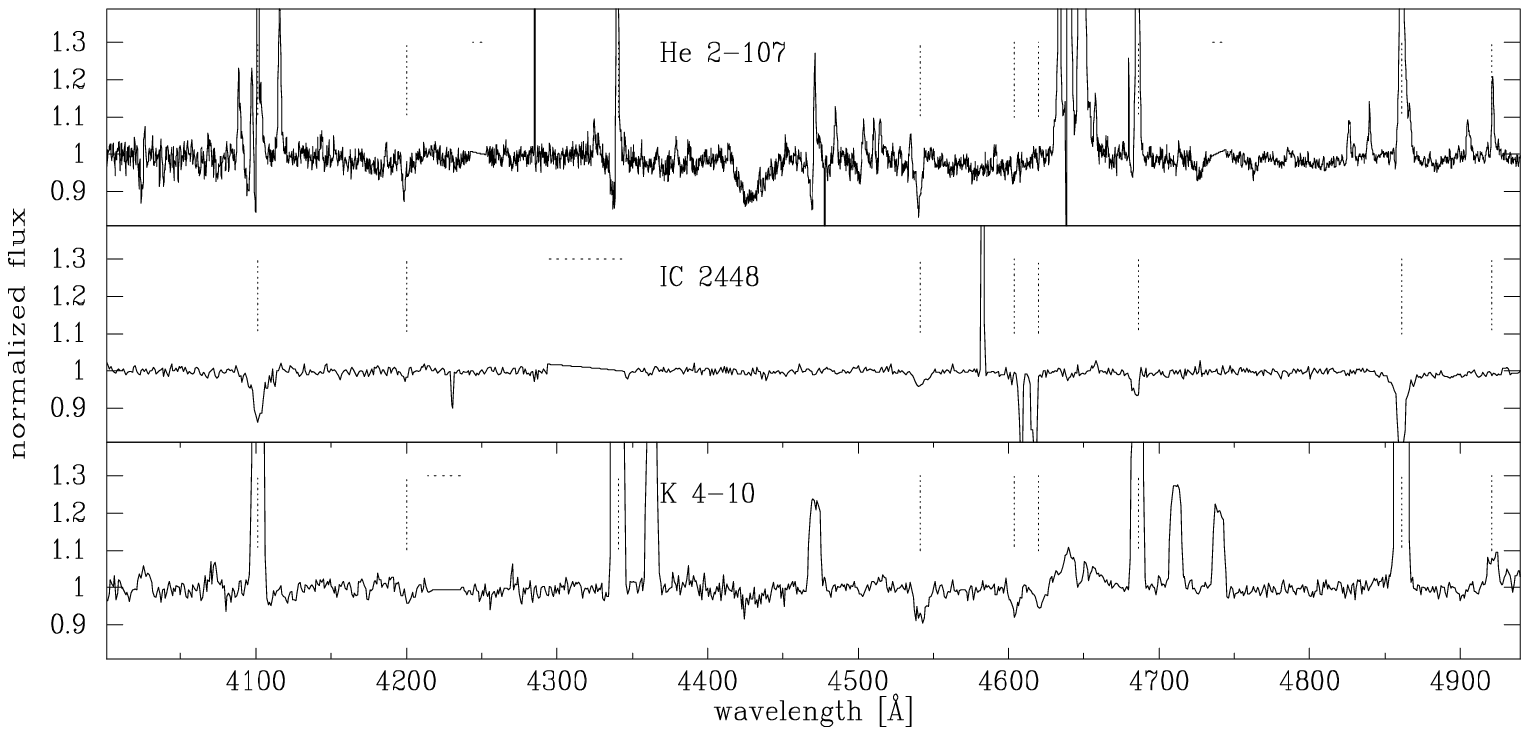}
   \includegraphics[width=0.95\textwidth]{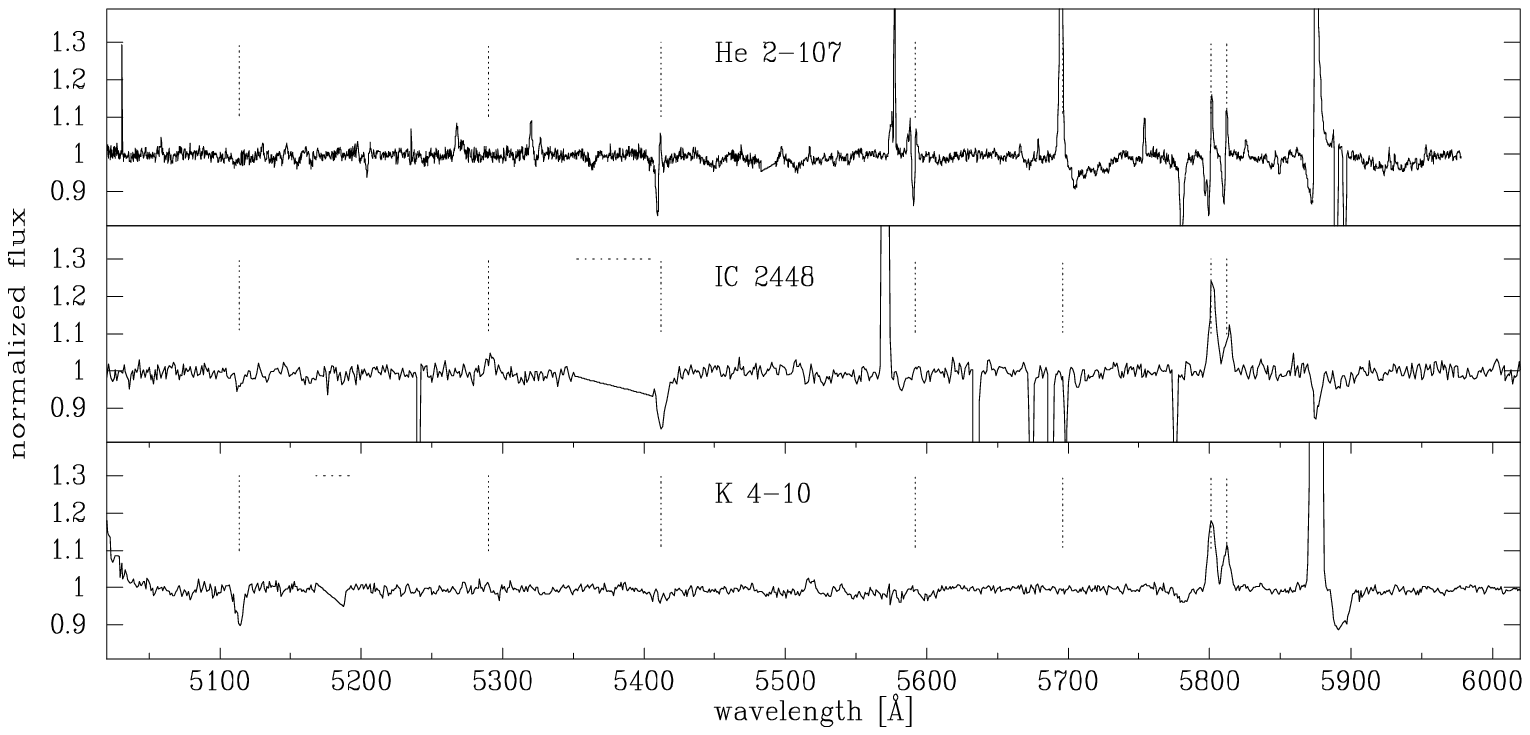}
     \caption[]{Same as Fig.~\ref{f01}.  
}
         \label{f04}
   \end{figure*}

\begin{figure*}
   \centering
   \includegraphics[width=0.95\textwidth]{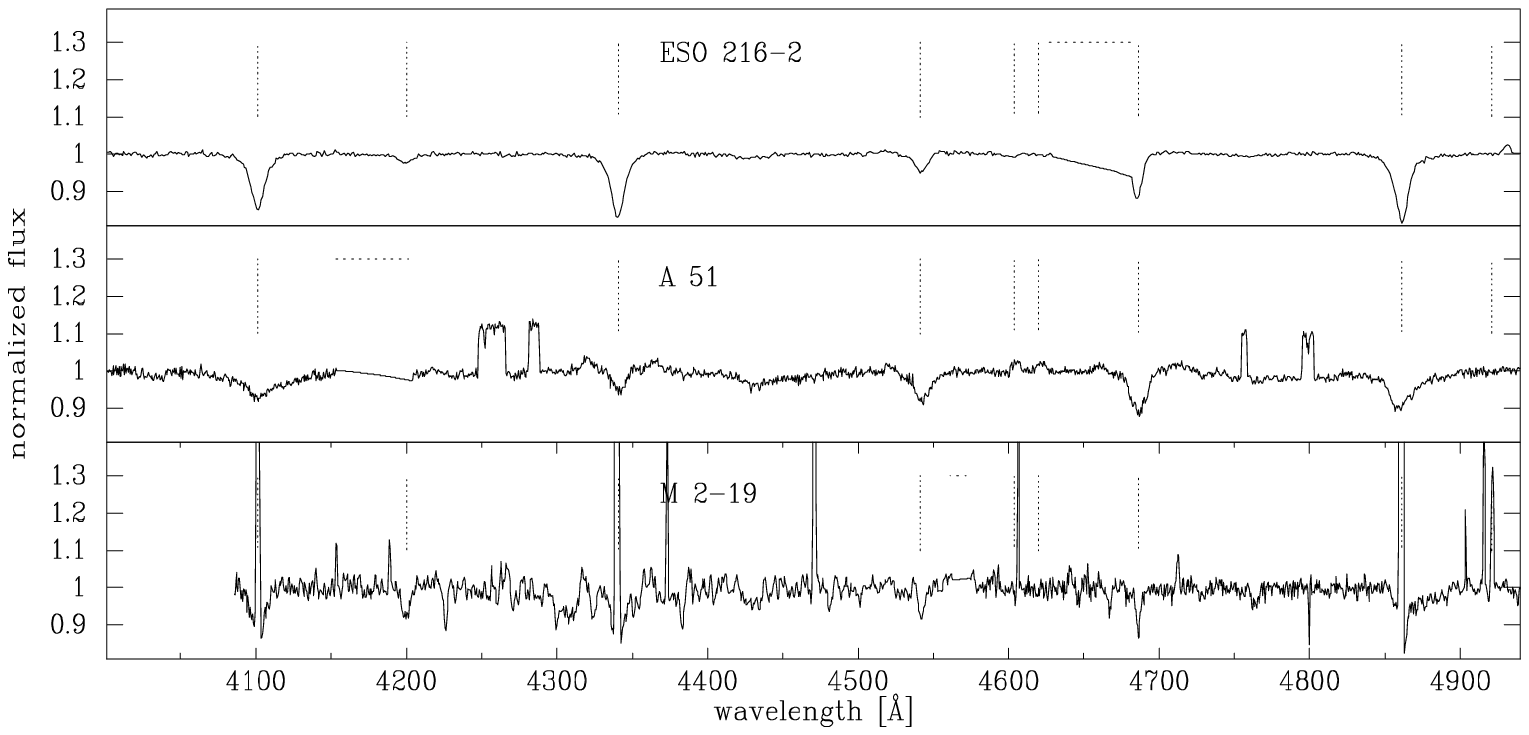}
   \includegraphics[width=0.95\textwidth]{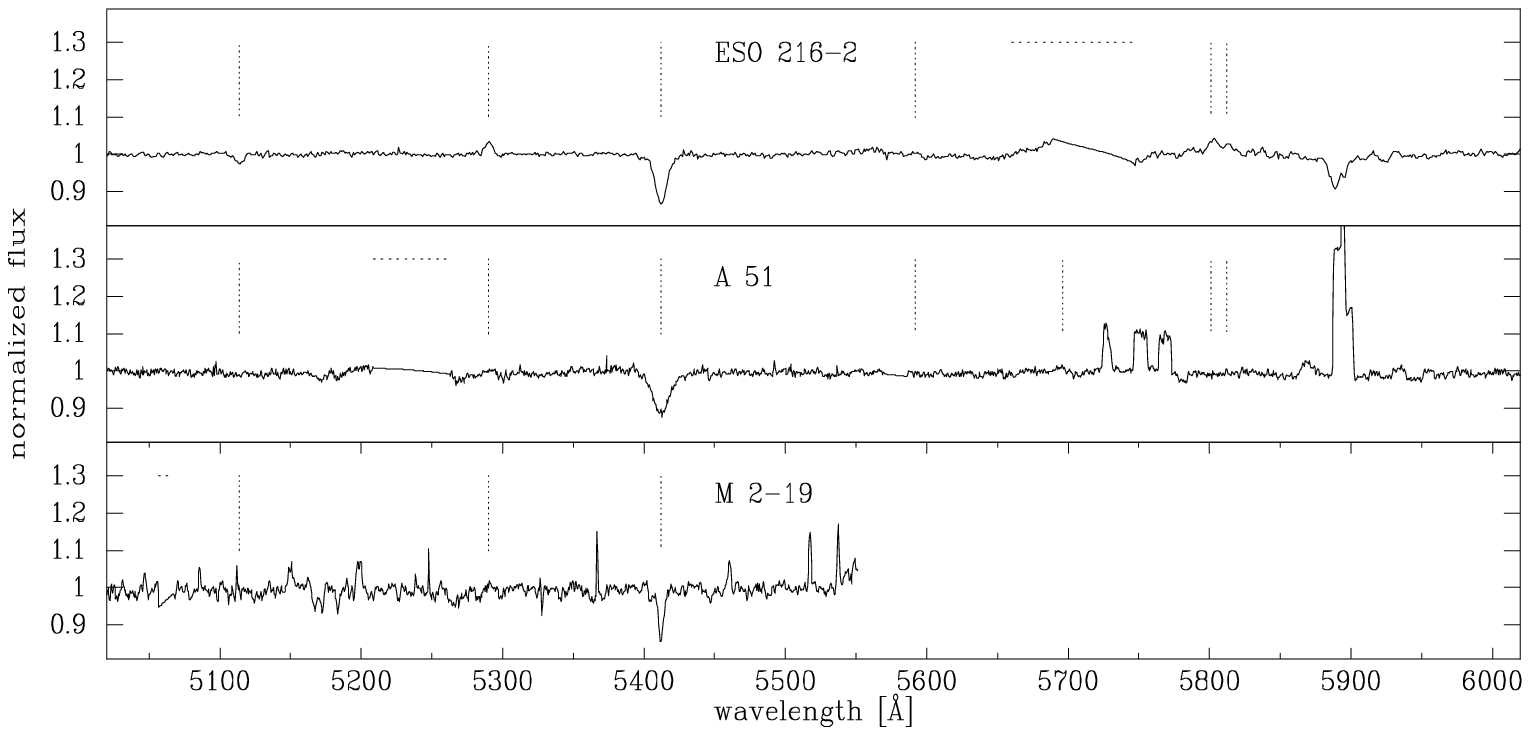}
       \caption[]{Same as Fig.~\ref{f01}. 
    }
         \label{f05}
   \end{figure*}

\begin{figure*}
   \centering
   \includegraphics[width=0.95\textwidth]{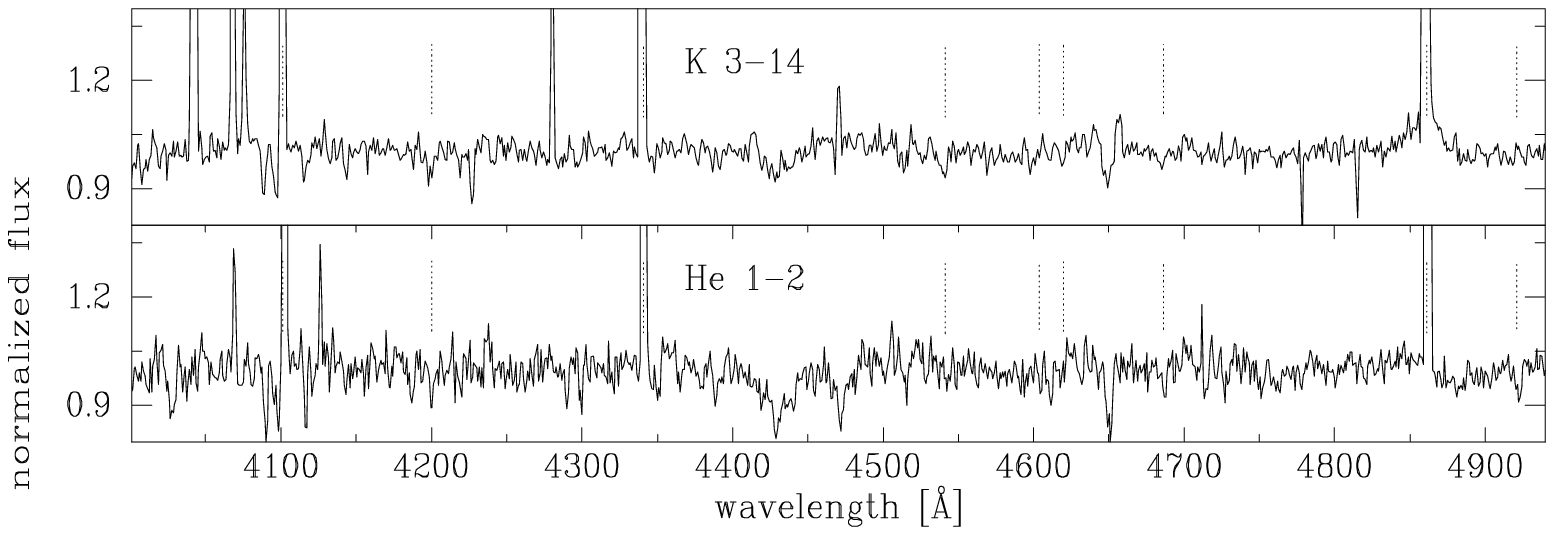}
   \includegraphics[width=0.95\textwidth]{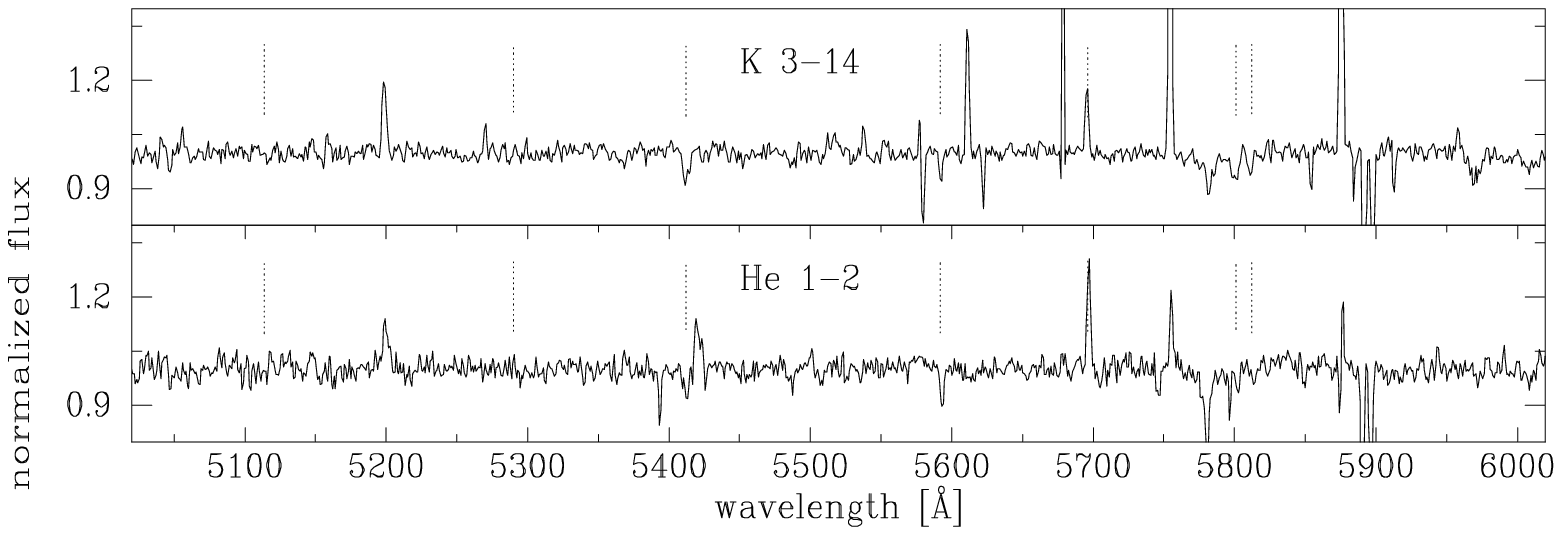}
      \caption[]{      Normalized spectra of  O-type CSPN 
                           (see Table~\ref{iones}),
                           in these objects we could not detect the Balmer serie.
                 The interstellar absorption bands at $\lambda$4428, the  complex at 5780 and 5890-6 are not indicated.
                 The most important spectral features (absorption and emission) identified are:
   \ion{He}{ii} $\lambda$4200, 4542, and 4686, 
   \ion{N}{v} $\lambda$4604-19 for the  blue spectral range (top panel), and same features as Fig.~\ref{f01} for the red spectral range (bottom panel).
}
         \label{f009}
   \end{figure*}

\begin{figure*}
   \centering
   \includegraphics[width=0.95\textwidth]{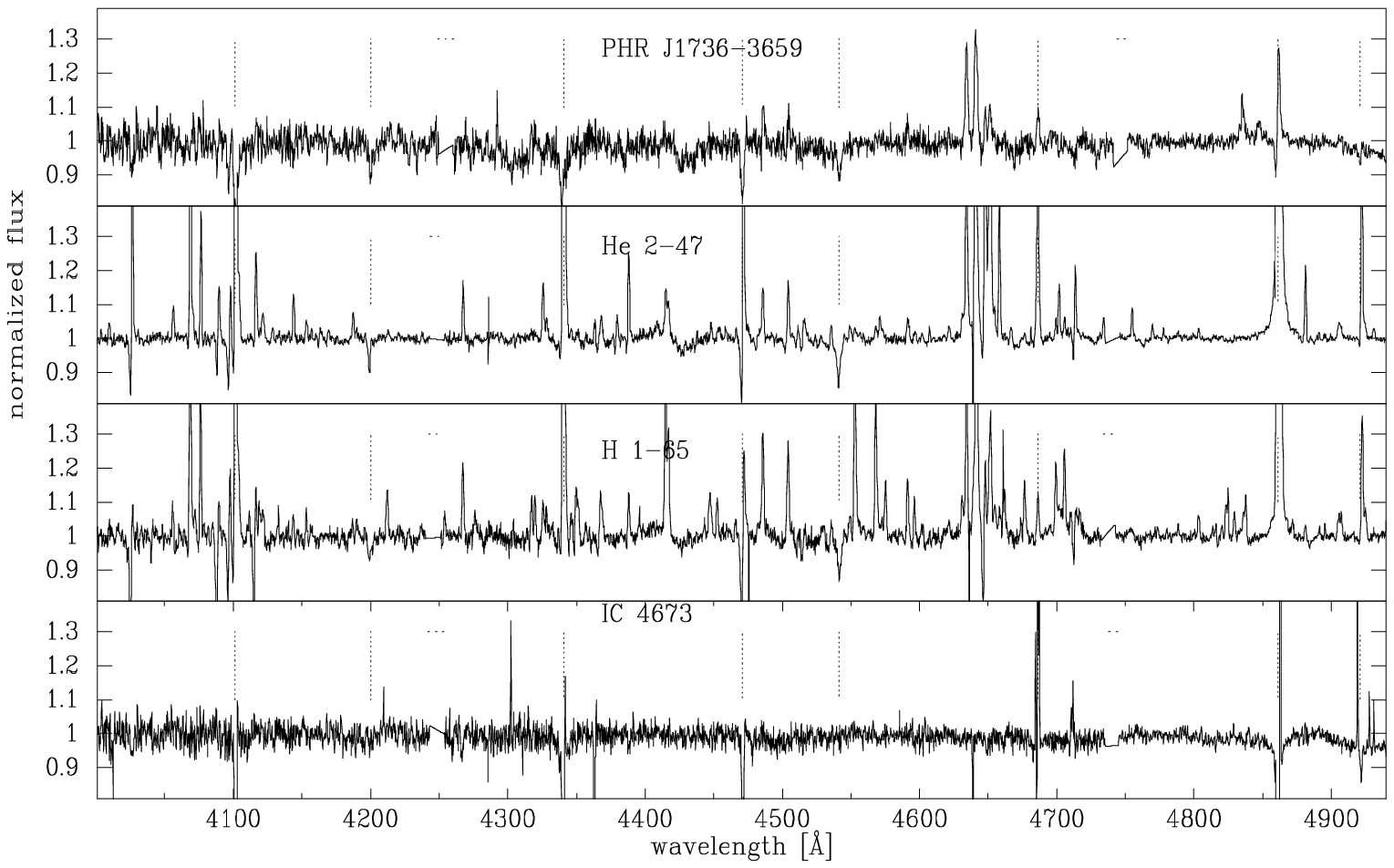}
   \includegraphics[width=0.95\textwidth]{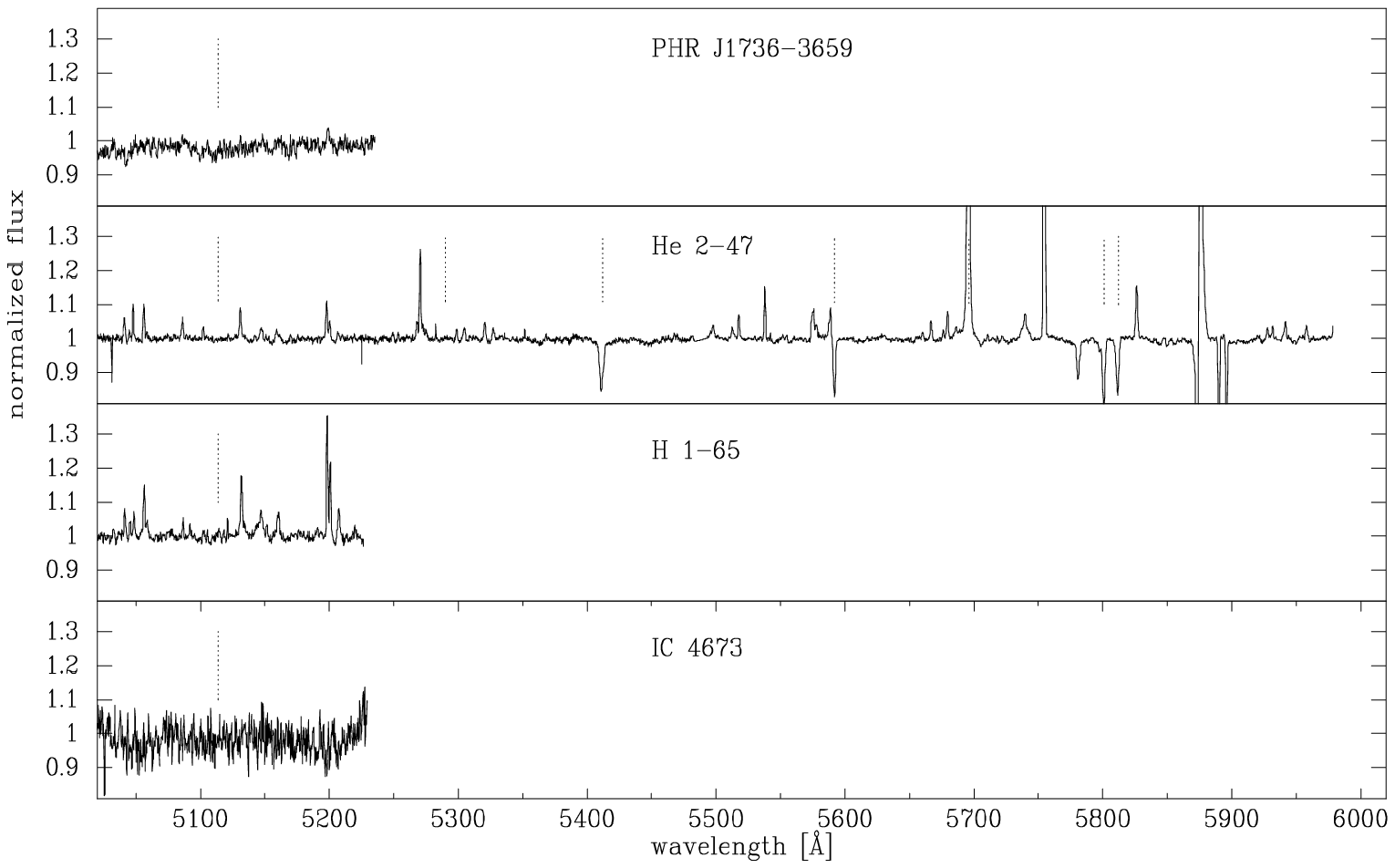}
      \caption[]{Normalized spectra of late O-type CSPN 
                           (see Table~\ref{iones}).
                The interstellar absorption bands at $\lambda$4428, the  complex at 5780 and 5890-6 are not indicated.
                 The most important spectral features (absorption and emission) identified are:
   H$\beta$, H$\gamma$, H$\delta$,
  \ion{He}{i} $\lambda$4471 and $\lambda$4921,
   \ion{He}{ii} $\lambda$4200, 4542, and 4686, for the blue spectral range (top panel), and the same spectral features as Fig.~\ref{f01} for the red spectral range (bottom panel). 
}
         \label{f09A}
   \end{figure*}

\begin{figure*}
   \centering
   \includegraphics[width=0.95\textwidth]{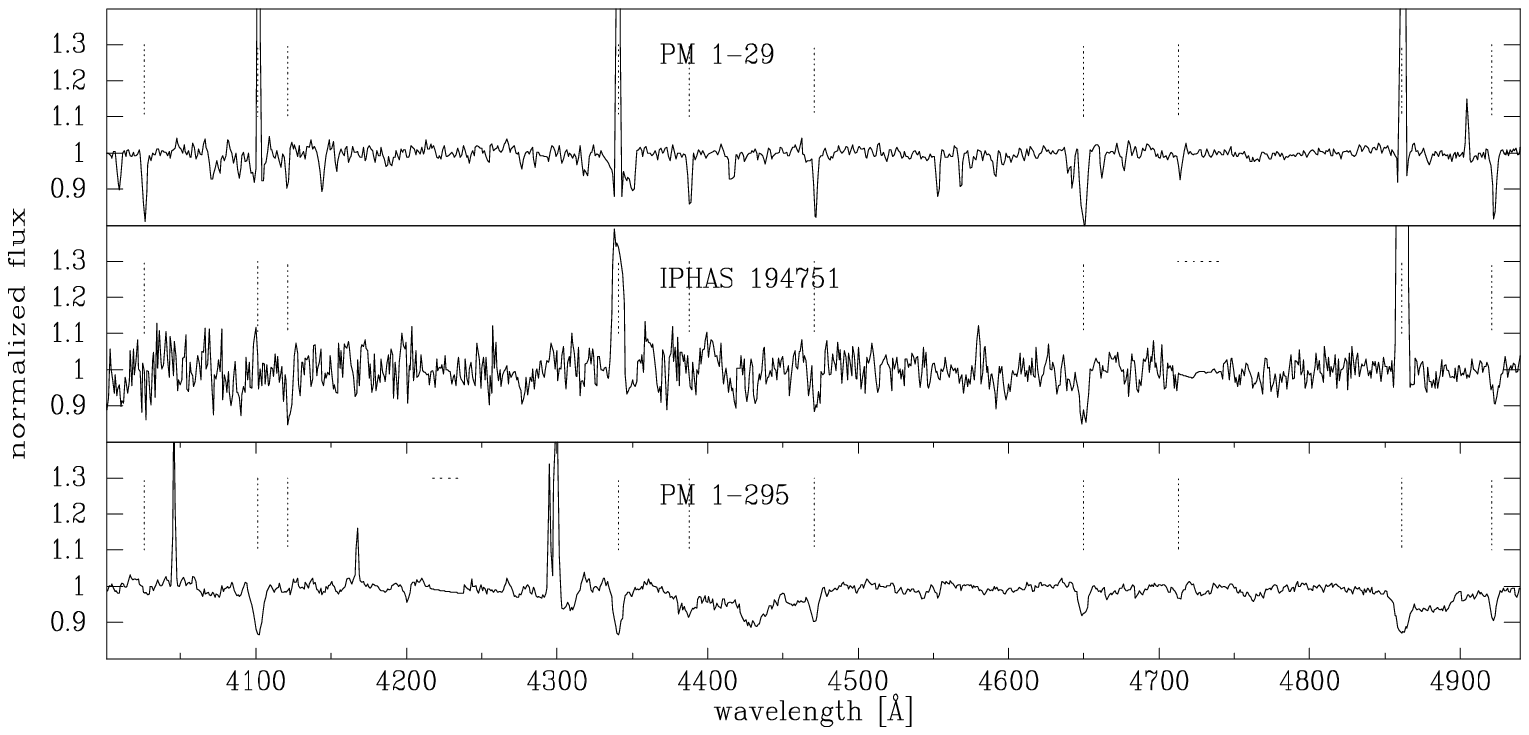}
   \includegraphics[width=0.95\textwidth]{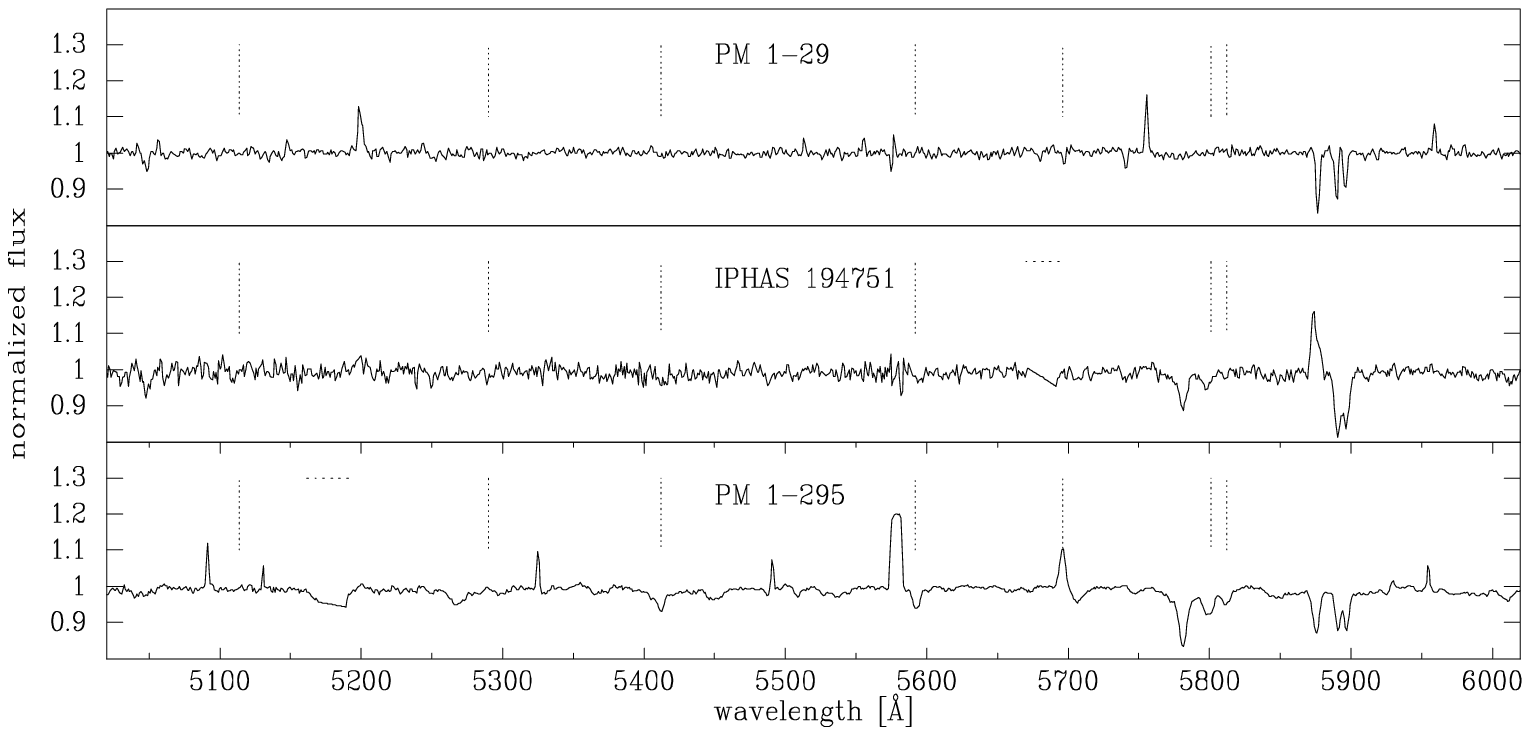}
      \caption[]{Normalized spectra of early B-type CSPN 
                           (see Table~\ref{iones}).
                 The interstellar absorption bands at $\lambda$4428, the  complex at 5780 and 5890-6 are not indicated.
                 The most important spectral features (absorption and emission) identified are:
   H$\beta$, H$\gamma$, H$\delta$,
   \ion{He}{i} $\lambda$4026, 4121, 4387, 4471, 4713 and 4921,
   \ion{C}{iii} $\lambda$ 4650, for the blue spectral range (top panel), and same features as Fig.~\ref{f01} for the red spectral range (bottom panel). 
}
         \label{f10}%%%%%%%%%%%%%%%%%%   Fig. 12
   \end{figure*}

\begin{figure*}
   \centering
   \includegraphics[width=0.95\textwidth]{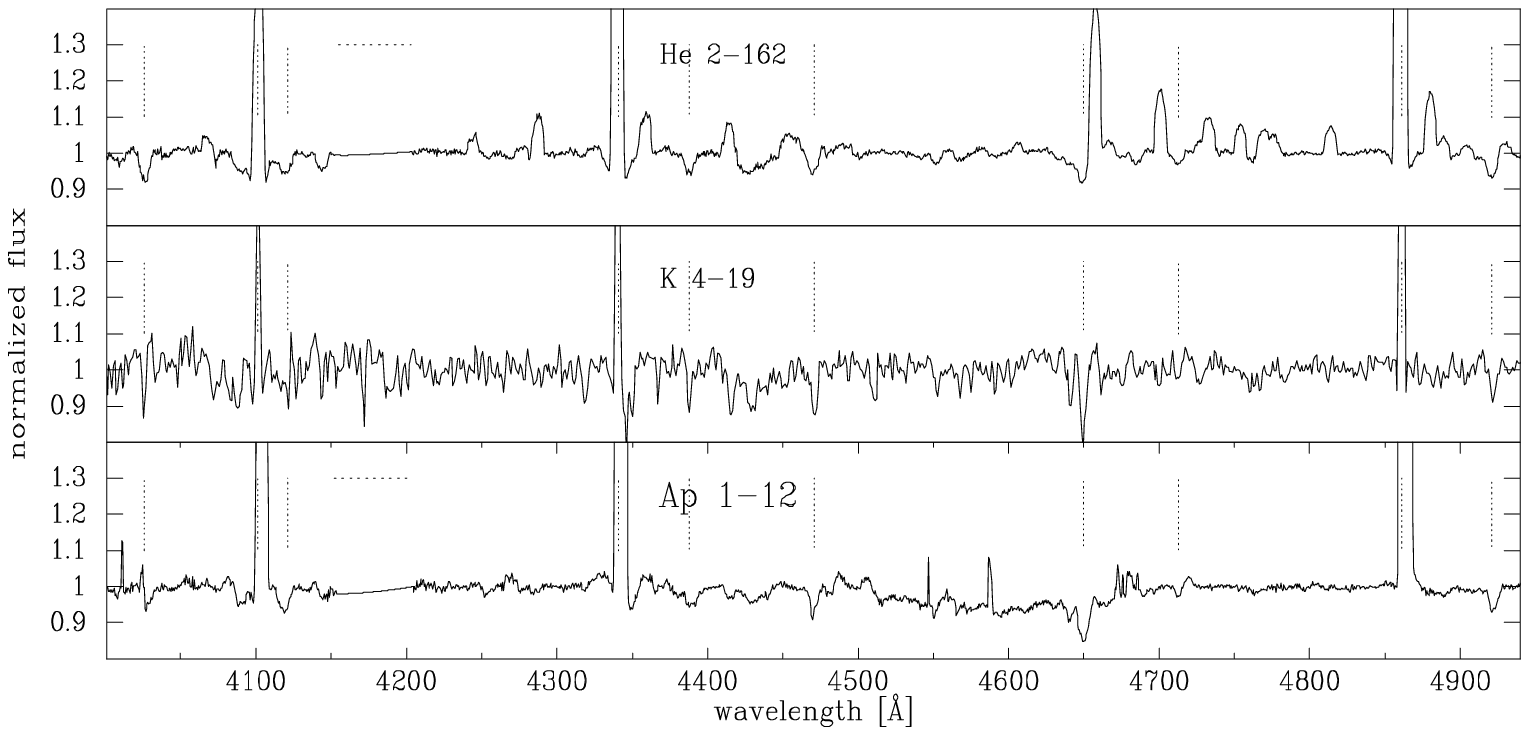}
   \includegraphics[width=0.95\textwidth]{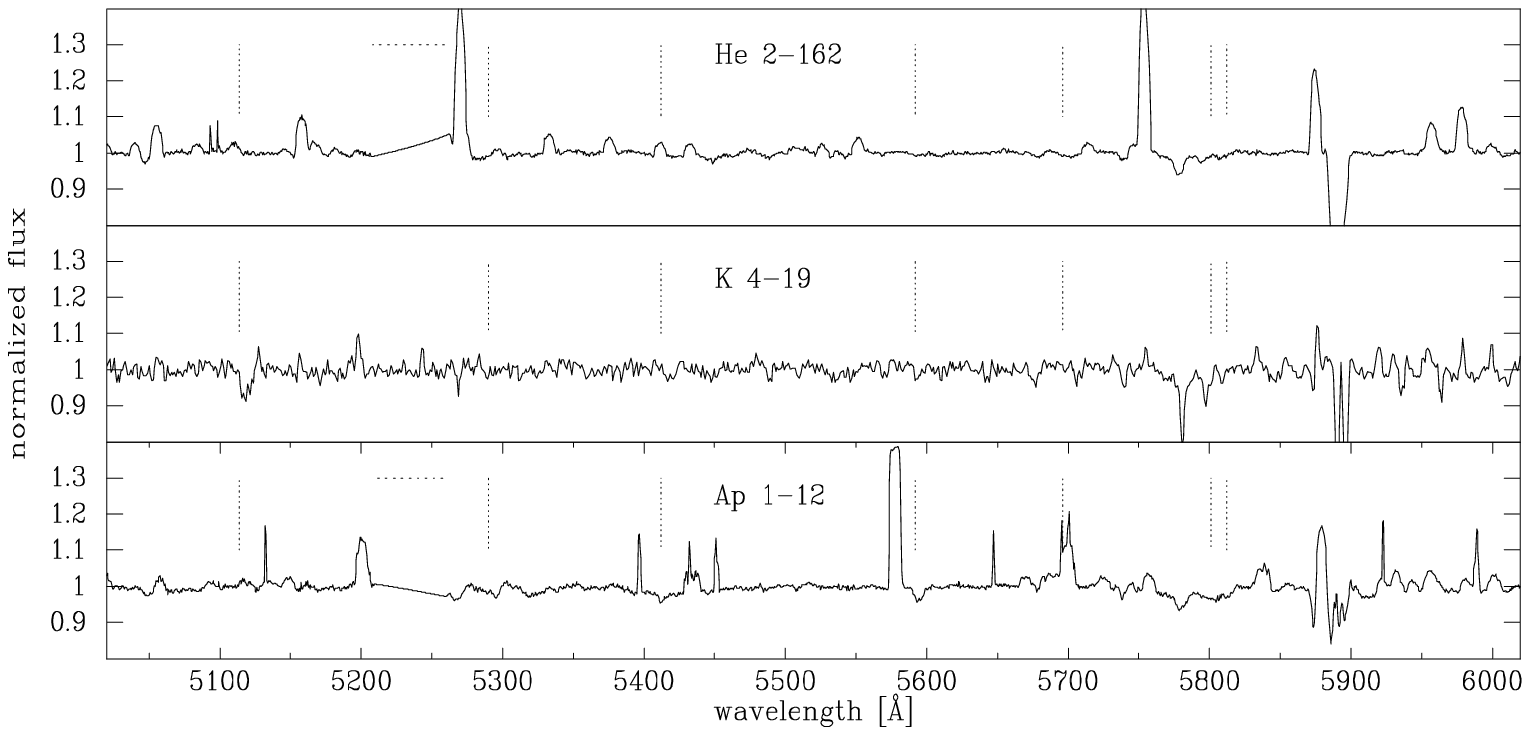}
      \caption[]{Same as Fig.~\ref{f10}.
}
         \label{f11}%%%%%%%%%%%%%%%%%%   Fig. 13
   \end{figure*}

\begin{figure*}
   \centering
   \includegraphics[width=0.95\textwidth]{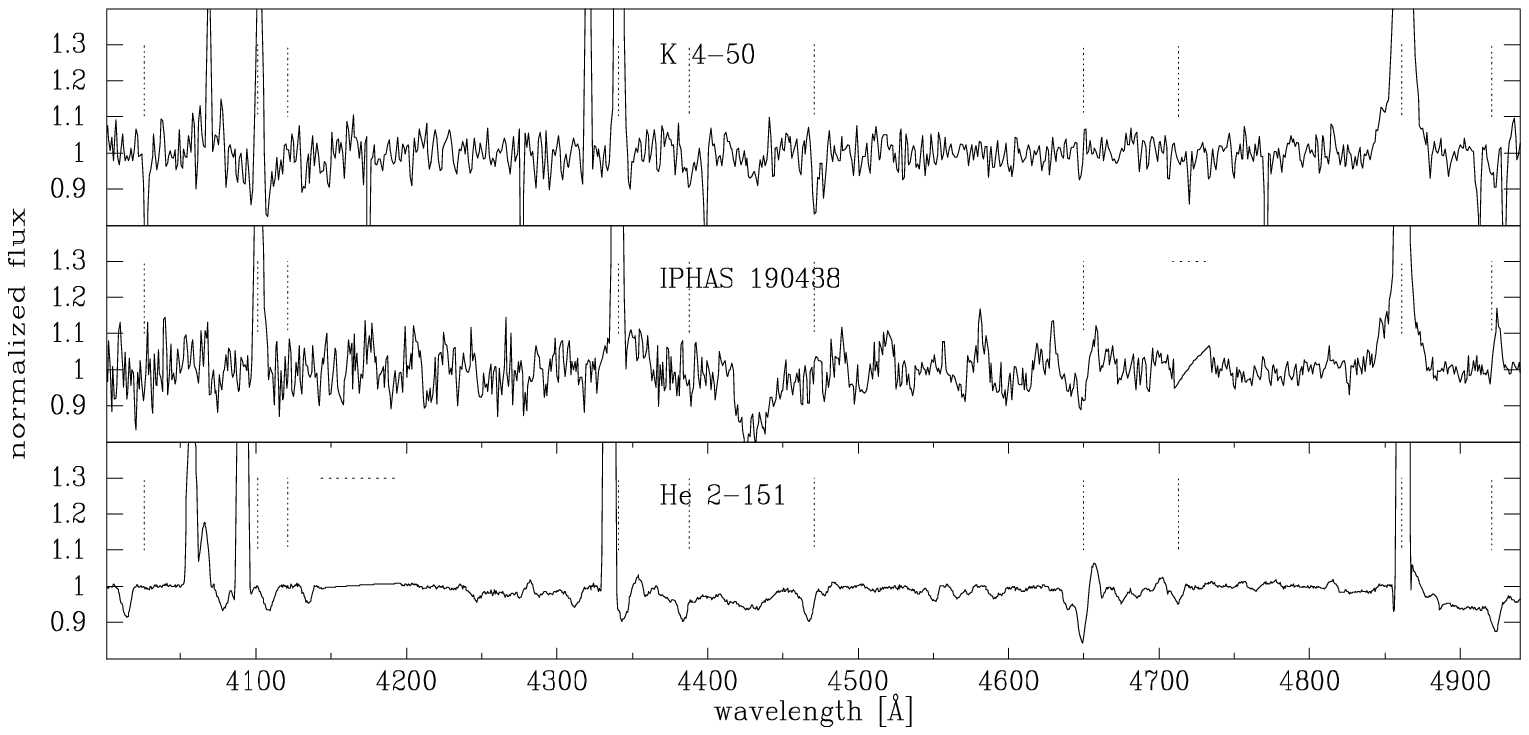}
   \includegraphics[width=0.95\textwidth]{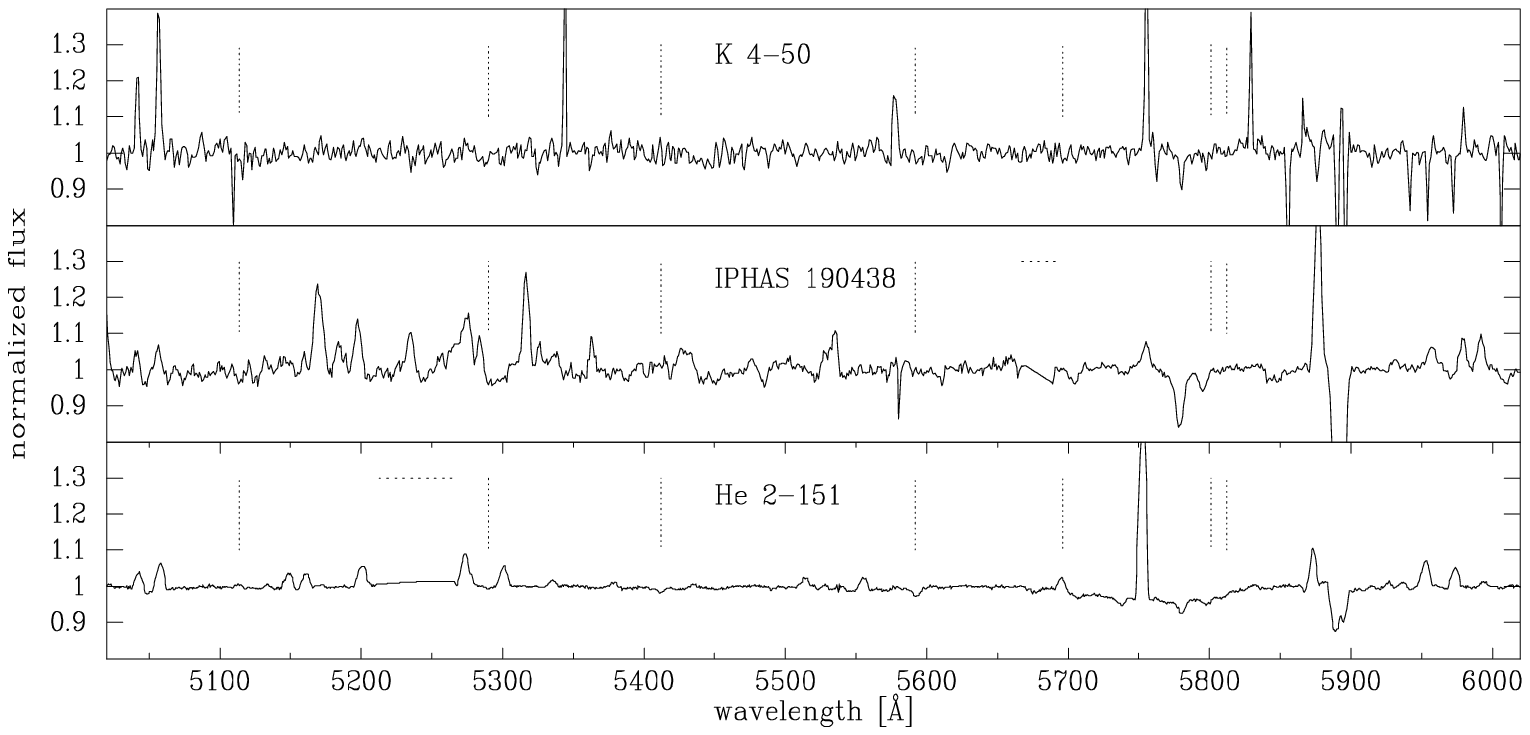}
      \caption[]{Same as Fig.~\ref{f10}.
}
         \label{f09}%%%%%%%%%%%%%%%%%%   Fig. 14
   \end{figure*}

\begin{figure*}
   \centering
   \includegraphics[width=0.95\textwidth]{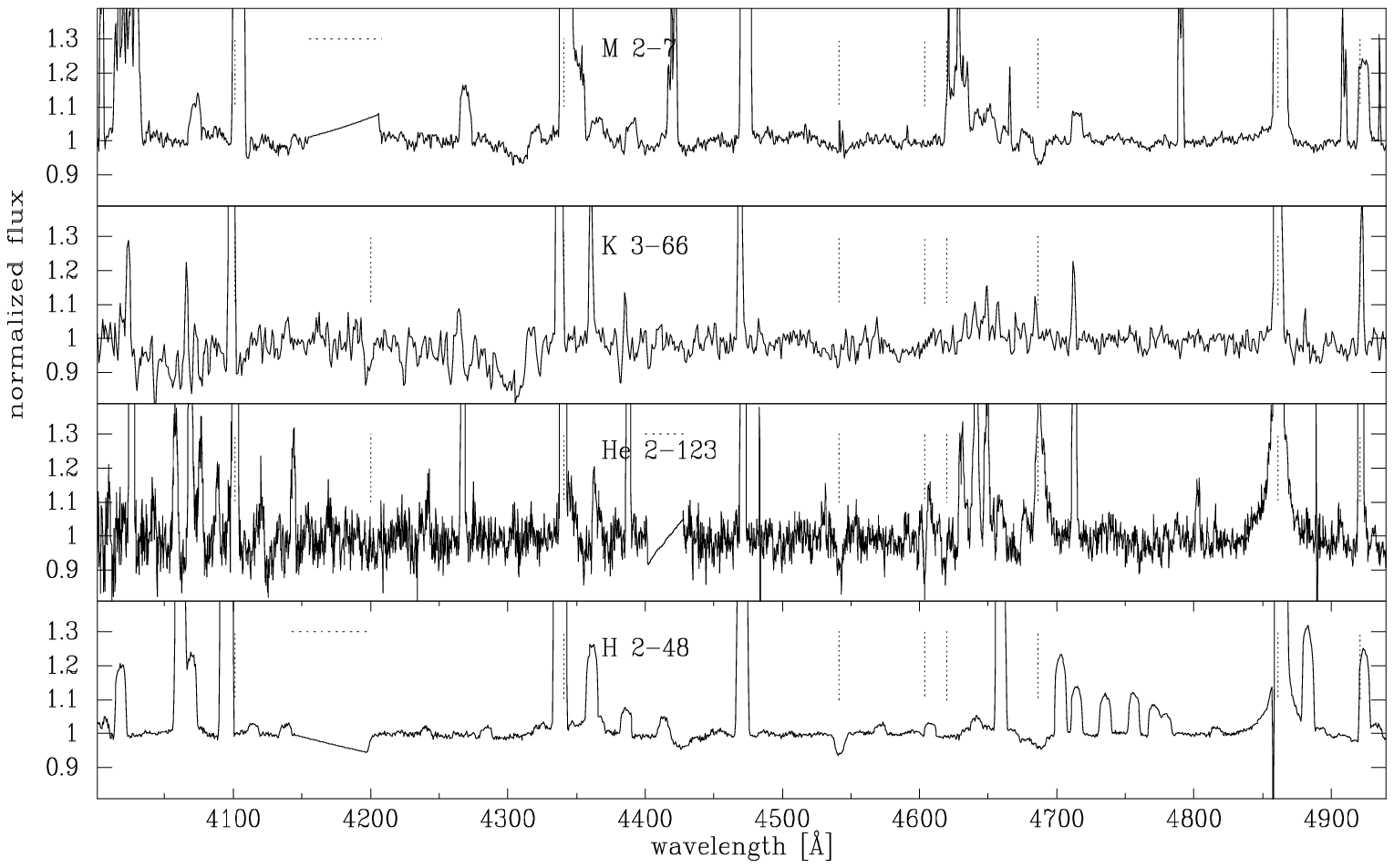}
   \includegraphics[width=0.95\textwidth]{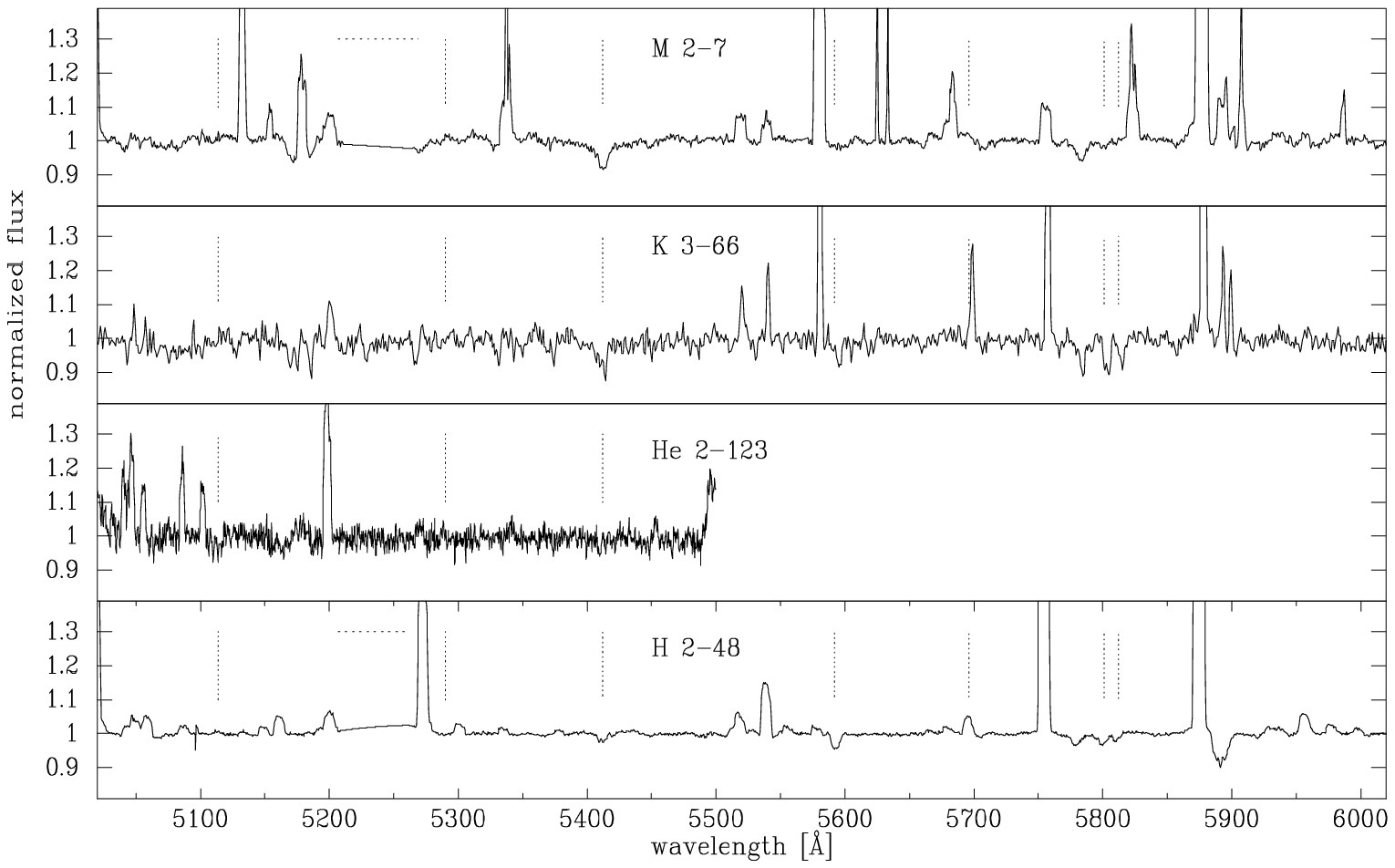}
      \caption[]{Normalized spectra of  O-type CSPN 
                           (see Table~\ref{iones}).
                           For these objects we only detect \ion{He}{ii} absorption lines.
   Blue and red spectral ranges features as Fig.~\ref{f01}.  
}
         \label{f14A} %%%%%%%%%%%%%%%%%%   Fig. 18
   \end{figure*}

\begin{figure*}
   \centering
   \includegraphics[width=0.95\textwidth]{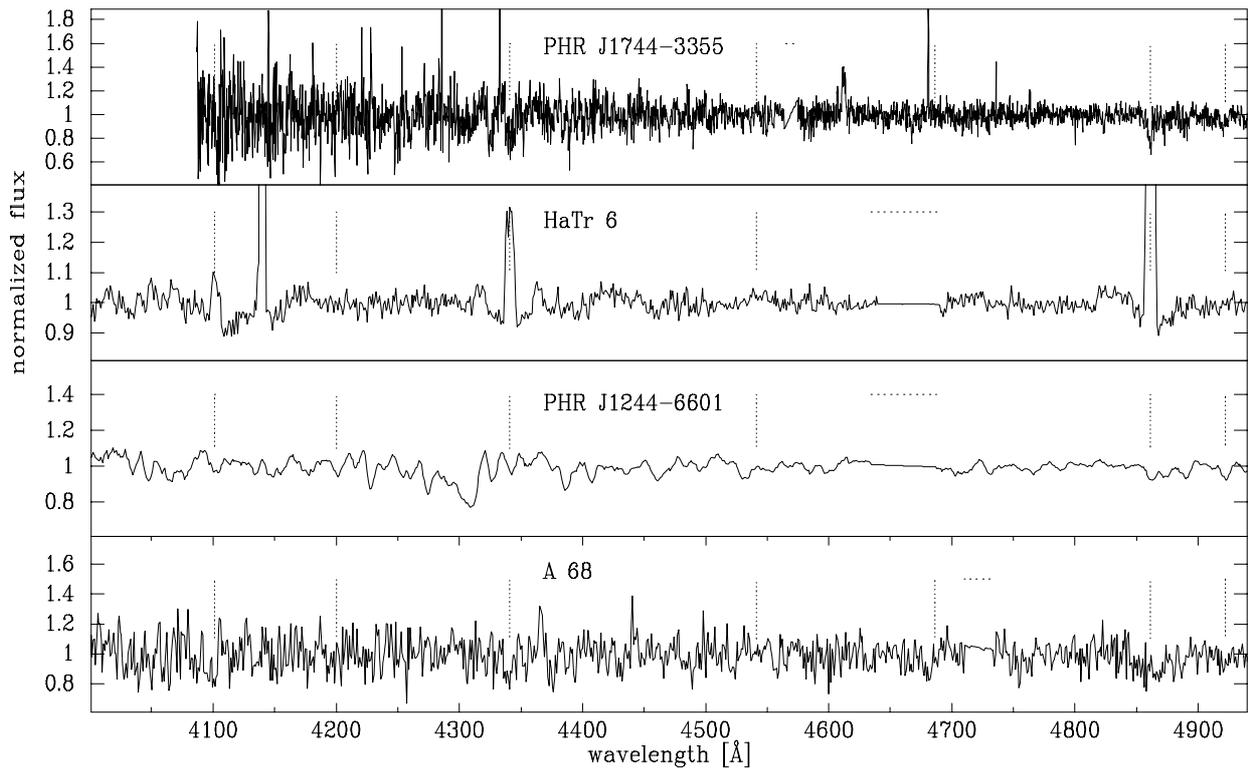}
   \includegraphics[width=0.95\textwidth]{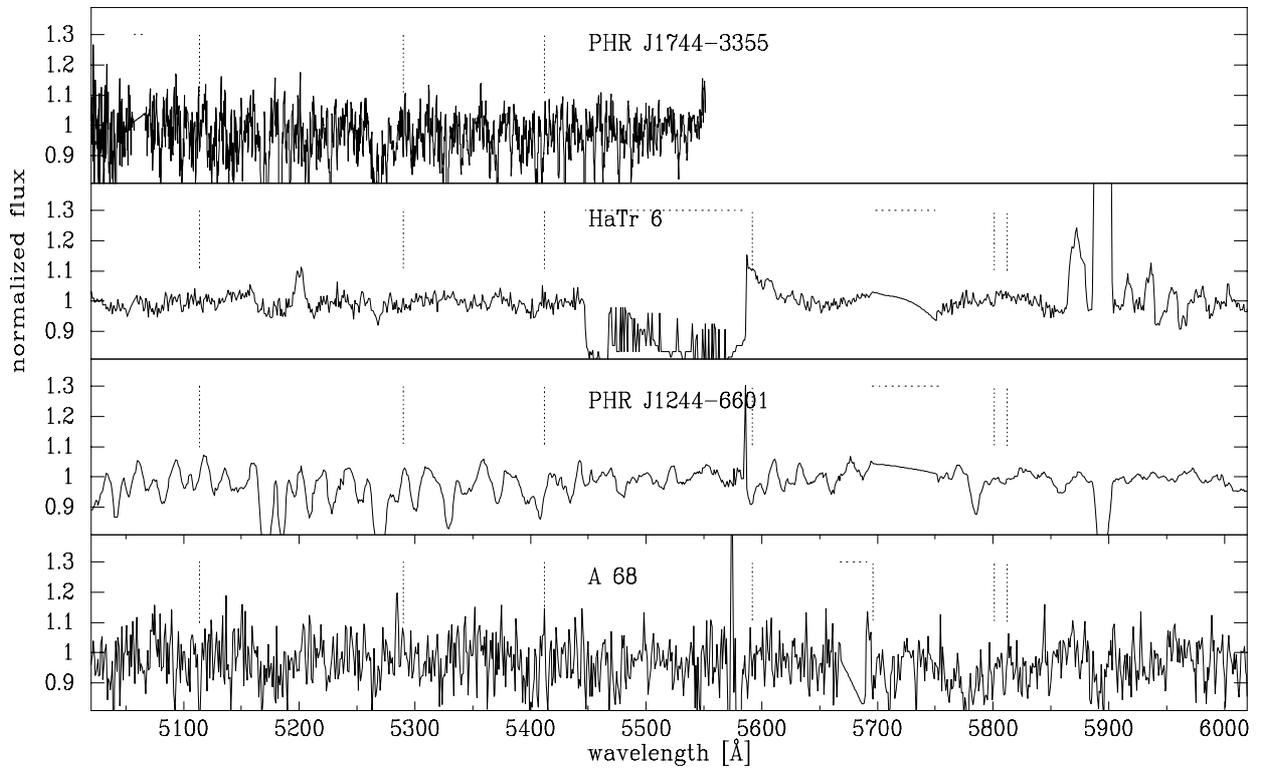}
      \caption[]{Normalized spectra of H-rich CSPN 
                           (see Table~\ref{iones}).
                           For these objects we only detect the Balmer series.
                 The interstellar absorption bands at $\lambda$4428, the  complex at 5780 and 5890-6 are not indicated.
                 Blue and red spectral ranges features as Fig.~\ref{f01}.
}
         \label{f16}%%%%%%%%%%%%%%%%%%   Fig. 20
   \end{figure*}

\begin{figure*}
   \centering
   \includegraphics[width=0.95\textwidth]{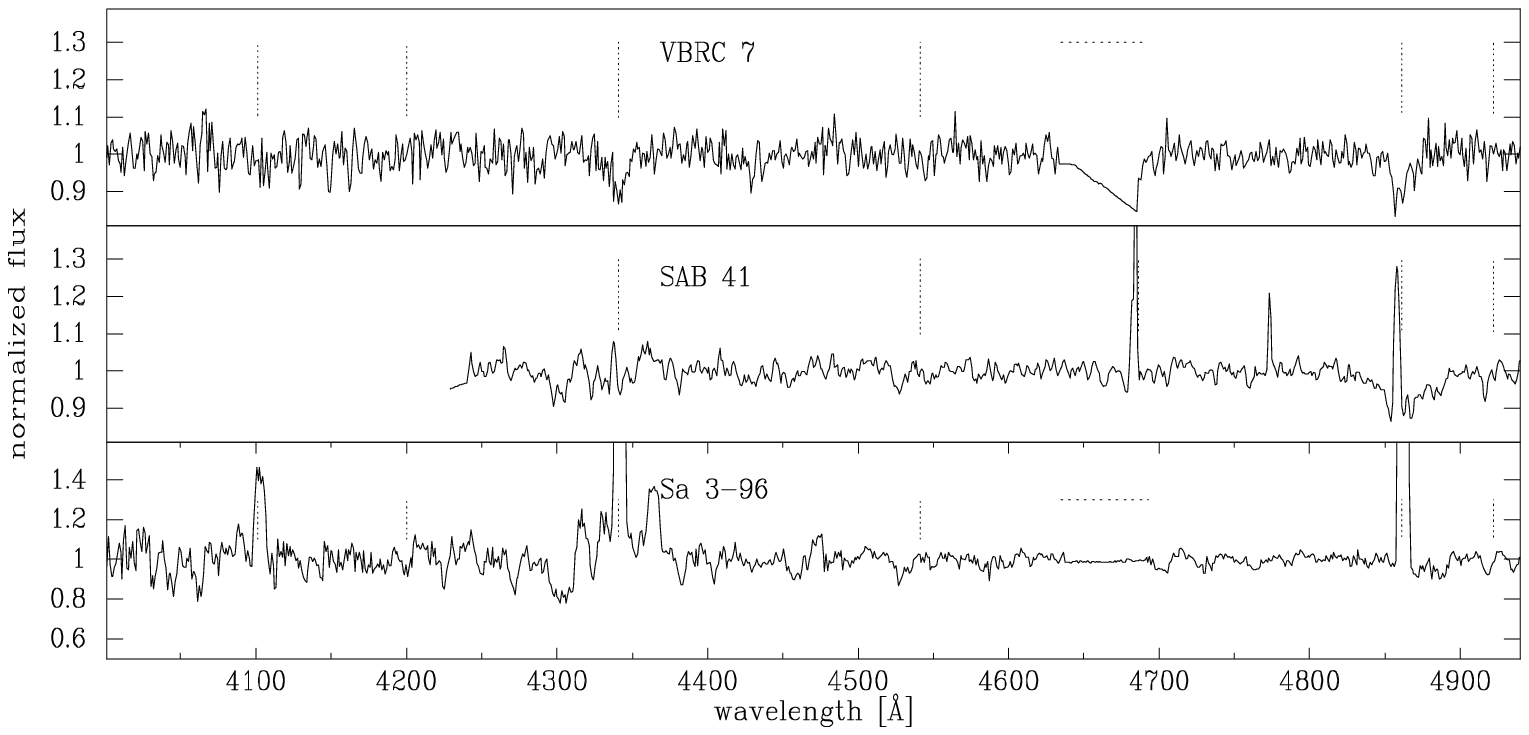}
   \includegraphics[width=0.95\textwidth]{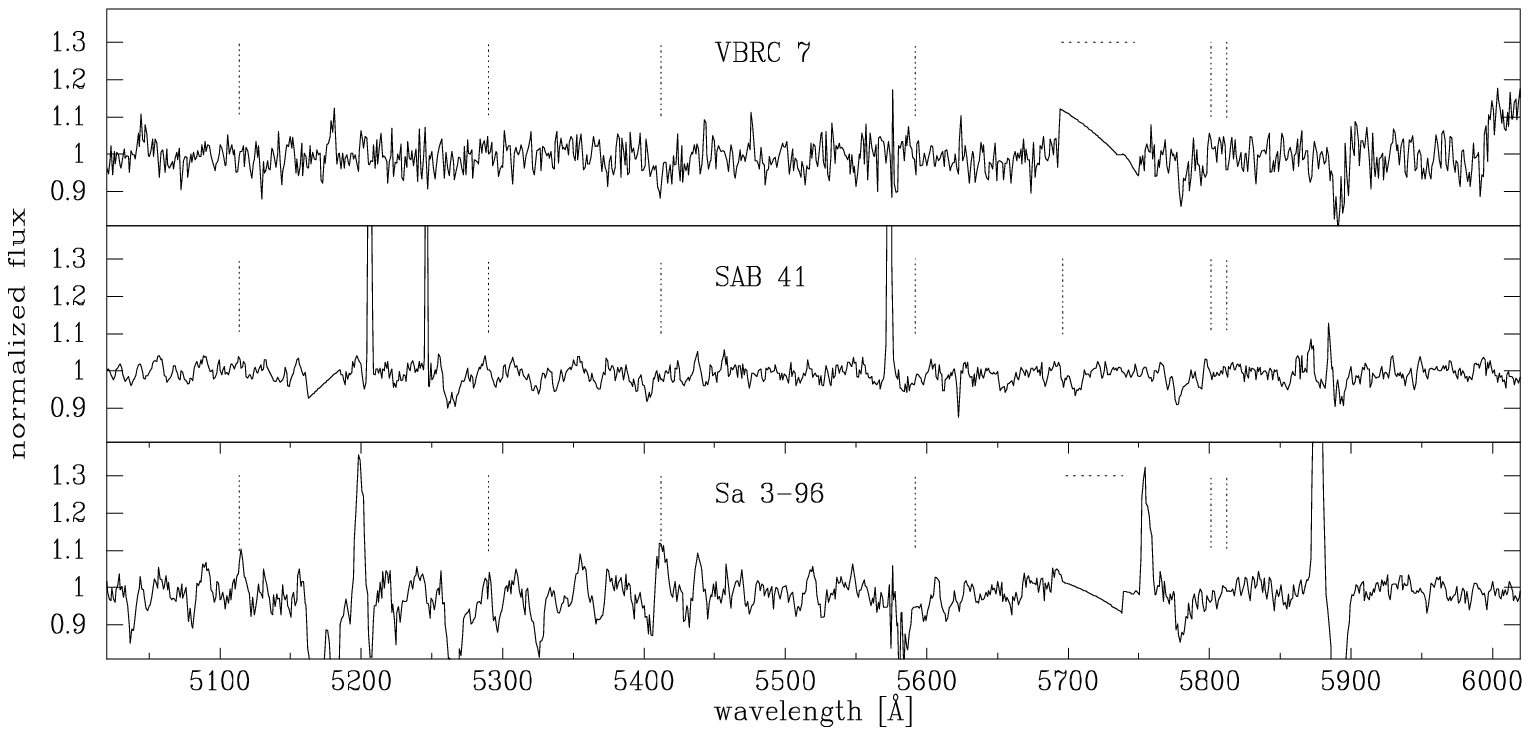}
      \caption[]{Same as Fig.~\ref{f16}.
      }
         \label{f16A} %%%%%%%%%%%%%%%%%%   Fig. 21
   \end{figure*}

\begin{figure*}
   \centering
   \includegraphics[width=0.95\textwidth]{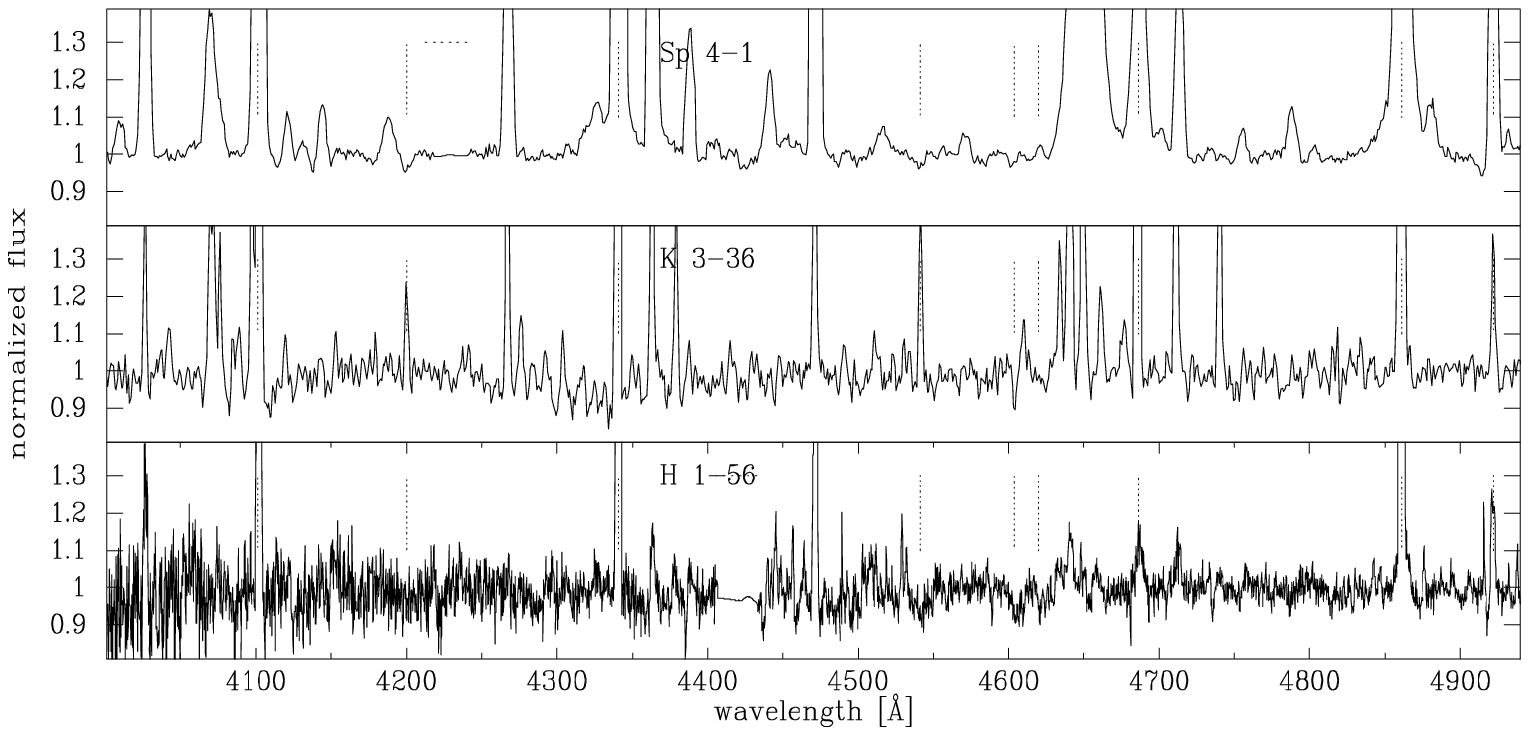}
   \includegraphics[width=0.95\textwidth]{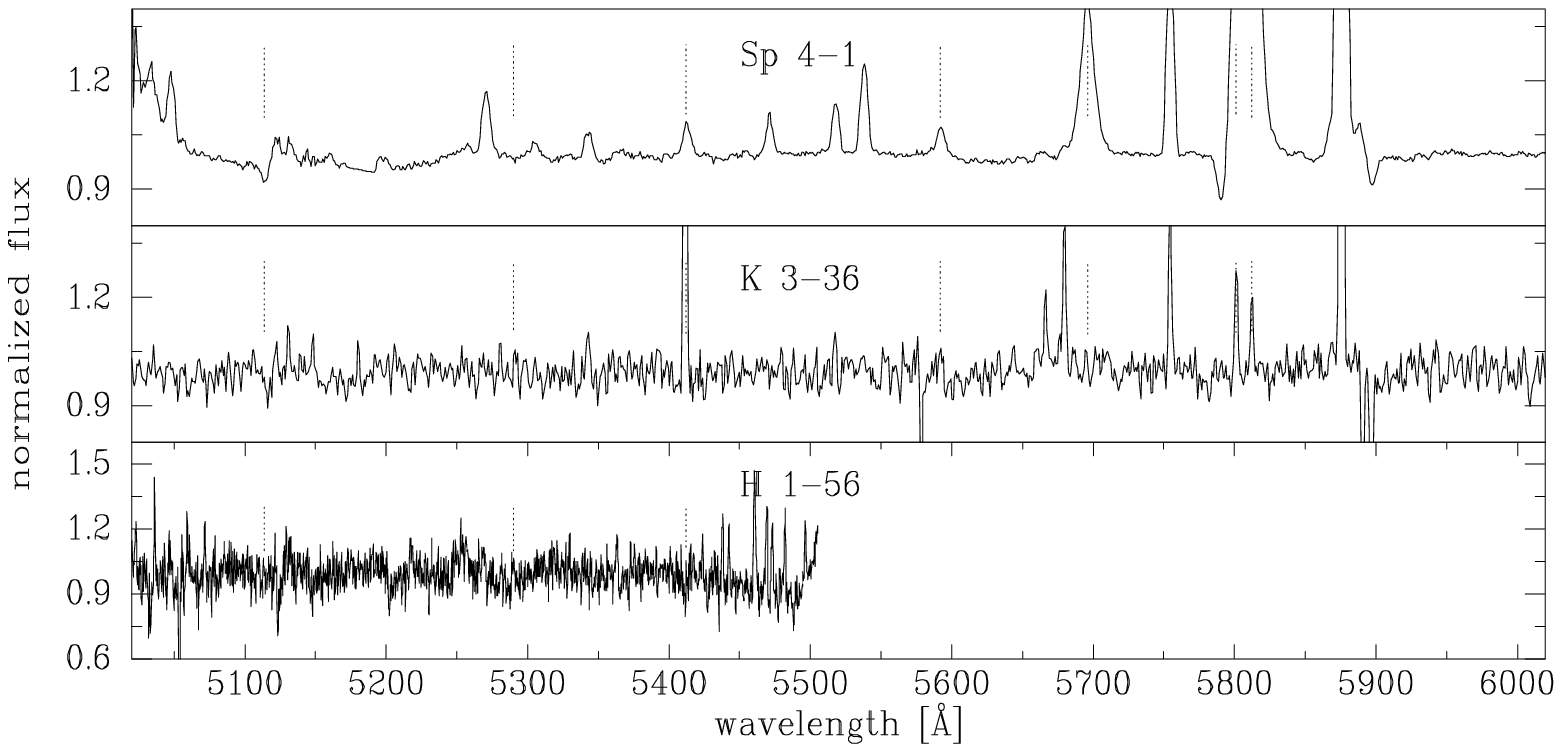}
      \caption[]{Normalized spectra of possible O-type CSPN 
                           (see Table~\ref{iones}).
  Blue and red spectral ranges features as Fig.~\ref{f01}.
}
         \label{f16B}%%%%%%%%%%%%%%%%%%   Fig. 22
   \end{figure*}

\begin{figure*}
   \centering
   \includegraphics[width=0.95\textwidth]{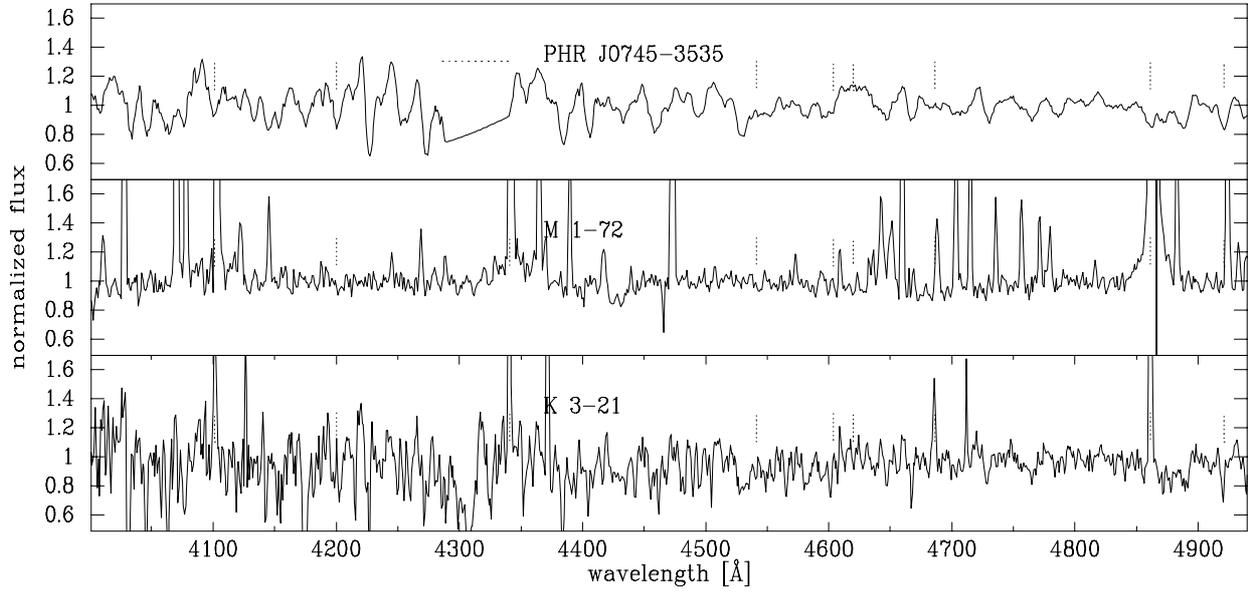}
   \includegraphics[width=0.95\textwidth]{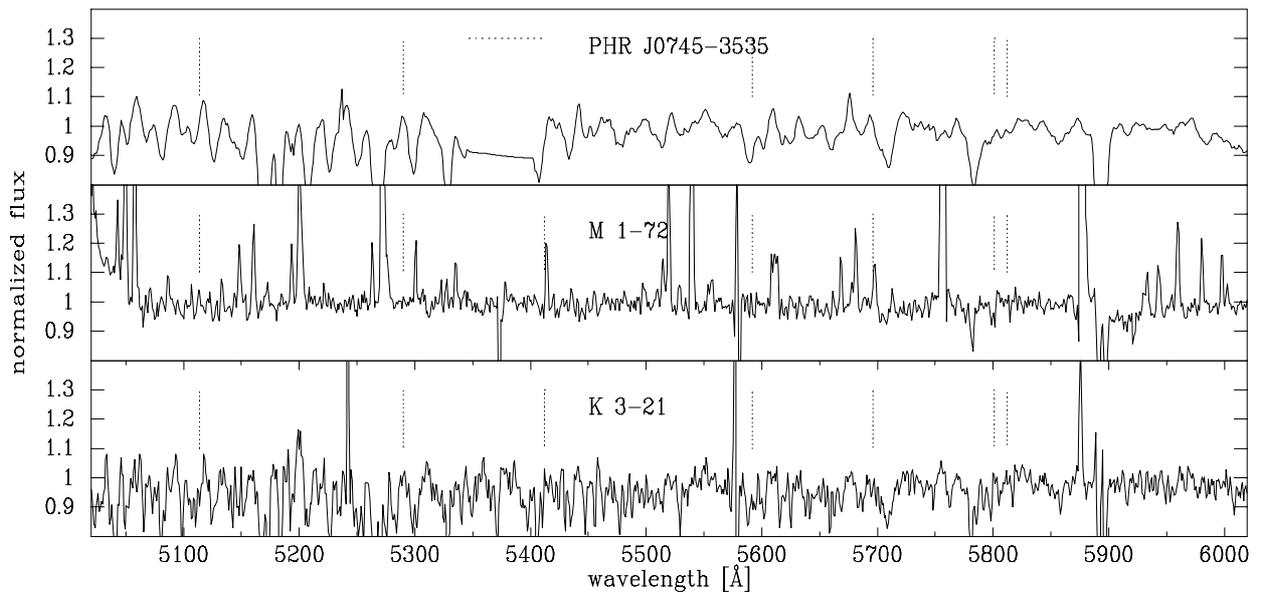}
      \caption[]{Normalized spectra of continuum CSPN 
                           (see Table~\ref{iones}).
  Blue and red spectral regions features as Fig.~\ref{f01}.
}
         \label{f17}%%%%%%%%%%%%%%%%%%   Fig. 23
   \end{figure*}

\begin{figure*}
   \centering
   \includegraphics[width=0.95\textwidth]{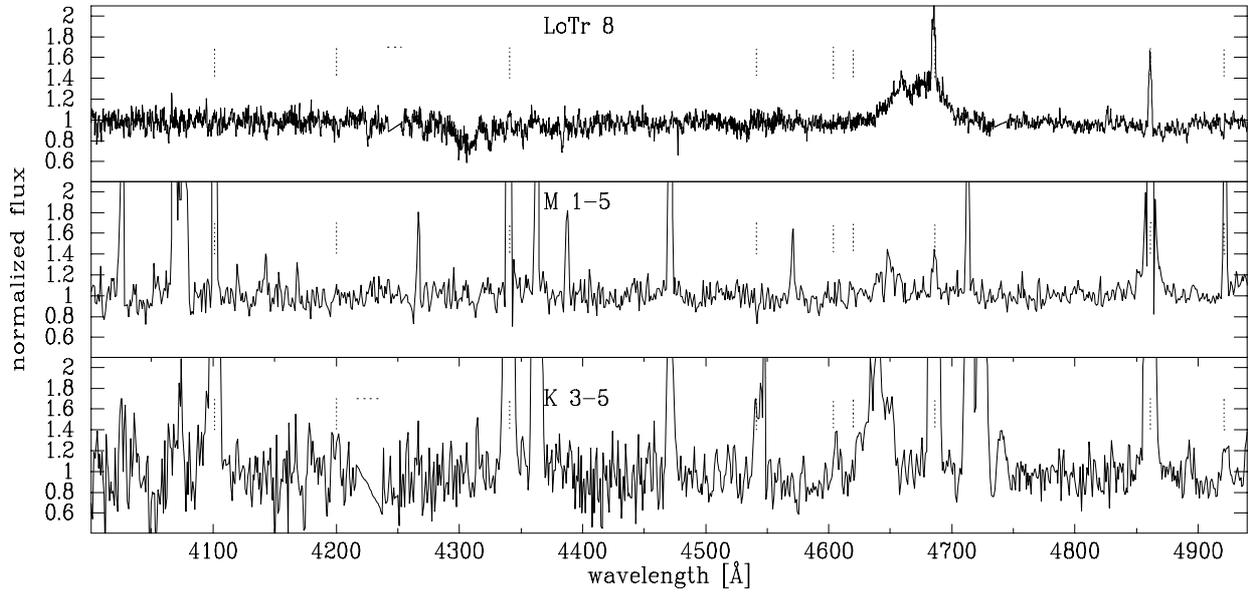}
   \includegraphics[width=0.95\textwidth]{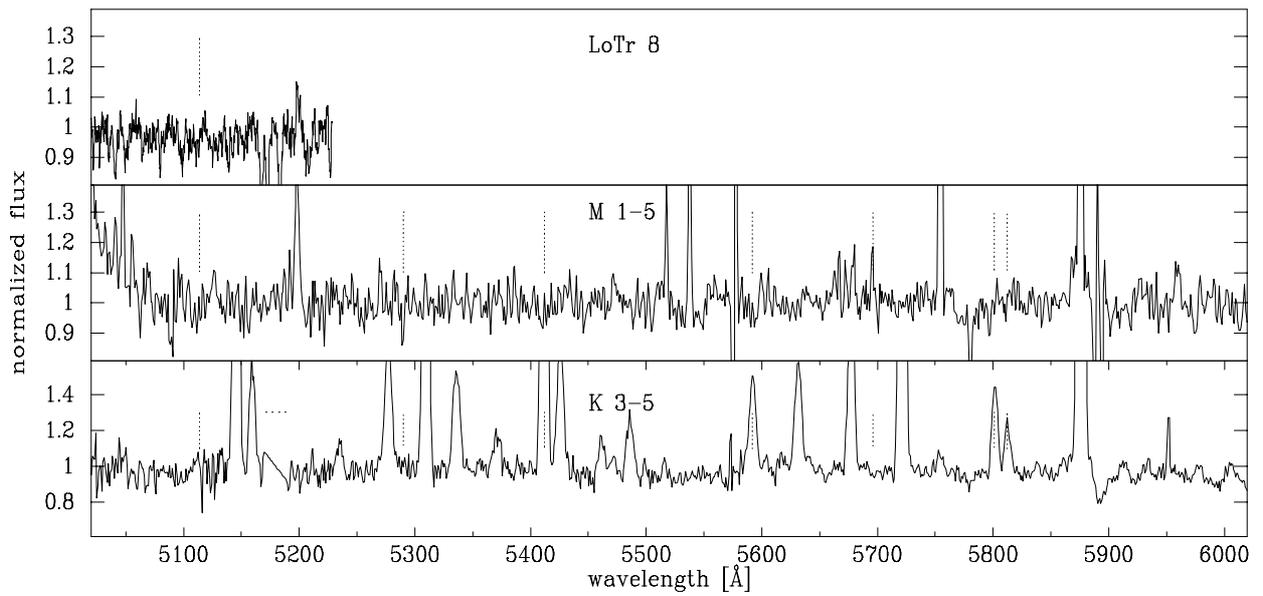}
      \caption[]{Normalized spectra of emission lines CSPN 
                           (see Table~\ref{iones}).
   Blue and red spectral regions features as Fig.~\ref{f01}. 
}
         \label{f18}%%%%%%%%%%%%%%%%%%   Fig. 24
   \end{figure*}

\textit{PN~G075.9$+$11.6} (AMU~1), Fig.~\ref{f01}:
Although originally classified as sdO \citep{2010MNRAS.409.1470O}, 
the recent discovery of a PN around it 
\citep{2013A&A...552A..25A}
leads us to prefer a classification O(H)3~V, 
more according to a PN central star.
In addition, in our spectra we detect \ion{O}{v} at 5114 \AA\
and \ion{N}{iv} at 5203\AA.
This feature was not reported by \cite{2013A&A...552A..25A}.
Moreover, the line 4686 is very intense in relation to 4542 and
4471  is absent. In this sense we prefer a qualifier `z'
\citep{2016AJ....152...31A}.

\begin{figure*}
   \centering
   \includegraphics[width=0.95\textwidth]{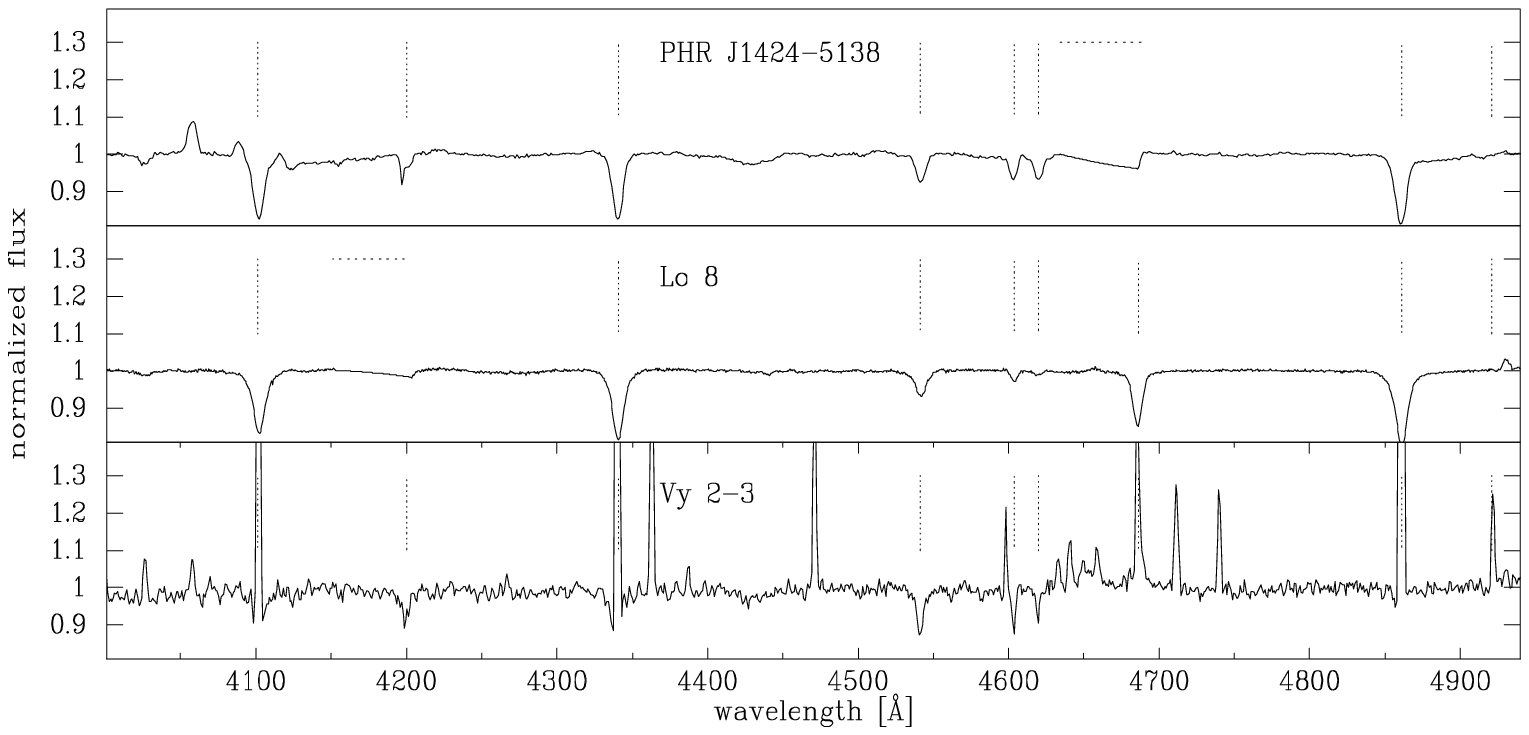}
   \includegraphics[width=0.95\textwidth]{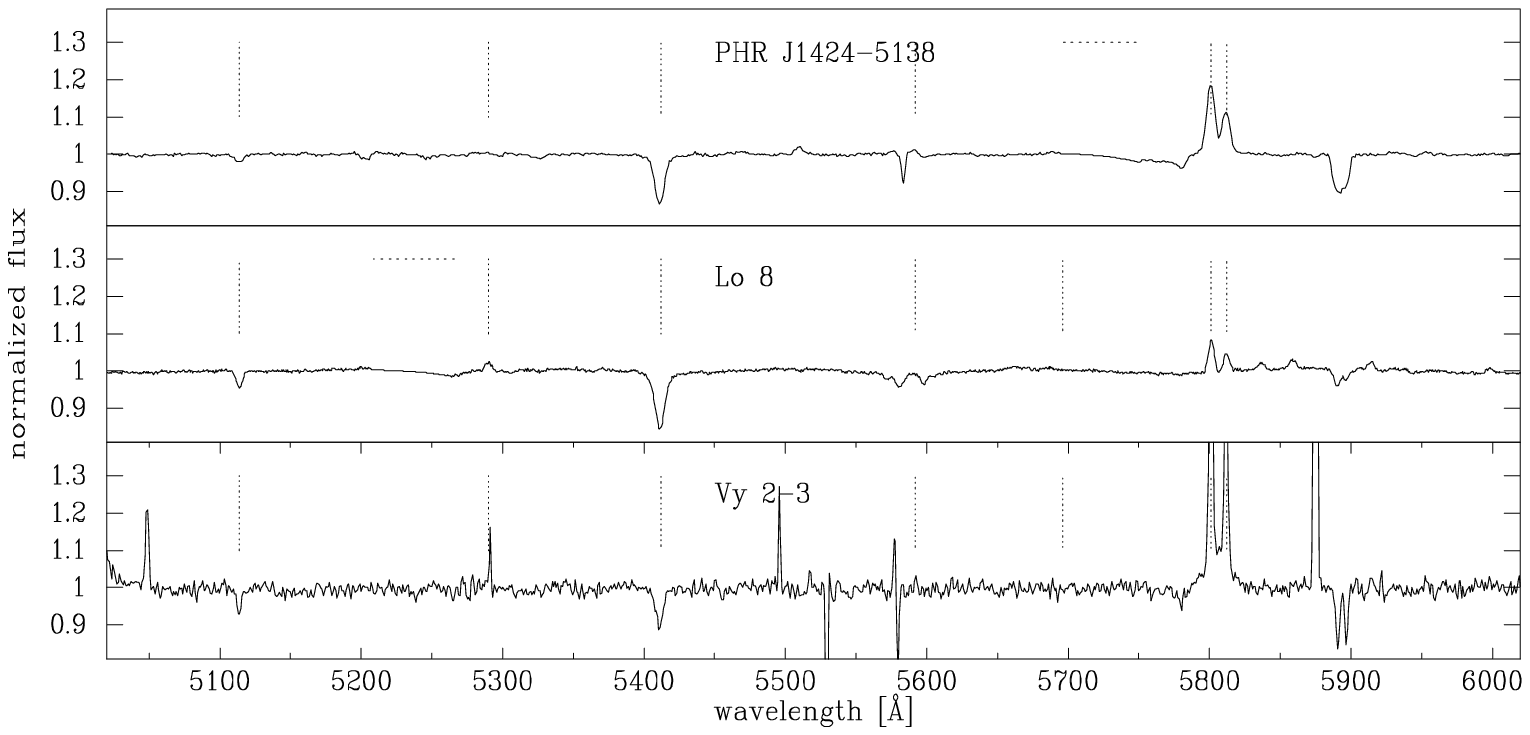}
      \caption[]{Same as Fig.~\ref{f01}. 
  }
         \label{f02}
   \end{figure*}

\textit{PN~G107.6$-$13.3} (Vy~2$-$3), Fig.~\ref{f02}:
The lines that we detect in the optical range are listed in
Table~\ref{iones}. In addition, the spectra display
clear absorption of \ion{O}{v} at 5114 \AA, and quite
doubtful of \ion{He}{i} at 4121 \AA.
\cite{2013A&A...553A.126G} reported two lines at far-UV (\ion{S}{vi} and \ion{O}{vi}),
both display P-Cygni profile. These features are compatible with our classification of
an early O(H)-type.

\textit{PN~G135.9$+$55.9} (TS~01), Fig.~\ref{f13}:
This CSPN shows a clear absorption of \ion{He}{ii} at 4686 \AA.
This feature leads us to reject a luminosity class I.
We rather prefer a classification O(H)~III-V.
The spectrum displays clear absorption at H$\gamma$ and \ion{He}{i} at 4471 \AA.
However, it is evident that the nebular lines are 
shifted with respect to the stellar ones, suggesting a binary core.
Multiplicity of this CSPN was reported by \cite{2005AIPC..804..173N}. On the other hand,
\cite{2004ApJ...616..485T} by means of
UV spectra, classified this object as WD/NS. Nevertheless,
we cannot give an accurate description of the spectral type and we 
 see no evidence of wide absorption lines.

\textit{PN~G184.0$-$02.1} (M~1$-$5), Fig.~\ref{f18}:
This is a young PN. It is difficult to infer some features of the CSPN.
We detect narrow emission lines only at 4686 and 5806 \AA. 
Although we can not be sure that these features come from the star.

\textit{PN~G199.4$+$14.3} (PM~1$-$29), Fig.~\ref{f10}:
There is a possibility that it is a proto-planetary nebula.
This object was studied by \cite{2006AstL...32..661A}, who
 classified the CSPN as an early B-type.

\begin{figure*}
   \centering
   \includegraphics[width=0.95\textwidth]{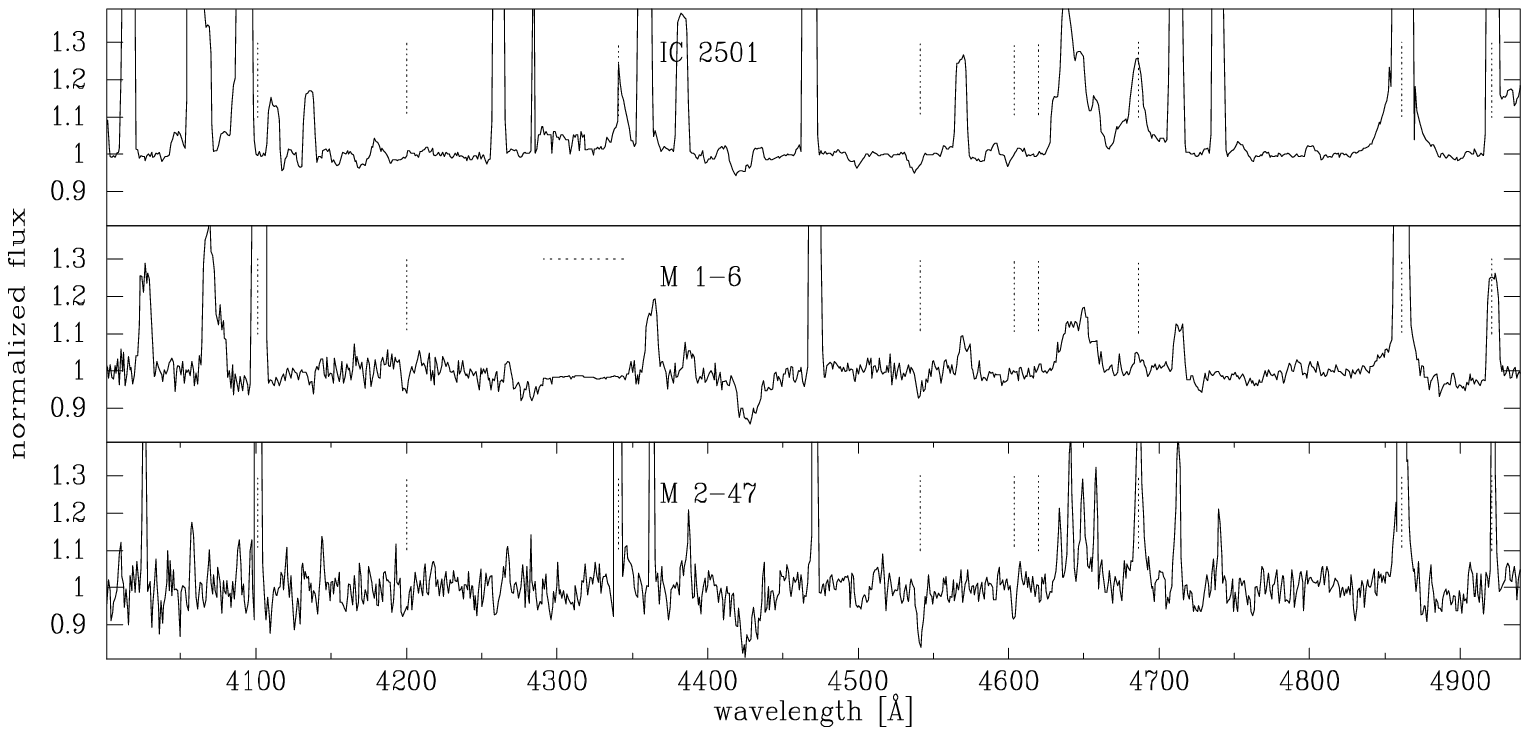}
   \includegraphics[width=0.95\textwidth]{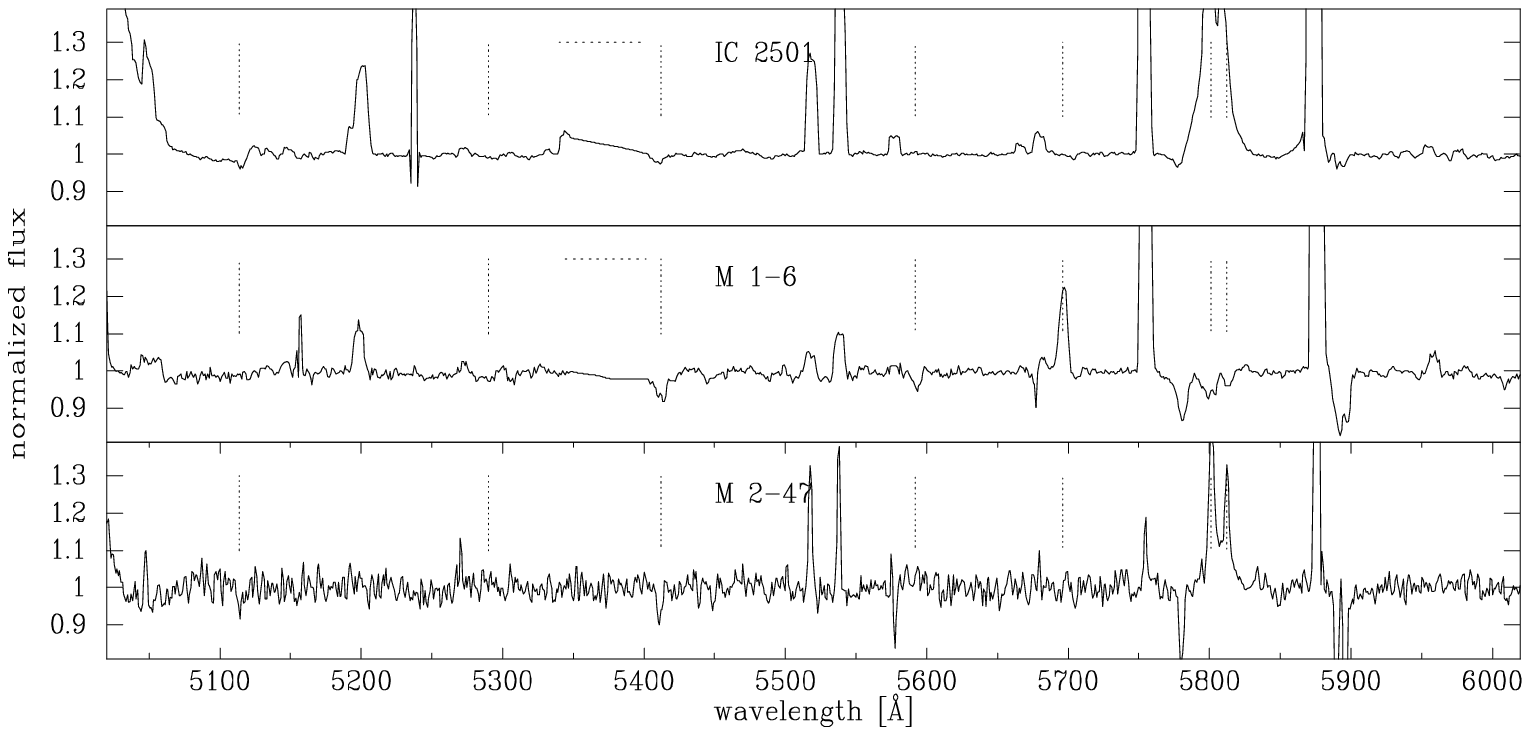}
      \caption[]{Normalized spectra of  O-type CSPN 
                           (see Table~\ref{iones}). In these objects we could not detect the Balmer series.
                 The interstellar absorption bands at $\lambda$4428, the  complex at 5780 and 5890-6 are not indicated.
                 The most important spectral features (absorption and emission) identified are:
\ion{He}{ii} $\lambda$4200, 4542, and 4686,  
   \ion{N}{v} $\lambda$4604-19, for the blue spectral range (top panel), and \ion{O}{v}   $\lambda$5114,
 \ion{O}{vi}  $\lambda$5290,
 \ion{He}{ii} $\lambda$5412,
 \ion{O}{iii} $\lambda$5592,
 \ion{C}{iii} $\lambda$5696,
 \ion{C}{iv}  $\lambda$5801-12, for the red spectral range (bottom panel).
}
         \label{f08}
   \end{figure*}

\textit{PN~G211.2$-$03.5} (M~1$-$6), Fig.~\ref{f08}:
This CSPN was classified as a possible [WC10-11] (see, Paper~II).
However,  we can now identify \ion{He}{ii} absorption lines.
The nebula is of low EC, so the nebular feature at 4686 \AA\ has to be very weak or absent.
In our spectra the line 4686 is weak, thus the stellar feature of \ion{He}{ii}
must be weak too. In this sense, we classified this CSPN as late OI.
In addition, absorption of \ion{O}{iii} at 5592 \AA\ is detected.

\begin{figure*}
   \centering
   \includegraphics[width=0.95\textwidth]{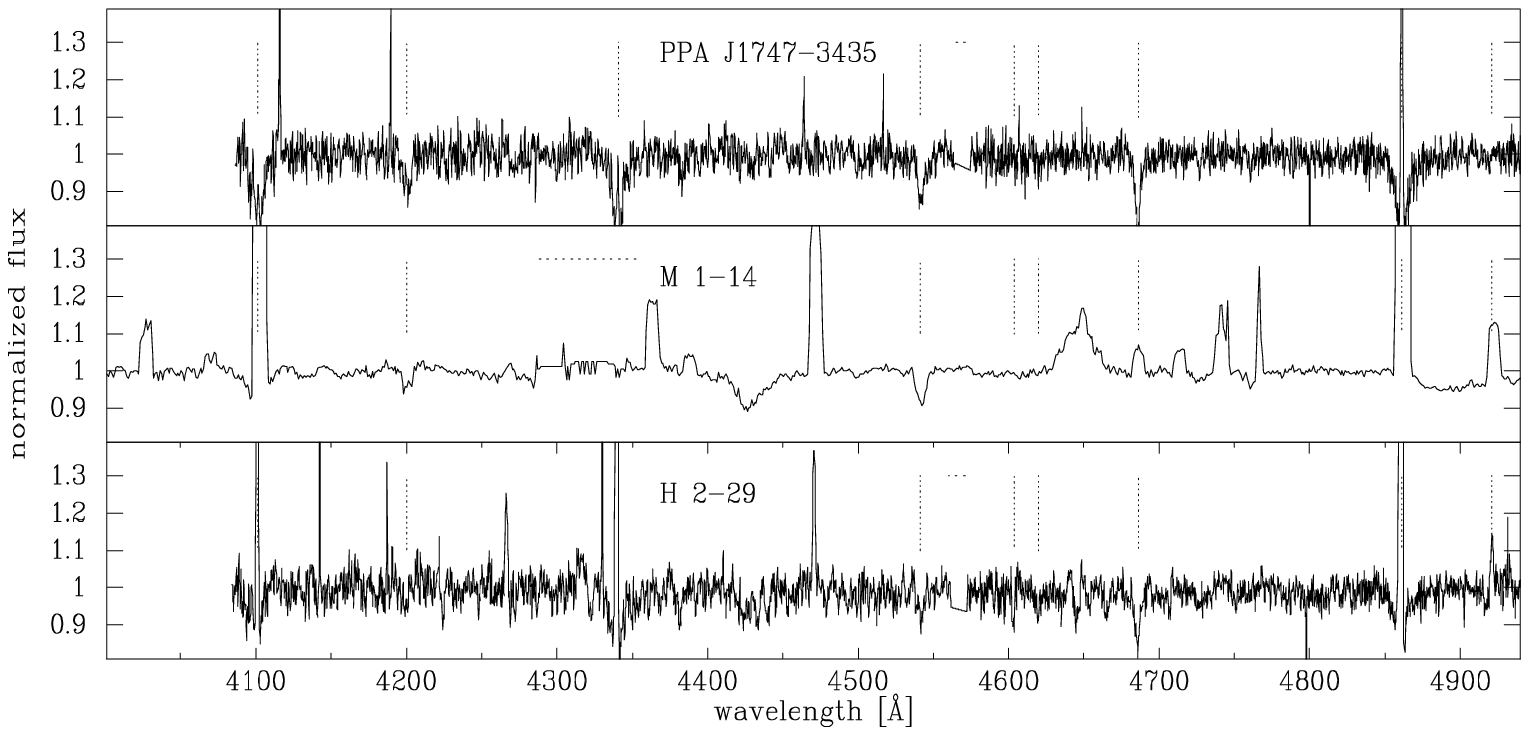}
   \includegraphics[width=0.95\textwidth]{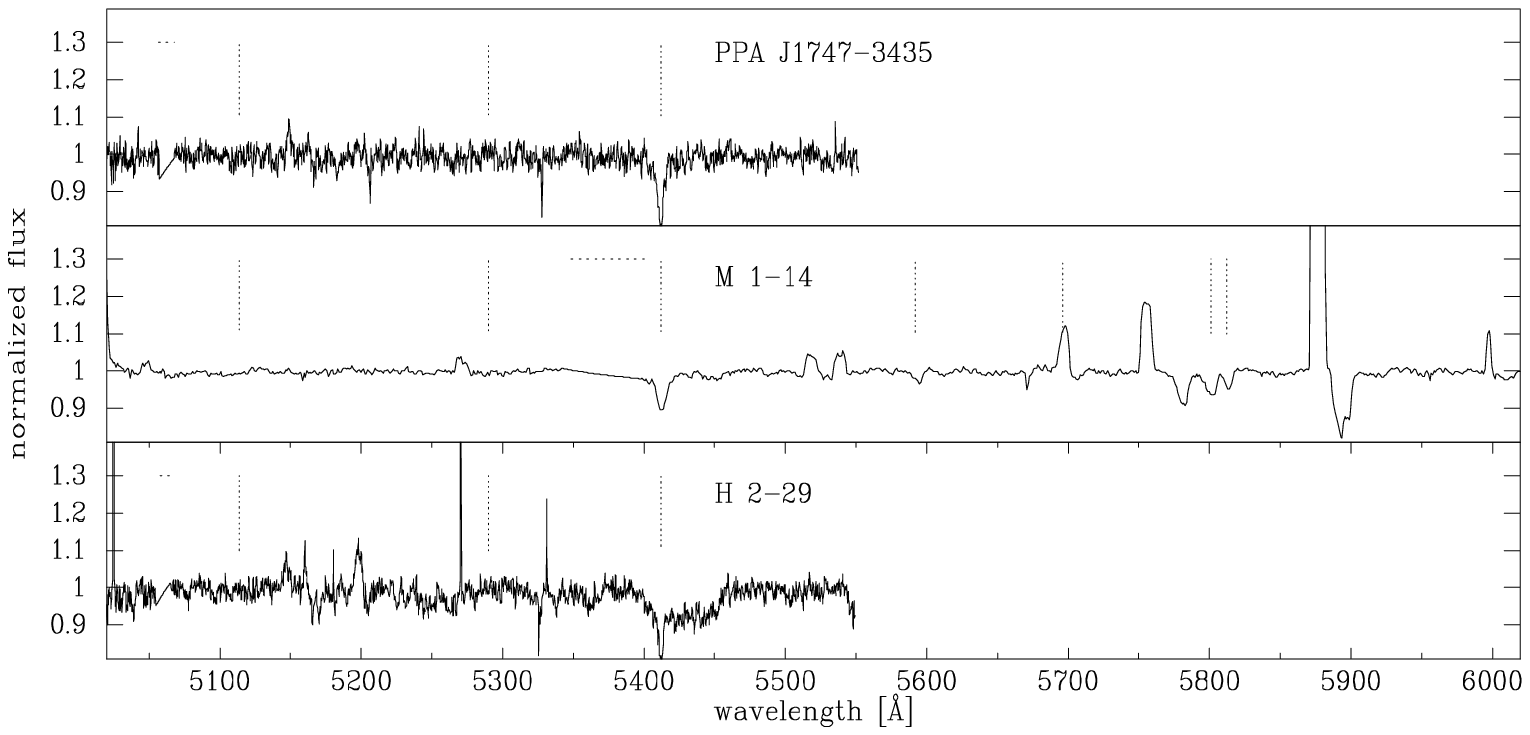}
      \caption[]{Normalized spectra of  O(H)-type CSPN 
                           (see Table~\ref{iones}).
                           In these objects was not possible identify the sub-type of the O-type star.
     Blue and red spectral ranges features as in Fig.~\ref{f01}.
}
         \label{f12}%%%%%%%%%%%%%%%%%%   Fig. 15
   \end{figure*}

\begin{figure*}
   \centering
   \includegraphics[width=0.95\textwidth]{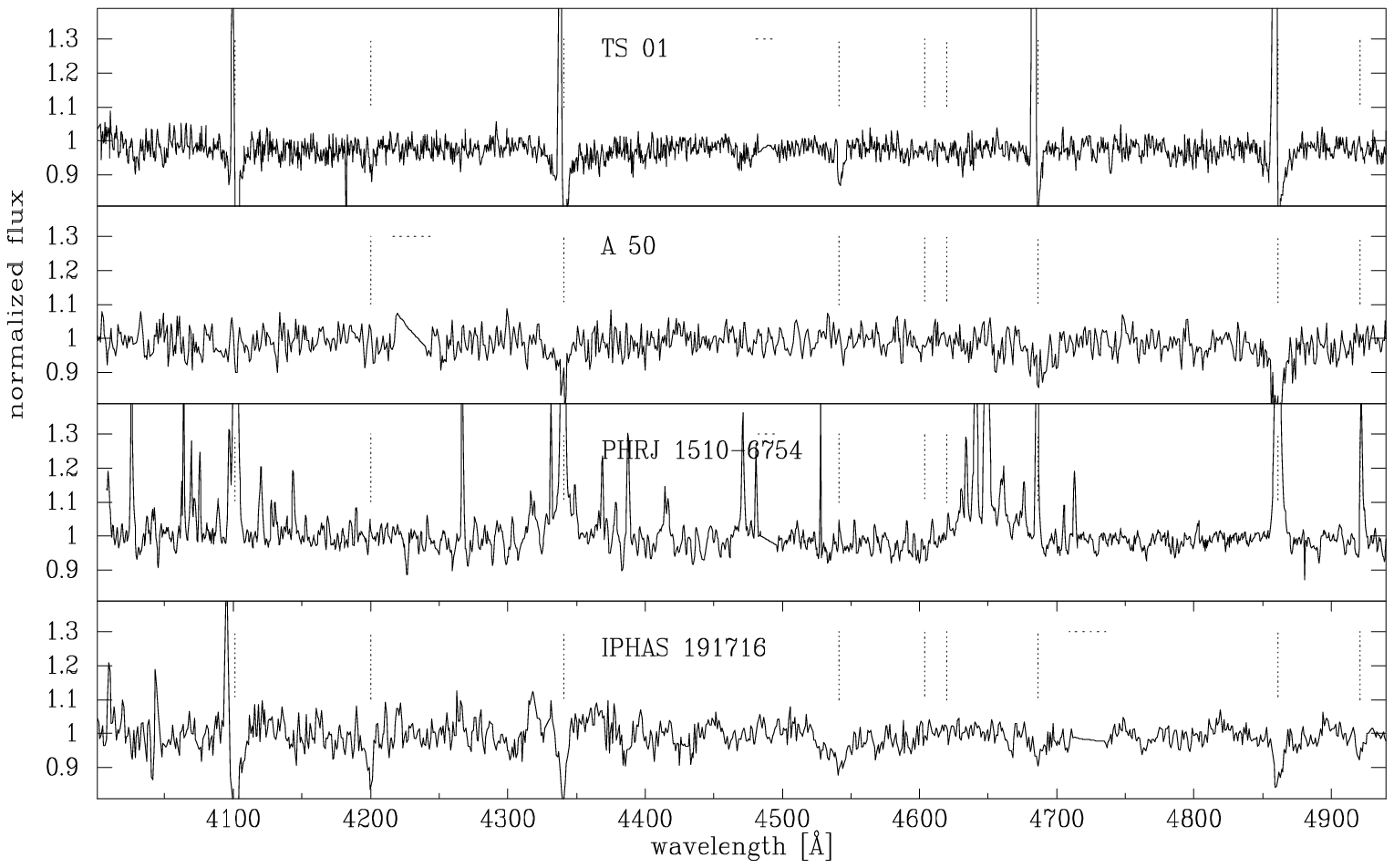}
   \includegraphics[width=0.95\textwidth]{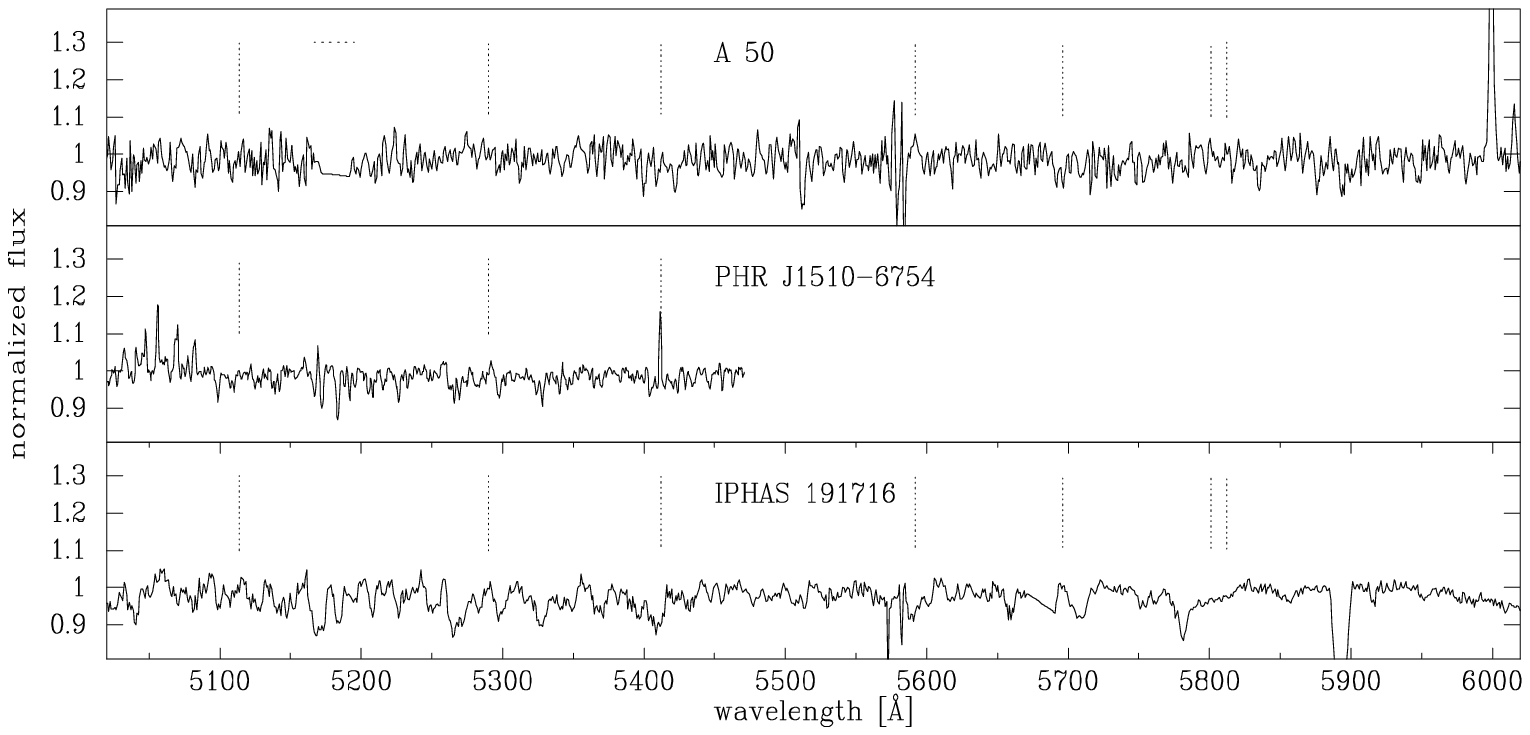}
      \caption[]{Same as Fig.~\ref{f12}.
}
         \label{f13}%%%%%%%%%%%%%%%%%%   Fig. 16
   \end{figure*}

\begin{figure*}
   \centering
   \includegraphics[width=0.95\textwidth]{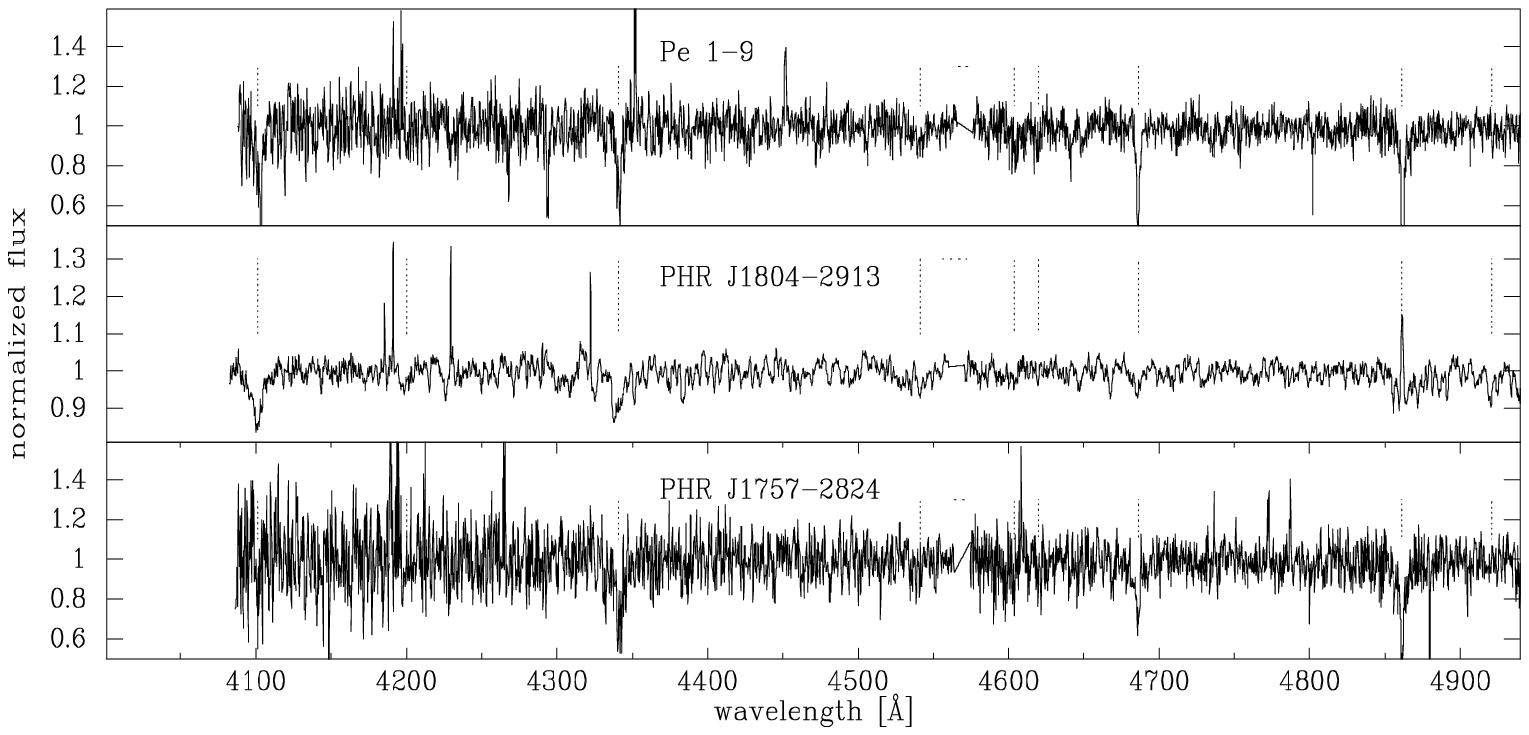}
   \includegraphics[width=0.95\textwidth]{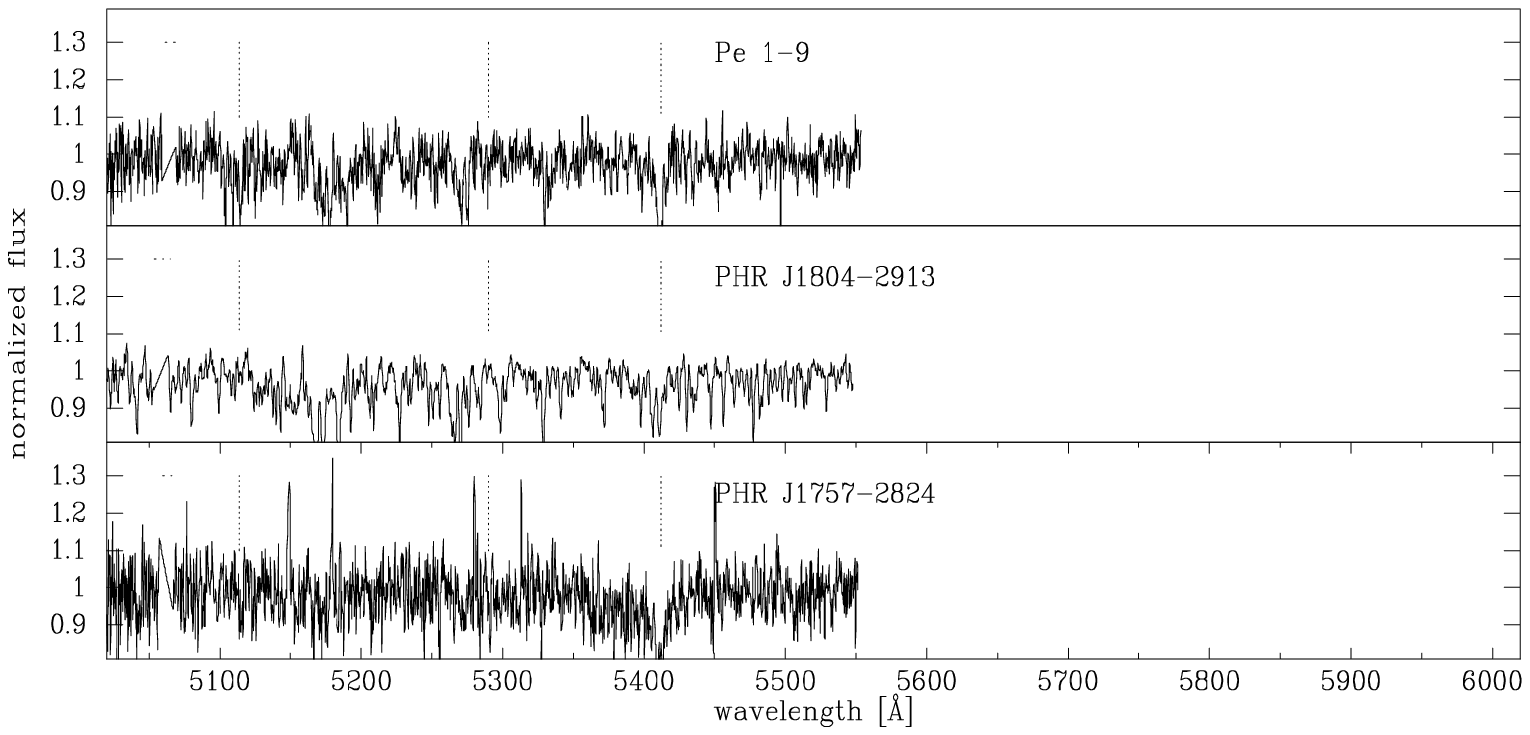}
      \caption[]{Same as Fig.~\ref{f12}.
}
         \label{f14} %%%%%%%%%%%%%%%%%%   Fig. 17
   \end{figure*}

\textit{PN~G234.9$-$01.4} (M~1$-$14), Fig.~\ref{f12}:
This CSPN was classified as O-type in Paper~II. Now, with a better
quality spectrum, we could detect the Balmer lines.
In addition we detect
absorption of \ion{O}{iii} at 5592 \AA\ and \ion{C}{iv}
at 5801-12 \AA. Both features are typical of a late O.
Also the spectrum displays
emission of \ion{C}{iii} at 5696 \AA\ (stellar origin?).

\begin{figure*}
   \centering
   \includegraphics[width=0.95\textwidth]{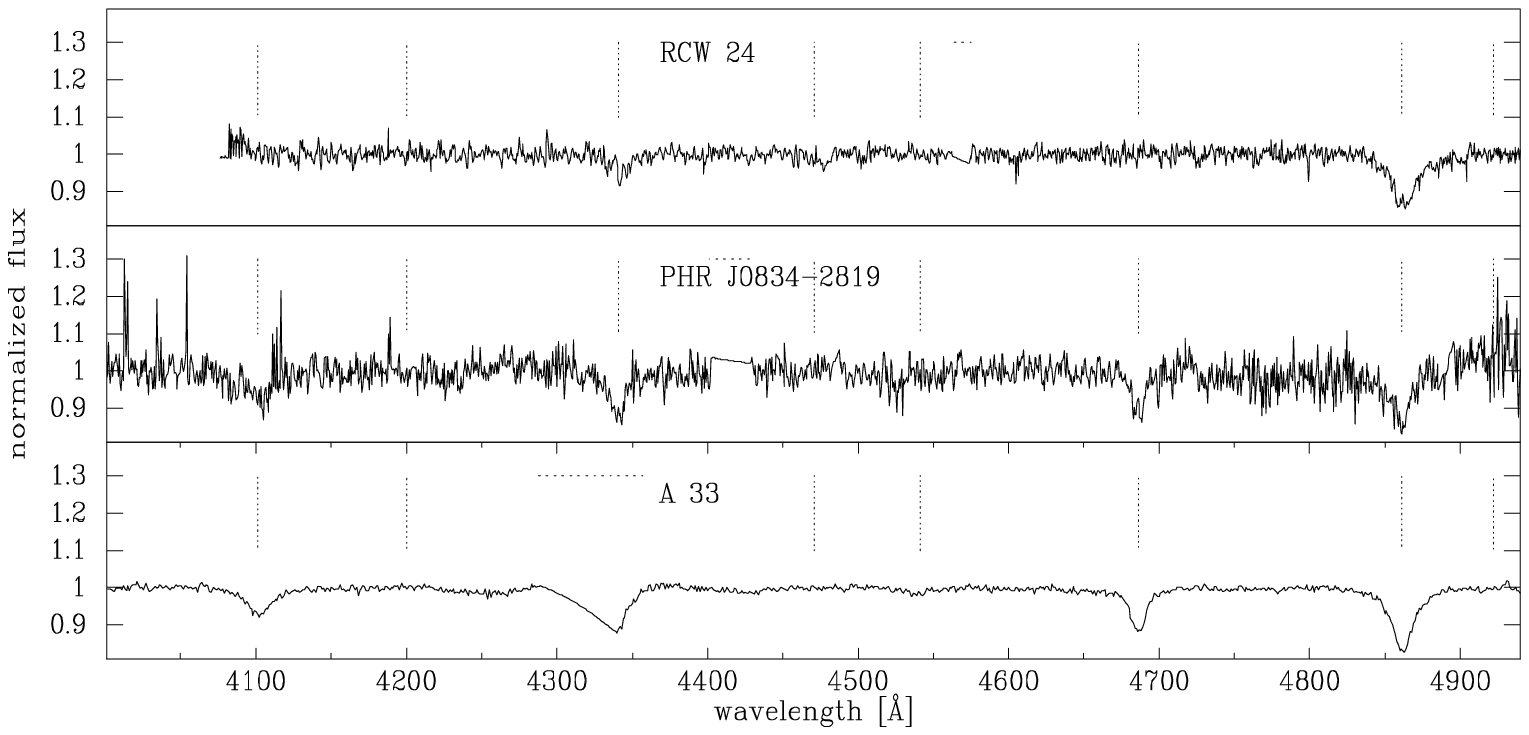}
   \includegraphics[width=0.95\textwidth]{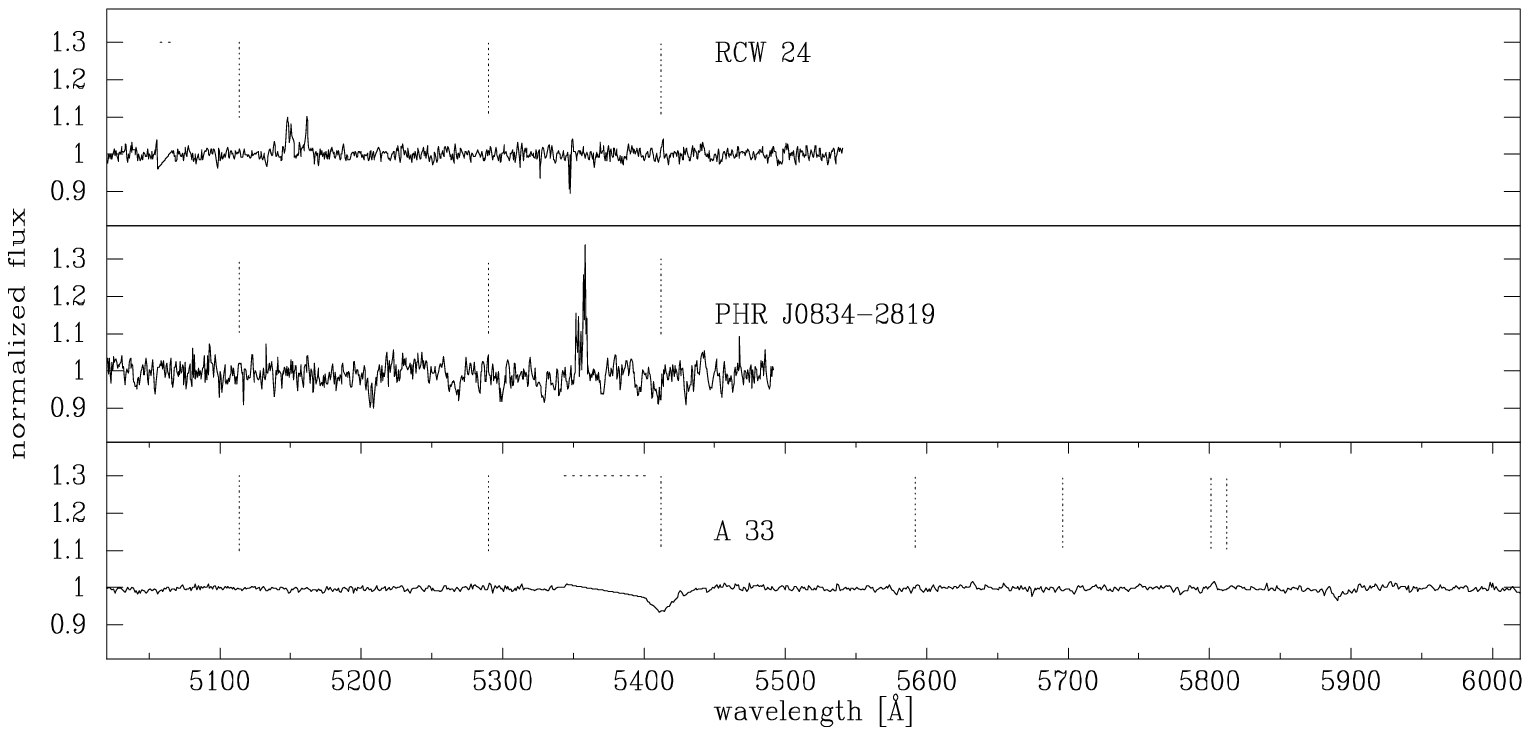}
      \caption[]{Normalized spectra of WD CSPN 
                           (see Table~\ref{iones}).
                           You should note the wide absorption lines.
                 The interstellar absorption bands at $\lambda$4428, the  complex at 5780 and 5890-6 are not indicated.
                 The most important spectral features (absorption and emission) identified are
   H$\beta$, H$\gamma$, H$\delta$,
   \ion{He}{i} $\lambda$4471 and $\lambda$4921,
   \ion{He}{ii} $\lambda$4200, 4542, and 4686, for the blue spectral range. Red range as Fig.~\ref{f01}.
}
         \label{f15}%%%%%%%%%%%%%%%%%%   Fig. 19
   \end{figure*}

\textit{PN~G238.0$+$34.8} (A~33), Fig.~\ref{f15}:
This object has wide absorption lines (FWHM of $\sim$20 \AA).
\ion{He}{ii} at 4200 \AA\ is not detected.
It is expected to be an evolved star, the ionizing object of an old PN as A~33.
\cite{1975MSRSL...9..271A} 
classified this CSPN as sd Op and  \cite{1991IAUS..145..375M} reported a spectral type O(H).
Our spectrum, of better quality,  allows us to determine that the FWHM of the absorption  
lines are comparable to WD stars. Our classification is similar to that obtained by \cite{1975MSRSL...9..271A}.
We confirm detection of H reported by \cite{1991IAUS..145..375M}.

\textit{PN~G249.8$+$07.1} (PHR~J0834$-$2819), Fig.~\ref{f15}:
The slit intersected  two stars lying at the geometrical centre of the PN. 
We confirmed that the CSPN is the brighter of the two. The other star is of a late type. 
The spectrum has a low S/N. Although the lines appear to be broad, it seems to be an evolved object. 
This is consistent with the nebula's large angular size.

\textit{PN~G258.5$-$01.3} (RCW~24), Fig.~\ref{f15}:
Our spectrum  displays broad lines of the Balmer series.
\cite{2006MNRAS.372.1081F} report an uncertain absorption feature near
4650 \AA\ that we do not detect, despite the high S/N  of our spectrum.

\textit{PN~G273.6$+$06.1} (HbDs~1), Fig.~\ref{f01}:
The CSPN of this nebula is a clear O(H)3.
Moreover, we detect
emission lines of \ion{N}{v} at 4603-19 \AA\ (in general this line appears in absorption),
\ion{C}{iv} at 4658, 5801-11 \AA, and strong \ion{O}{vi} at 5290 \AA.
All these indicate that this is an extremely hot star.
Moreover, the line 4686 is very intense in relation to 4542 and
4471  is absent. In this sense we prefer a qualifier `z'
\citep{2016AJ....152...31A}.

\textit{PN~G281.0$-$05.6} (IC~2501), Fig.~\ref{f08}:
\cite{2013A&A...553A.126G} report three lines at far-UV (\ion{S}{vi}, \ion{O}{vi} and \ion{P}{v}),
all display P-Cygni profile. These features are compatible with an early O-type star.
In addition, we observe that the
nebular emissions are shifted with respect to 
 the stellar features (binary?).

\textit{PN~G285.6$-$02.7} (He~2$-$47), Fig.~\ref{f09A}:
The spectrum has a high S/N.
There is absorption of \ion{O}{iii} at 5592 \AA.
The feature of \ion{He}{i} at 4471, 4713, 4921 and 5876 \AA, presents a P-Cygni profile.
Absorption of the Balmer series can be inferred.
It is possible that \ion{He}{ii} at 4026 \AA\ presents a P-Cygni profile.
In addition, there are strong emission of \ion{C}{ii}, \ion{C}{iii}, and 
\ion{C}{iv} near 4650 \AA\ that could be of stellar origin.
In Paper~II we classified this CSPN as
[WC10-11], however our new high quality spectrum displays key
absorption lines of a late O-type star.
It is not possible to be sure that the emission of \ion{C}{iii} at
5696 \AA\ is stellar.

\textit{PN~G285.7$-$14.9} (IC~2448), Fig.~\ref{f04}:
Nebular subtraction is not good, for example in the strongest lines at [\ion{O}{iii}]. 
Nevertheless, the absorption of \ion{N}{v} at 4604-20 \AA\ and the emission 
of \ion{C}{iv} at 5801-12 \AA\ are clear features typical of an early O-type star.

\textit{PN~G288.7$+$08.1} (ESO~216$-$2), Fig.~\ref{f05}:
This is a poorly studied object.
We classified the CSPN of this ring nebula as early O-type star.
The absorption at 4686 \AA\ rejects a luminosity class I, and
 absence of \ion{N}{v} at 4603-19 \AA\  suggests an object later than O(H)4.
Moreover, emission of \ion{O}{vi} at 5290 \AA\ indicates a hot star.

\textit{PN~G302.2$-$03.1} (PHR~J1244$-$6601), Fig.~\ref{f16}:
The bright central star probably does not correspond to the CSPN.
The S/N of the spectrum is high. Balmer lines are narrow, suggesting an evolved star.

\textit{PN~G310.3$+$24.7} (Lo~8), Fig.~\ref{f02}:
Spectra display emission of \ion{C}{iv} at 4658 \AA\ and \ion{O}{vi} at 5290 \AA, 
together with absorption of \ion{O}{v} at 5114 \AA.
These are features of a very hot star.
In addition, there are some absorption features of unknown origin at
4441,
5572 (\ion{O}{v}?),
5581 (\ion{O}{v}?), and
5598\AA\ (\ion{O}{v}?).

\textit{PN~G312.6$-$01.8} (He~2$-$107), Fig.~\ref{f04}:
Strong emission of \ion{C}{ii}, \ion{C}{iii}, and \ion{C}{iv} around 4650 \AA, that could be of stellar origin.
The nebular EC is low, so there should be no nebular emission at 4686 \AA.
Thus this feature have to be stellar, in this sense we
assign to this star a luminosity class  I.
Significant P-Cygni profile at \ion{He}{i} lines.
Besides lines reported in Table~\ref{iones} the spectrum displays an absorption of \ion{N}{iv}
at 5203 \AA.
This line is present in early O-type stars of population I
\citep{2002RMxAC..14...16G}.
This object was classified as [WC10-11] in Paper~II.
Nevertheless the present high quality spectrum leads us to reject such a classification.

\textit{PN~G315.4$-$08.4} (PHR~J1510$-$6754), Fig.~\ref{f13}:
The CSPN of this true planetary nebulae has been identified by \cite{2006MNRAS.373...79P}.
It has a large angular size and low surface brightness. In this case we hoped to find an evolved CSPN, such as a WD.
However, the spectrum of this CSPN is dominated by narrow emission lines, especially \ion{N}{iii} and \ion{C}{iii},
and an absorption feature is seen at H$\beta$.
The very narrow absorption, which may be covered by the emission, suggests a low surface gravity.
It is surprising that having strong emission lines, these are so narrow.
Perhaps this star is binary and we see the emission arising from another part of the system, 
such as is seen in cataclysmic variables.
Nevertheless, the All-Sky Automated Survey for SuperNovae\footnote{http://www.astronomy.ohio-state.edu/~assassin/}
(ASAS-SN \citealt{2014ApJ...788...48S}) 
does not show photometric variation.

A search of the literature revealed two cool [WC] that display similar spectra:
\cite{1985ApJ...289..342G} and \cite{1990A&A...234..435H}. However, in these cases 
both objects are compact, so it is difficult to
be sure if the lines are nebular or stellar.
Another object that displays a similar spectrum is HD120678,
a hot star of population I.
Undoubtedly, this object requires a more detailed study.

\textit{PN~G315.7$+$05.5} (LoTr~8), Fig.~\ref{f18}:
Wide emission of \ion{C}{iv} and \ion{He}{ii} at 4658 and 4686 \AA\ respectively.
Clear features of a [WR] star, which is consistent with the high EC of the nebula.

\textit{PN~G317.2$+$08.6} (PHR~J1424$-$5138), Fig.~\ref{f02}:
This is a poorly studied object.
The spectra display clear features of an early O(H)-type star.
The strong emission of \ion{N}{iv} at 4058 \AA\ is more compatible with a
luminosity class I.
In addition, the spectrum displays an 
intense absorption of \ion{N}{iv} at 5203 \AA.

\textit{PN~G323.9$+$02.4} (He~2$-$123), Fig.~\ref{f14A}:
The spectrum of this CSPN is of low S/N.
The clearest feature is \ion{He}{ii} at 4542 \AA.
In addition, \ion{N}{v} at 4603-19 \AA\ is detectable,
this feature is typical of early O-type stars.
Nevertheless, the line at 4603 \AA\ shows a profile of emission plus absorption.

\textit{PN~G323.6$-$04.5} (WKK136$-$337), Fig.~\ref{f03}:
Strong absorption of \ion{O}{v} at 5114 \AA, a feature typical of hot objects.
In addition, emission of \ion{Si}{iv} and \ion{C}{iv} at 4089 and 4658 \AA\ respectively.
According to \cite{2011ApJS..193...24S}, emission at 4658\AA\ is not present in O-type stars
of population I.
In addition the spectra display 
absorption of unknown origin at 4120-25 \AA\ (\ion{O}{v}?).

\begin{figure*}
   \centering
   \includegraphics[width=0.95\textwidth]{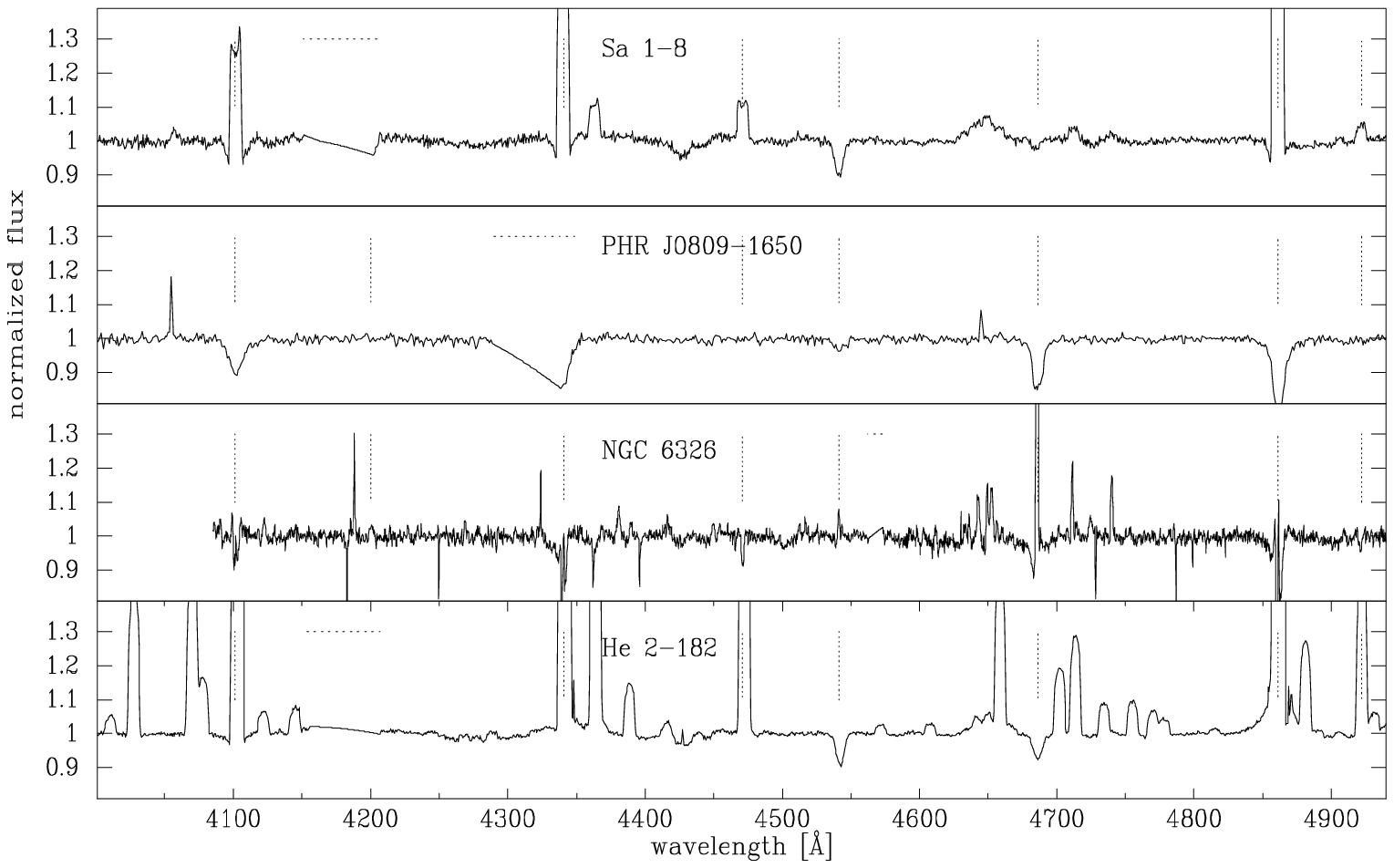}
   \includegraphics[width=0.95\textwidth]{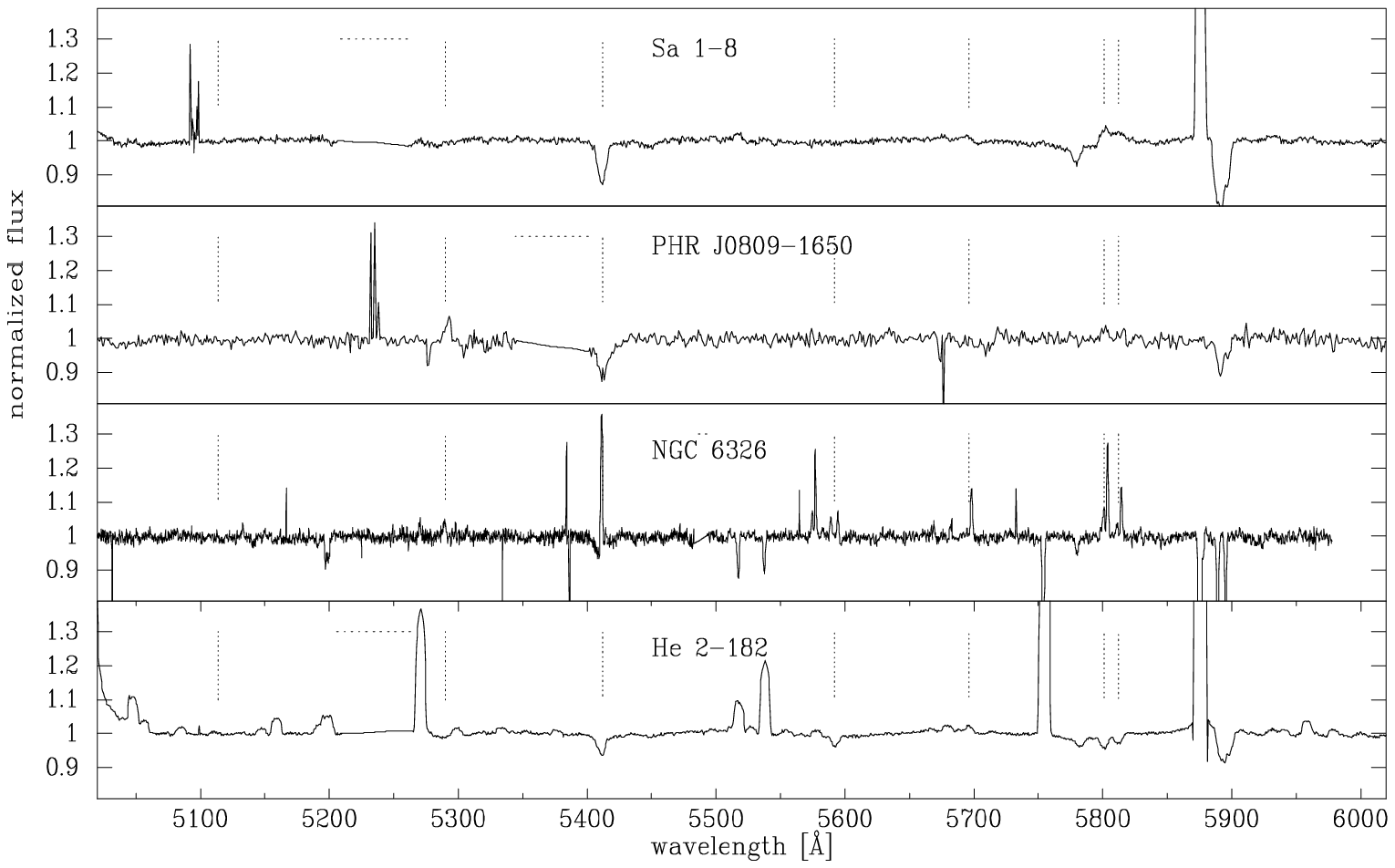}
      \caption[]{Normalized spectra of intermediate/late O-type CSPN 
                           (see Table~\ref{iones}). Spectral features for blue and red spectral regions as Fig.~\ref{f09A}.
}
         \label{f09B}
   \end{figure*}

\textit{PN~G325.8$-$12.8} (He~2$-$182), Fig.~\ref{f09B}:
This is an intermediate O(H)-type star.
It is possible to infer an stellar absorption at H$\delta$.
In addition, clear absorption of \ion{O}{iii} at 5592 \AA\ is detected.

\textit{PN~G326.0$-$06.5} (He~2$-$151), Fig.~\ref{f09}:
In this object's spectra, the absorption at H$\epsilon$ is clear.
In addition, absorption of \ion{O}{iii} at 5592 \AA\ is detected.
P-Cygni profiles are detected at 5016 and 5876 \AA.

\begin{figure*}
   \centering
   \includegraphics[width=0.95\textwidth]{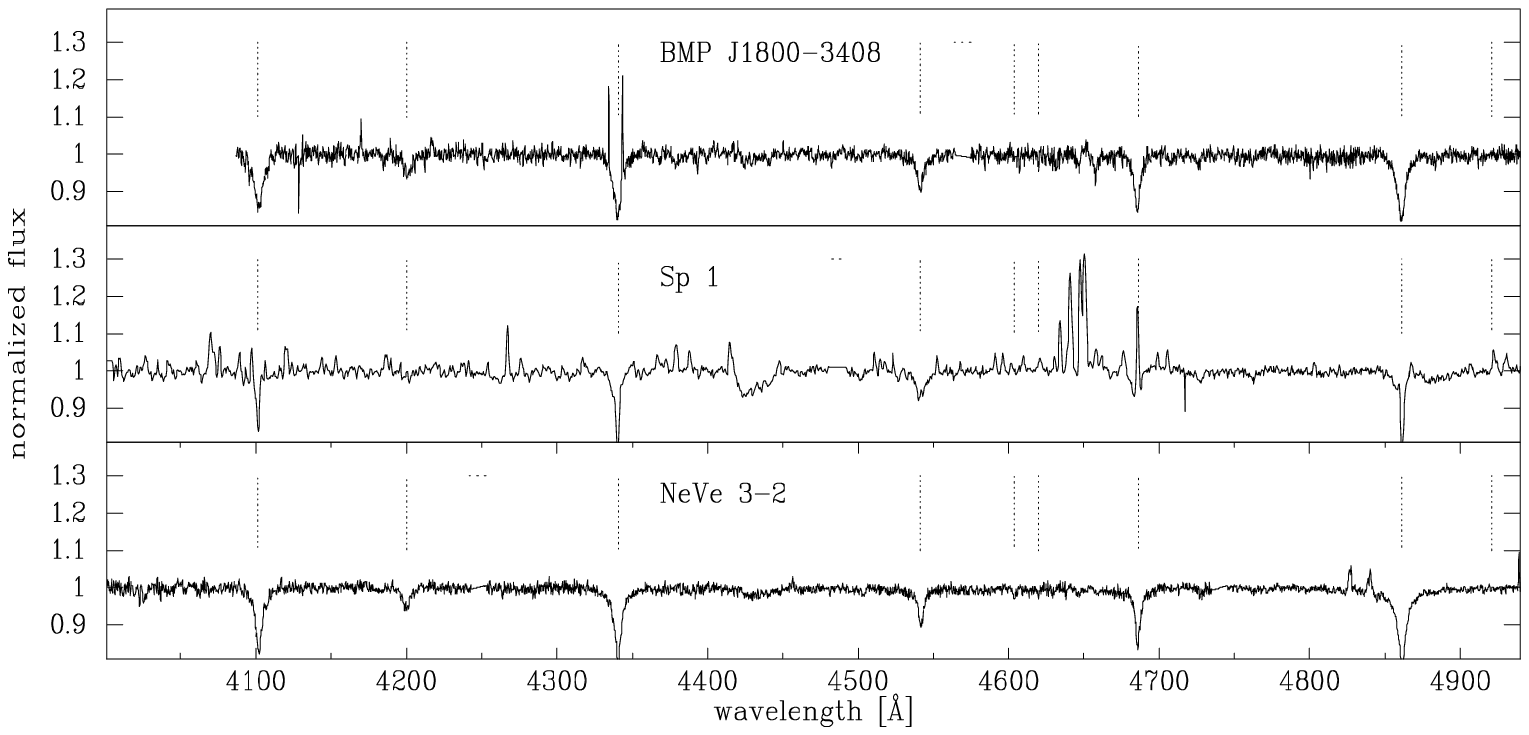}
   \includegraphics[width=0.95\textwidth]{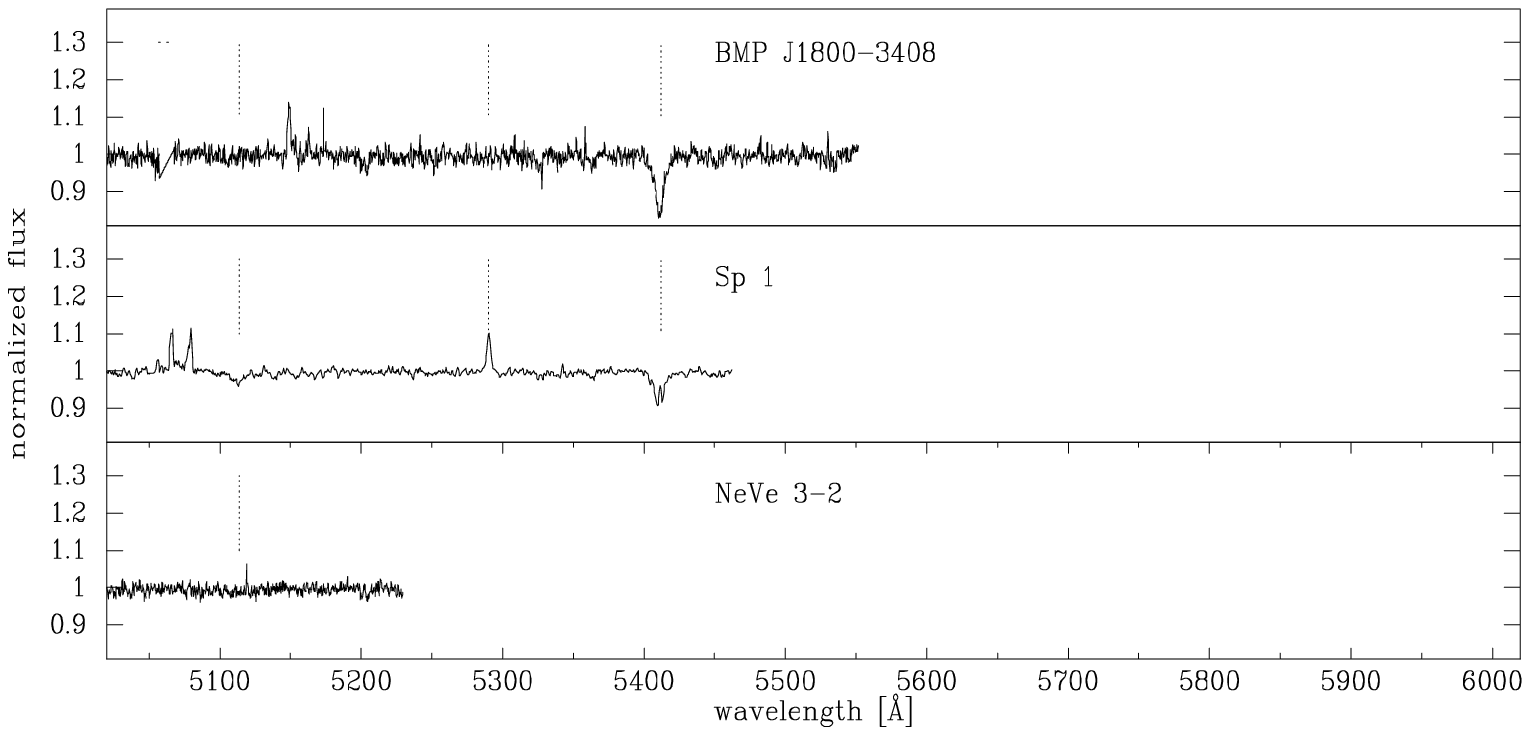}
      \caption[]{Same as Fig.~\ref{f01}. 
  }
         \label{f06}
   \end{figure*}

\textit{PN~G326.4$+$07.0} (NeVe~3$-$2), Fig.~\ref{f06}:
This is a poorly studied object.
The spectrum of this CSPN displays clear features of an early
O-type star. In addition, a strong absorption of \ion{N}{iv} at 5203 \AA\ is detectable.

\textit{PN~G329.0$+$01.9} (Sp~1), Fig.~\ref{f06}:
The subtraction of the nebular component is very good.
There is emission of \ion{O}{vi} at 5290 \AA\ and possible absorption of \ion{O}{v} at 5114 \AA.
These are typical of early O-type stars, which is consistent with the absence of \ion{He}{i}.
Emission lines of \ion{N}{iii} at 4634-40-42 \AA\ and \ion{C}{iii} at 4647-50-51 \AA,
 lead us to prefer an `(fc)' qualifier.
However, the \ion{C}{iii} features are more intense than those of \ion{N}{iii}, 
and the feature at 4686 \AA, with emission plus absorption. This is very strange.
\cite{1988A&A...190..113M} classified this object as O(H), and reported emission lines in
the region around 4650\AA.
Particularly interesting is the emission of \ion{Si}{iv} at 4630-55\AA\
that we detect too.
It is in good agreement with the classification that we propose.

The apparent asymmetric shape in H$\beta$ and H$\gamma$ absorptions 
(blue smooth wing and narrow red),  
is caused by the \ion{He}{ii} absorptions at 4859 and 4338 \AA\ respectively.
This feature is reported by \cite{1988A&A...190..113M}.

\textit{PN~G329.5$+$01.7} (VBRC~7), Fig.~\ref{f16A}:
It is a poorly studied object.
Our nebular subtraction is not good, however,
the Balmer absorption seems real.
We cannot detect any other stellar features.

\textit{PN~G331.4$-$03.5} (He~2$-$162), Fig.~\ref{f11}:
Absorption lines that display the spectrum are compatible with a B0 star. 
Moreover, this classification agrees with the low T$_{eff}$ reported by
\cite{1992A&A...260..329M}.
In addition, P-Cygni profile is detect at 4650 \AA.

\textit{PN~G338.1$-$08.3} (NGC~6326), Fig.~\ref{f09B}:
We performed a nebular subtraction, although with not very good results. 
This makes the 4471 \AA \ and 5806 \AA, features uncertain.
We infer that emissions around 4650 \AA\ are of stellar origin.

\textit{PN~G349.3$-$04.2} (Lo~16), Fig.~\ref{f03}: 
We classified this CSPN as early O(H)-type.
In addition we detect emission of \ion{O}{vi} at 5290 \AA,
which is consistent with a hot star.
Moreover, the
stellar emission of \ion{C}{iii} and \ion{N}{iii} at 4650 \AA\
suggests  an  ((fc)) qualifier.
The line at 4542\AA\ is
clearly wider than the other \ion{He}{ii} lines.

\textit{PN~G353.7$+$06.3} (M~2$-$7), Fig.~\ref{f14A}:
The spectrum only displays a couple of absorption lines of \ion{He}{ii}.
In addition, a wide absorption line of unknown origin at 5176 \AA\ can be detected.

\textit{PN~G354.5$-$03.9} (SAB~41), Fig.~\ref{f16A}:
The only stellar feature that we can detect of this CSPN is
 a strong and wide absorption (32 \AA) at H$\beta$.
According to \cite{2009A&A...505..249M} the nucleus of this 
planetary nebula is likely a binary system.

\textit{PN~G357.0$-$04.4} (PHR~J1756$-$3342), Fig.~\ref{f03}:
This weak object was classified as likely planetary nebulae
according to the MASH \citep{2008MNRAS.384..525M} catalogue.
The central star of this object is clearly  an early O(H)-type star, nevertheless,
its spectra displays  wide lines of \ion{He}{ii} and H.  
In particular, the line 4542 is wider than 4686.

\textit{PN~G357.1$-$05.3} (BMP~J1800$-$3408), Fig.~\ref{f06}: 
We do not detect nebular emission.
We note clear absorption of \ion{N}{iv} at 5203 \AA\
and subtle absorption of \ion{C}{iv} at 4658 \AA. 
Both are features of a hot O-type star.
Moreover, I(4686)$>$I(4542) so the luminosity class is V.

\textit{PN~G357.6$-$03.3} (H~2$-$29), Fig.~\ref{f12}:
Nebular emission masks the H lines, however, the absorption at H$\delta$ is clear.
There is an absorption of unknown origin at 4382 \AA.
On the other hand, according to the described in section \ref{ovz},
this CSPN could be an early O-type star.

\textit{PN~G358.7$-$03.0} (K~6$-$34), Fig.~\ref{f07}: 
The spectrum displays emission
 of \ion{Si}{iv} at 4116 \AA, that suggests an early O(H)-type star.
The absence of \ion{N}{v} rejects an O(H)3 star.

%%%%%%%%%%%%%%%%%%%%%%%%%%%%%%%%%%%%%%%%%%%%%%%%%%%%%%%%%%%%%%%%%
%%%%%%%%%%%%%%%%%%%%%%%%%%%%%%%%%%%%%%%%%%%%%%%%%%%%%%%%%%%%%%%%%

\subsection{Discussion}

We were able to determine sub-types for the first time for
many CSPN. Sometimes it was possible to apply
some non-standard criteria which should be discussed
in future works.
For example, ten objects from our sample were
previously classified by \cite{1991IAUS..145..375M}
as O(H) CSPN.
These objects were included in our sample with the goal of
improve such classification by adding a sub-type.
In all cases
(A~33, A~51, HbDs~1, He~2-151, He~2-162, He~2-182,
IC~2448, Lo~8, Sa~4-1, and Sp~1)
our spectra confirm the presence of H and \ion{He}{ii}.
In this sense, we agree with classification reported 
by \cite{1991IAUS..145..375M}.

As many times the lines used as criteria for the classification of Pop~I stars 
are contaminated by nebular lines, it might be useful to analyse the behaviour 
of other stellar lines. To that purpose, we analysed some identified
absorption lines (see Section~\ref{red-arm}).

Some lines, noted in a few spectra, remained unidentified.
These features are, in general, of low S/N, and in the two dimensional images are not clearly detectable.
Nevertheless, for those objects in which we did not subtract the nebular 
contribution, we are sure that the line is real and not a residual of the subtraction. 
Also, they were observed in more than just one spectra,
which reinforces the fact that these features are not artifacts.
We discarded them as diffuse interstellar bands and
telluric features as we did not find them in 
the works of \citet{1995ARA&A..33...19H}, nor as
telluric features in \citet{2003A&A...407.1157H}. 
In order to identify these features with known
ions, we used \citet{1945CoPri..20....1M} and the
NIST Atomic Spectra Database \citep{NIST_ASD}.
No transitions correspond to ions of the \ion{C}{iv}, \ion{O}{v}, or \ion{O}{vi} 
that are typical ions of hot stars.

It is widely known that the spectra of CSPN and
Population I stars are very similar, specially the 
WR and [WR] stars which are indistinguishable.
Perhaps the object M~1$-$67 is a good example that shows 
the similarities between these two groups of stars \citep{1991A&A...249..518C}, and we also
recall the case of HD~151853 which was 
considered in the literature as either a post-AGB or a
massive star \citep{2015A&A...584A...7G}.
As far as we know, there are few studies analysing whether O-type stars 
of population I and II display different spectrum.
The reason for this could be in the lack of population II O-type stars with 
precise spectral classification performed, that is, sub-types determined. 
Our current improved database of CSPN spectra allows us to perform some analysis in this direction.

%%%%%%%%%%%%%%%%%%%%%%%%%%%%%%%%%%%%%%%%%%%%%%%%%%%%%%%%%%%%%%%%%%%%%%%%%
%%%%%%%%%%%%%%%%%%%%%%%%%%%%%%%%%%%%%%%%%%%%%%%%%%%%%%%%%%%%%%%%%%%%%%%%%

\section{Summary and conclusions}\label{conclu}

We have obtained and analysed high quality optical spectra  
using GEMINI and INT telescopes.
We performed a qualitative determination of the spectral types of 78 CSPN, which represents an 
increase of 18\% of the objects catalogued in Paper~I.
The quality of these spectra was essential in order to be able 
to provide sub-types to the spectral classification. 
We were not able to do this in our previous work - Paper~II.

Fifty CSPN spectra were classified for the first time.
Five objects previously classified as WELS
(H~1$-$56, He~2$-$123, M~1$-$46, NGC~6326 and Sp~4$-$1)
were re-classified as O-type stars,
reaffirming the results obtained in Paper~III.
We have detected several CSPN with uncommon spectral types.
For example, IPHASX~J190438.6$+$021424 and PHR~J1510$-$6754
were classified as B[e] and O(H)e, respectively.
These objects require more detailed observations 
since they might be transitional objects.
Ap~1$-$12, H~1$-$65, He~2$-$47, He~2$-$107, and M~1$-$6, previously classified as H-poor, were re-classified as H-rich.
K~3$-$66, previously classified as continuous,
was re-classified as O-type star.
This fact reinforces the hypothesis of \cite{1981Msngr..26....7K} that these objects are expected to be H-rich.
The classification of another twelve CSPN, previously classified as just O(H),
were improved by adding a sub-type and, in some cases, their luminosity class.
Finally, in seven objects we identified P-Cygni profile at \ion{He}{i} lines. 
Six of these CSPN are late O-type. On the other hand, in nine early O-type
CSPN we found evidence that they are not undergoing mass loss.
According to the evolution sequence of H-rich CSPN,
these evolve from late O-type to early O-type. In this sense,
it is more likely that an object experiences mass loss in the
earliest stages of its evolution, that is, late O-type. This is in complete
agreement with that we have found.

It is surprising that, in this large spectral sample of CSPN, only one
object was classified as [WR], especially since our selection criteria for
the sample was based, mainly, on the apparent brightness of the star. 
This may be due to a selection effect.

The spectral range of our data, which includes 5000--6000 \AA, allowed us to study the behaviour of several lines of stellar origin, and explored they as indicators of sub-types of the OB(H)-type CSPN.
The importance of these lines is that they are easily identifiable (they are not contaminated by nebular features), their origin is rarely nebular, and they fall in the spectral range in which modern CCDs are most sensitive.
The results are promising but further observations and analysis are
needed to define
more robust classification criterion with these lines.

The catalogue of Paper~I lists 20 O(H)-type CSPN with sub-type determined.
With this work plus the object of Paper~III, we have added 45 CSPN to 
the list. 
This improved sample will allow population analysis to be carried out. For example,
contrasting the observed distribution of sub-types 
with those predicted from a theoretical sample of low- and intermediate- mass 
stars evolving as recently proposed by \citet{2016A&A...588A..25M}.

Finally, with respect to 
the ratio H-rich/H-poor CSPN, the simplest evolutionary models predict that there 
should be no H-poor CSPN because the star leaves the AGB before it runs out the photospheric hydrogen. 
Then other explanations are required: born-again or evolution in an interacting binary system. 
In this sense, empirical information on the percentage of H-poor is basic data to verify 
whether the predictions of the most complex models are fulfilled.
These present new spectral classifications, together with those of Paper~III, 
helps us to determine with better precision, the ratio between
H-rich and H-poor CSPN.
In Paper~I we reported H-rich/H-poor $=$ 1.44 (reject the wels stars).
Adding these new classifications to the catalogue of Paper~I
we found that the ratio H-rich/H-poor $=$ 2.07.
Clearly, our current ratio is different to the one obtained in Paper~I.

We hope that the spectroscopic data presented here 
will be useful as a guide for future 
observations that help us to understand better  the final stages 
of stellar evolution for ordinary mass stars.

%%%%%%%%%%%%%%%%%%%%%%%%%%%%%%%%%%%%%%%%%%%%%%%%%%%%%%%%%%%%%%%%%%%%%%%%%
%%%%%%%%%%%%%%%%%%%%%%%%%%%%%%%%%%%%%%%%%%%%%%%%%%%%%%%%%%%%%%%%%%%%%%%%%

\begin{acknowledgements}
We would like to thank our anonymous referee whose critical remarks helped us to substantially improve this paper.
Part of this research was supported by grant PIP 112-201201-00298 (CONICET),
and by grant SeCyT UNC project NRO PIP: 30820150100067CB.
Based on observations obtained at the Gemini Observatory, which is operated by the Association of 
Universities for Research in Astronomy, Inc., under a cooperative agreement with the NSF on behalf 
of the Gemini partnership: the National Science Foundation (United States), the National Research Council 
(Canada), CONICYT (Chile), Ministerio de Ciencia, Tecnología e Innovación Productiva (Argentina), 
and Ministério da Ciência, Tecnologia e Inovação (Brazil).  RPA acknowledges funding from the European
Research Council under the European Union’s Seventh Framework Programme (FP/2007-2013) / ERC Grant Agreement no. 320964 (WDTracer).
This research has made use of SAO
Image DS9, developed by Smithsonian Astrophysical Observatory. 
This research made use of the SIMBAD database, operated at the CDS, Strasbourg, France.
W.W. would like to thank Roberto Méndez.
\end{acknowledgements}

%\begin{thebibliography}{}
\bibliographystyle{aa}
\bibliography{aa3}
%\end{thebibliography}

%%%%%%%%%%%%%%%%%%%%%%%%%%%%%%%%%%%%%%%%%%%%%%%%%%%%%%%%%%%%%%%%%%%%%%%%
%%%%%%%%%%%%%%%%%%%%%%%%%%%%%%%%%%%%%%%%%%%%%%%%%%%%%%%%%%%%%%%%%%%%%%%%
%%%%%%%%%%%%%%%%%%appendix
%%%%%%%%%%%%%%%%%%%%%%%%%%%%%%%%%%%%%%%%%%%%%%%%%%%%%%%%%%%%%%%%%%%%%%%%
%%%%%%%%%%%%%%%%%%%%%%%%%%%%%%%%%%%%%%%%%%%%%%%%%%%%%%%%%%%%%%%%%%%%%%%%

\clearpage
\appendix

\onecolumn

\section{Atlas of spectra}
\label{fig-apendi}

%%%%%%%%%%%%%%%%%%%%%%%%%%%%%%%%%%%%%%%%%%%%%%%%%%%%%%%%%%%%%%%%%%%%%%%%%%%%%%%%%%%%%%%%%%%%%%%%%%%%%%%%%%%%%%%%%
%%%%%%%%%%%%%%%%%%%%%%%%%%%%%%%%%%%%%%%%%%%%%%%%%          estrellas O 
%%%%%%%%%%%%%%%%%%%%%%%%%%%%%%%%%%%%%%%%%%%%%%%%%%%%%%%%%%%%%%%%%%%%%%%%%%%%%%%%%%%%%%%%%%%%%%%%%%%%%%%%%%%%%%%%%

%%%%%%%%%%%%%%%%%%%%%%%%%%%%%%%%%%%%%%%%%%%%%%%%%%%%%%%%%%%%%%%%%%%%%%%%%%%%%%%%%%%%%%%%%%%%%%%%%%%%%%%%%%%%%%%%%
%%%%%%%%%%%%%%%%%%%%%%%%%%%%%%%%%%%%%%%%%%%%%                         B
%%%%%%%%%%%%%%%%%%%%%%%%%%%%%%%%%%%%%%%%%%%%%%%%%%%%%%%%%%%%%%%%%%%%%%%%%%%%%%%%%%%%%%%%%%%%%%%%%%%%%%%%%%%%%%%%%

%%%%%%%%%%%%%%%%%%%%%%%%%%%%%%%%%%%%%%%%%%%%%%%%%%%%%%%%%%%%%%%%%%%%%%%%%%%%%%%%%%%%%%%%%%%%%%%%%%%%%%%%%%%%%%%%%
%%%%%%%%%%%%%%%%%%%%%%%%%%%%%%%%%%%%%%%%%%%%%                         O(H)
%%%%%%%%%%%%%%%%%%%%%%%%%%%%%%%%%%%%%%%%%%%%%%%%%%%%%%%%%%%%%%%%%%%%%%%%%%%%%%%%%%%%%%%%%%%%%%%%%%%%%%%%%%%%%%%%%

%%%%%%%%%%%%%%%%%%%%%%%%%%%%%%%%%%%%%%%%%%%%%%%%%%%%%%%%%%%%%%%%%%%%%%%%%%%%%%%%%%%%%%%%%%%%%%%%%%%%%%%%%%%%%%%%%
%%%%%%%%%%%%%%%%%%%%%%%%%%%%%%%%%%%%%%%%%%%%%                         O
%%%%%%%%%%%%%%%%%%%%%%%%%%%%%%%%%%%%%%%%%%%%%%%%%%%%%%%%%%%%%%%%%%%%%%%%%%%%%%%%%%%%%%%%%%%%%%%%%%%%%%%%%%%%%%%%%

%%%%%%%%%%%%%%%%%%%%%%%%%%%%%%%%%%%%%%%%%%%%%%%%%%%%%%%%%%%%%%%%%%%%%%%%%%%%%%%%%%%%%%%%%%%%%%%%%%%%%%%%%%%%%%%%%
%%%%%%%%%%%%%%%%%%%%%%%%%%%%%%%%%%%%%%%%%%%%%                         WD
%%%%%%%%%%%%%%%%%%%%%%%%%%%%%%%%%%%%%%%%%%%%%%%%%%%%%%%%%%%%%%%%%%%%%%%%%%%%%%%%%%%%%%%%%%%%%%%%%%%%%%%%%%%%%%%%%

%%%%%%%%%%%%%%%%%%%%%%%%%%%%%%%%%%%%%%%%%%%%%%%%%%%%%%%%%%%%%%%%%%%%%%%%%%%%%%%%%%%%%%%%%%%%%%%%%%%%%%%%%%%%%%%%%
%%%%%%%%%%%%%%%%%%%%%%%%%%%%%%%%%%%%%%%%%%%%%                         H-rich
%%%%%%%%%%%%%%%%%%%%%%%%%%%%%%%%%%%%%%%%%%%%%%%%%%%%%%%%%%%%%%%%%%%%%%%%%%%%%%%%%%%%%%%%%%%%%%%%%%%%%%%%%%%%%%%%%

%%%%%%%%%%%%%%%%%%%%%%%%%%%%%%%%%%%%%%%%%%%%%%%%%%%%%%%%%%%%%%%%%%%%%%%%%%%%%%%%%%%%%%%%%%%%%%%%%%%%%%%%%%%%%%%%%
%%%%%%%%%%%%%%%%%%%%%%%%%%%%%%%%%%%%%%%%%%%%%%%          posible O
%%%%%%%%%%%%%%%%%%%%%%%%%%%%%%%%%%%%%%%%%%%%%%%%%%%%%%%%%%%%%%%%%%%%%%%%%%%%%%%%%%%%%%%%%%%%%%%%%%%%%%%%%%%%%%%%%

%%%%%%%%%%%%%%%%%%%%%%%%%%%%%%%%%%%%%%%%%%%%%%%%%%%%%%%%%%%%%%%%%%%%%%%%%%%%%%%%%%%%%%%%%%%%%%%%%%%%%%%%%%%%%%%%%
%%%%%%%%%%%%%%%%%%%%%%%%%%%%%%%%%%%%%%%%%%%%%%%          continuo
%%%%%%%%%%%%%%%%%%%%%%%%%%%%%%%%%%%%%%%%%%%%%%%%%%%%%%%%%%%%%%%%%%%%%%%%%%%%%%%%%%%%%%%%%%%%%%%%%%%%%%%%%%%%%%%%%

%%%%%%%%%%%%%%%%%%%%%%%%%%%%%%%%%%%%%%%%%%%%%%%%%%%%%%%%%%%%%%%%%%%%%%%%%%%%%%%%%%%%%%%%%%%%%%%%%%%%%%%%%%%%%%%%%
%%%%%%%%%%%%%%%%%%%%%%%%%%%%%%%%%%%%%%%%%%%%%%%          wr
%%%%%%%%%%%%%%%%%%%%%%%%%%%%%%%%%%%%%%%%%%%%%%%%%%%%%%%%%%%%%%%%%%%%%%%%%%%%%%%%%%%%%%%%%%%%%%%%%%%%%%%%%%%%%%%%%

%%%%%%%%%%%%%%%
%%%  Tablas
%%%%%%%%%%%%%%%

\begin{table}[h]
\caption{Details of the spectra of CSPN observed with the INT.
The PA and exptime columns are the same as \ref{sample}.}  
\label{sample-b}      
\centering                          
\begin{tabular}{l l c c c}        
\hline\hline                 
PN G              & Name                & Date                    & PA [$^\circ$]        & exptime [s] \\     
\hline                        
038.4$-$03.3	  &  	K~4-19 	  	&  	07-28-2014	  &  	205	  &  	3$\times$900	\\
039.5$-$02.7	  &  	M~2-47	  	&  	09-01-2016    &  	0 	  &  	3$\times$2100	\\
042.0$+$05.4	  &  	K~3-14 	  	&  	07-12-2016	  &     300	  &  	3$\times$1800	\\
044.3$-$05.6	  &  	K~3-36 	  	&  	09-01-2016	  &  	57 	  &  	3$\times$2100	\\
047.1$+$04.1	  &  	K~3-21	  	&  	07-13-2016	  &  	300	  &  	2$\times$1800	\\
054.4$-$02.5	  &  	M~1-72  	&  	05-21-2015	  &  	305	  &  	3$\times$2100	\\
055.6$+$02.1	  &  	He~1-2  	&  	07-13-2016	  &  	290	  &  	3$\times$1800	\\
107.6$-$13.3	  &  	Vy~2-3	  	&  	08-17-2016	  &  	310	  &  	2$\times$1200	\\
167.4$-$09.1	  &  	K~3-66	  	&  	01-10-2014	  &  	120	  &  	3$\times$1800	\\
184.0$-$02.1	  &  	M~1-5	  	&  	11-12-2016	  &  	74	  &  	1$\times$1200	\\
199.4$+$14.3	  &  	PM~1-29		&  	02-25-2016	  &  	64	  &  	3$\times$1200	\\
\ \ \ \ \ \ \ \ --- &   K~4-50          &       11-08-2014        &      0        &     2$\times$1500   \\  
\hline                                   
\end{tabular}
\tablefoot{The object K~4-50 (PK~208$+$01.1) was classified as possible PN by \citet{1992secg.book.....A}.}
\end{table}

%\longtab{
\begin{longtable}{lllccccc}

\caption{\label{sample} Details of the spectra of CSPNs observed with GMOS. The PA column indicates the slit position angle in degrees, `exptime' is the integration time 
(underexposed spectra is denoted by $^{\dagger}$),
 and `bin' is the binning applied to the CCDs.           
}\\
\hline\hline
PN G & Name & Programme ID & Grating & Spectral range [\AA] & PA [$^\circ$]& exptime [s] & bin [px] \\
\hline
\endfirsthead
\caption{continued.}\\
\hline\hline
PN G & Name & Programme ID & Grating & Spectral range [\AA] & PA [$^\circ$]& exptime [s] & bin [px]  \\
\hline
\endhead
\hline
\endfoot
000.2$-$01.9 	&	M~2-19		&	GS-2008B-Q-65	&	B1200	&	4090-5540	&	90	&	3$\times$1336  & 2$\times$2	\\
001.7$-$04.6	&	H~1-56		&	GS-2016A-Q-74	&	B1200	&	3950-5500	&	76	&	3$\times$500$^{\dagger}$  &  1$\times$1	\\
001.8$-$02.0	&	PHR~J1757-2824	&	GS-2009A-Q-35	&	B1200	&	4100-5530	&	120	&	2$\times$1950$^{\dagger}$	&  1$\times$2\\
001.8$-$03.7	&	PHR~J1804-2913	&	GS-2009A-Q-35	&	B1200	&	4100-5530	&	90	&	1$\times$1200$^{\dagger}$  &  1$\times$2	\\
003.3$-$04.6	&	Ap~1-12		&	GS-2016A-Q-101	&	B600	&	3500-6260	&	71	&	3$\times$800  & 1$\times$1	\\
003.5$-$02.4	&	IC~4673		&	GS-2011A-Q-65	&	B1200	&	3830-5200  	&	125	&	1$\times$1800 & 1$\times$2	\\
005.0$+$03.0 	&	Pe~1-9		&	GS-2009A-Q-35	&	B1200	&	4100-5530	&	45	&	1$\times$2700$^{\dagger}$	& 1$\times$2 \\
007.8$-$04.4	&	H~1-65		&	GS-2011A-Q-65	&	B1200	&	4000-5200	&	270	&	1$\times$900  &  1$\times$2	\\
011.3$-$09.4	&	H~2-48		&	GS-2016A-Q-101	&	B600	&	3500-6260	&	90	&	3$\times$900$^{\dagger}$  &  1$\times$1	\\
012.1$-$11.2    &   PPA~J1855-2328 & GS-2015A-Q-98  &   B600    &   3750-6750   &   70  &       2$\times$1000	& 2$\times$2  \\
013.1$+$05.0    &   Sa~3-96       & GS-2015AQ-98    &   B600    &   3660-6750   &   55  &   3$\times$600$^{\dagger}$	& 2$\times$2 \\
014.8$-$08.4	&	SB~20		&	GS-2015AQ-98	&	B600	&	3660-6750	&	90	&	3$\times$1200	& 2$\times$2 \\
016.4$-$01.9 	&	M~1-46		&	GS-2016A-Q-101	&	B600	&	3500-6260	&	64	&	3$\times$600  &  1$\times$1	\\
017.6$-$10.2	&	A~51		&	GS-2016A-Q-101	&	B600	&	3500-6260	&	239	&	3$\times$1000  & 1$\times$1	\\
020.7$-$05.9	&	Sa~1-8		&	GS-2016A-Q-101	&	B600	&	3500-6260	&	239	&	3$\times$900  & 1$\times$1	\\
034.3$+$06.2	&	K~3-5		&	GN-2016A-Q-97	&	B600	&	3500-6000	&	309	&	3$\times$900$^{\dagger}$ & 2$\times$2	\\
036.4$-$01.9	&	IPHAS~190438$^{(A)}$ & GN-2015A-Q-405	&	B600	&	3850-6600	&	0	&	3$\times$600 & 2$\times$2 \\
039.0$-$04.0	&	IPHAS~191716$^{(B)}$ &GN-2015A-Q-405	&	B600	&	3850-6600	&	123	&	1$\times$600$^{\dagger}$ & 2$\times$2	\\
051.3$+$01.8	&	PM~1-295	&	GN-2016A-Q-97	&	B600	&	3500-6000	&	282	&	3$\times$1000  & 2$\times$2	\\
052.2$+$07.6	&	K~4-10		&	GN-2016A-Q-97	&	B600	&	3500-6000	&	266	&	3$\times$1000 &  2$\times$2	\\
060.0$-$04.3	&	A~68		&	GN-2015A-Q-405	&	B600	&	3850-6600	&	120	&	3$\times$1000$^{\dagger}$  &  2$\times$2	\\
066.8$+$02.9	&	IPHAS~194751$^{(c)}$ &GN-2015A-Q-405	&	B600	&	3850-6600	&	80	&	3$\times$1000 & 2$\times$2	\\
068.7$+$14.8	&	Sp~4-1		&	GN-2016A-Q-97	&	B600	&	3500-6000	&	37	&	3$\times$900  & 2$\times$2 	\\
075.7$+$35.8	&	Sa~4-1		&	GN-2015A-Q-405	&	B600	&	3850-6600	&	280	&	2$\times$1200  & 2$\times$2	\\
075.9$+$11.6	&	AMU~1		&	GN-2012A-Q-124	&	B1200	&	4000-5500	&	147	&	4$\times$360  &  2$\times$2	\\
078.5$+$18.7	&	A~50		&	GN-2016A-Q-97	&	B600	&	3500-6000	&	215	&	2$\times$1000$^{\dagger}$  &  2$\times$2	\\
135.9$+$55.9	&	TS~01		&	GN-2006A-Q-82	&	B1200	&	3800-5000	&	25	&	8$\times$700  & 2$\times$2	\\
211.2$-$03.5	&	M~1-6		&	GS-2015B-Q-103	&	B600	&	3500-6400	&	258	&	5$\times$900  & 2$\times$2	\\
234.9$-$01.4	&	M~1-14		&	GS-2015B-Q-103	&	B600	&	3500-6400	&	0	&	2$\times$770  &  2$\times$2	\\
236.9$+$08.6	&	PHR~J0809-1650 	&	GS-2015B-Q-103	&	B600	&	3500-6400	&	294	&	4$\times$770  &	 2$\times$2 \\
238.0$+$34.8	&	A~33		&	GS-2015B-Q-103	&	B600	&	3500-6400	&	207	&	5$\times$450  & 2$\times$2	\\
249.8$+$07.1	&	PHR~J0834-2819 	&	GS-2016B-Q-65	&	B1200	&	4000-5500	&	50	&	3$\times$950  & 2$\times$2	\\
250.3$-$05.4	&	PHR~J0745-3535	&	GS-2015B-Q-103	&	B600	&	3500-6400	&	265	&	4$\times$750  & 2$\times$2	\\
258.5$-$01.3 	&	RCW~24		&	GS-2008B-Q-65	&	B1200	&	4090-5540	&	3	&	1$\times$2400 &	 2$\times$2 \\
273.6$+$06.1	&	HbDs~1		&	GS-2015B-Q-103	&	B600	&	3500-6400	&	0	&	3$\times$400  & 2$\times$2	\\
281.0$-$05.6	&	IC~2501		&	GS-2015B-Q-103	&	B600	&	3500-6400	&	210	&	3$\times$900  & 2$\times$2	\\
285.6$-$02.7	&	He~2-47		&	GS-2011A-Q-65	&	B1200	&	3900-5200$^{(D)}$	&	90	&  1$\times$1200 & 1$\times$2 \\
285.7$-$14.9	&	IC~2448		&	GS-2015B-Q-103	&	B600	&	3500-6400	&	61	&	1$\times$770  &  2$\times$2	\\
288.7$+$08.1	&	ESO~216-2	&	GS-2015AQ-98	&	B600	&	3750-6750	&	90	&	3$\times$1000 &	 2$\times$2 \\
302.2$-$03.1    &       PHR~J1244-6601  &       GS-2015AQ-98    &       B600    &       4000-6750       &       183     &  3$\times$1000$^{\dagger}$ &2$\times$2	\\
310.3$+$24.7	&	Lo~8		&	GS-2016A-Q-101	&	B600	&	3500-6260	&	277	&	3$\times$600  & 1$\times$1	\\
312.6$-$01.8	&	He~2-107	&	GS-2011A-Q-65	&	B1200	&	3900-5200$^{(E)}$ 	&	90	&	1$\times$1800 & 1$\times$2	\\
315.4$-$08.4	&	PHR~J1510-6754	&	GS-2012A-Q-85	&	B1200	&	4000-5450	&	337	&	2$\times$600$^{\dagger}$  &	  2$\times$2 \\
315.7$+$05.5	&	LoTr~8		&	GS-2011A-Q-65	&	B1200	&	3830-5200  	&	134	&	1$\times$2400  & 1$\times$2	\\
317.2$+$08.6    &   PHR J1424-5138    &       GS-2015A-Q-98   &       B600    &       3750-6750       &    282   &   3$\times$600 &	2$\times$2\\
323.6$-$04.5	&	WKK136-337 	&	GS-2011A-Q-65	&	B1200	&	3800-5200 &	90	&	1$\times$900  &	 1$\times$2\\
323.9$+$02.4	&	He~2-123	&	GS-2016A-Q-74	&	B1200	&	3950-5500 &	179	&	3$\times$500  &  1$\times$1	\\
325.8$-$12.8	&	He~2-182	&	GS-2016A-Q-101	&	B600	&	3500-6260	&	310	&	3$\times$900  &  1$\times$1	\\
326.0$-$06.5	&	He~2-151	&	GS-2016A-Q-101	&	B600	&	3500-6260	&	322	&	3$\times$600 &  1$\times$1	\\
326.4$+$07.0	&	NeVe~3-2	&	GS-2011A-Q-65	&	B1200	&	3830-5200  	&	225	&	1$\times$1200 & 1$\times$2	\\
329.0$+$01.9	&	Sp~1		&	GS-2011A-Q-91	&	B1200	&	4000-5460	&	90	&	1$\times$480 & 2$\times$2	\\
329.5$+$01.7    &   VBRC 7      &   GS-2015A-Q-98   &   B600    &   4000-6750   &   320 &   3$\times$1200 &	2$\times$2 \\
331.4$-$03.5	&	He~2-162	&	GS-2016A-Q-101	&	B600	&	3500-6260	&	270	&	3$\times$900  &  1$\times$1	\\
332.8$-$16.4    &   HaTr~6      &   GS-2015A-Q-98   &  B600     & 4000-5500     &  270  & 3$\times$1000$^{\dagger}$	 &  2$\times$2\\
338.1$-$08.3	&	NGC~6326	&	GS-2009A-Q-35	&	B1200	&	4510-5960	&	90	&	1$\times$1800$^{\dagger}$  & 1$\times$2	\\
349.3$-$04.2 	&	Lo~16		&	GS-2009A-Q-35	&	B1200	&	4510-5960	&	100	&	1$\times$1800 & 1$\times$2	\\
352.1$-$02.6	&	PHR~J1736-3659	&	GS-2011A-Q-65	&	B1200	&	3830-5200  	&	90	&	1$\times$900  & 1$\times$2	\\
353.7$+$06.3	&	M~2-7		&	GS-2016A-Q-101	&	B600	&	3500-6260	&	90	&	3$\times$900  &  1$\times$1	\\
354.5$-$03.9 	&	SAB~41		&	GS-2008B-Q-65	&	B600	&	4260-7030	&	149	&	1$\times$1800$^{\dagger}$  & 2$\times$2	\\
355.3$-$03.2	&	PPA~J1747-3435	&	GS-2009A-Q-35	&	B1200	&	4100-5530	&	135	&	1$\times$2400$^{\dagger}$  &  1$\times$2	\\
355.6$-$02.3	&	PHR~J1744-3355	&	GS-2009A-Q-35	&	B1200	&	4100-5530	&	168	&	2$\times$1950$^{\dagger}$ &	 1$\times$2\\
357.0$-$04.4	&	PHR~J1756-3342	&	GS-2009A-Q-35	&	B1200	&	4100-5530	&	180	&	1$\times$2400 &  1$\times$2	\\
357.1$-$05.3	&	BMP~J1800-3408	&	GS-2009A-Q-35	&	B1200	&	4100-5530	&	35	&	1$\times$2400 &	 1$\times$2 \\
357.6$-$03.3 	&	H~2-29		&	GS-2009A-Q-35	&	B1200	&	4100-5530	&	120	&	2$\times$1950$^{\dagger}$  &  1$\times$2	\\
358.7$-$03.0	&	K~6-34		&	GS-2009A-Q-35	&	B1200	&	4100-5530	&	90	&	1$\times$2400  & 1$\times$2	\\
359.1$-$02.3 	&	M~3-16		&	GS-2008B-Q-65	&	B1200	&	4090-5540	&	90	&	4$\times$1200  & 1$\times$2	\\
\hline 
\end{longtable}
\tablefoot{$^{(A)}$IPHASX J190438.6$+$021424, $^{(B)}$IPHASX J191716.4$+$033447,   $^{(c)}$IPHASX J194751.9$+$311818, $^{(D)}$and 4500-5900\AA, $^{(E)}$and 4500-5900\AA.}
%}

\newpage

%\longtab{
\begin{longtable}{cccccccccccc}
\caption{\label{iones} Spectral classification and key lines of CSPNs sample.
The N.S. column indicates whether or not it was possible to 
subtract the nebular component.
The letters A and E indicate whether the ion was in absorption or 
emission, respectively. An undetected ion was denoted (-), 
and N/D means no data available in the corresponding spectral range.
}\\
\hline\hline
 N & Name           & S.T. & \ion{He}{ii} & \ion{He}{ii} & \ion{He}{ii} & \ion{He}{i} & H$\beta$ & \ion{N}{v} & \ion{C}{iv} & N.S. & Fig. \\
   &                 &      & 4542 & 4686 & 5412 & 4471 & 4861     & 4603-19    &  5801-12    &      &         \\
\hline
\endfirsthead
\caption{continued.}\\
\hline\hline
 N  & Name           & S.T. & 4542 & 4686 & 5412 & 4471 & 4861 & 4603-19 & 5801-12  & N.S. & Fig. \\
\hline
\endhead
\hline
\endfoot
1  &   A~33                     & DAO              &  A  & A  & A  & -          & A         & -     & -    & y & \ref{f15} 	\\
2  &   A~50                     & O(H)~III-V  & -  & A  & -  & -  & A & - & -  & y & \ref{f13} 	\\
3  &   A~51                   & O(H)3-5~Vz    & A     & A     & A    & -          &    A   & E          & -         & n & \ref{f05} 	\\
4  &   A~68                     & H-rich       & -     & -     & -   & -        & -       & -         & -                 & y & \ref{f16} 	\\
5  &   AMU~1                     & O(H)3~Vz     & A      & A     & A   & -        & A       & A         & N/D               & y & \ref{f01} 	\\
6  &   Ap~1-12                    &  B0~I-III     & -    & -    & A    & A          & -        & -          & A             & n & \ref{f11} 	\\
7  &   BMP~J1800-3408             & O(H)4-5~Vz        & A    & A    & A    & -             & A           &  -     &  N/D  & n & \ref{f06} 	\\
8  &   ESO~216-2               & O(H)4-5~V    & A   & A    & A    & -           & A       & -         & E   & y & \ref{f05} 	\\
9  &   H~1-56                    & O?            & A?   & -    & A?   & -          & -        &  -         & N/D           & n  & \ref{f16B} 	\\
10 &   H~1-65                    & O(H)8-9~I        & A    & E   & N/D  & A        & -        &  -          & N/D          & n & \ref{f09A} 	\\
11 &   H~2-29                     & O(H)             & A    & A    & A    & -             & A           & -      &  N/D  & n & \ref{f12} 	\\
12 &   H~2-48                    & O~III-V      & A    & A    & A    & -         & -         & -          & A             & n & \ref{f14A} 	\\
13 &   HaTr~6                     &  H-rich     & -    & -    & -   & -            & A             & -   & -            &  n & \ref{f16} 	\\
14 &   HbDs~1                     & O(H)3~Vz          & A   & A   &  A   & -          & A    & E            & E     & y & \ref{f01} 	\\
15 &   He~1-2                     & O9 - B0~I & - & - & - & A & - & - & -  & n & \ref{f009}       \\
16 &   He~2-47                     & O(H)7-9~I      & A    & E    & A    &  ?         & -        &  -         & A   & n & \ref{f09A} 	\\
17 &   He~2-107                     & O(H)4~Ifc       & A   & E     & A   & A        & -        & A?          & A+E? & y & \ref{f04} 	\\
18 &   He~2-123                     & O3-4          & A    & -    & -    & -          & -         & A?         & N/D & n & \ref{f14A} 	\\
19 &   He~2-151                     & B0             & A    & A    & A   & A           & A?        & -          & A       & n & \ref{f09} 	\\
20 &   He~2-162                     & B0~I-III        & -    & A    & -   & A            & A             & -   & A        & n & \ref{f11} 	\\
21 &   He~2-182                     & O(H)4-8~III-V  & A    & A    &  A   & -           & -        & -          & A       & n & \ref{f09B} 	\\
22 &   IC~2448                     & O(H)3~III-V      & A   & A    & A    & -          & A        & A         & E   & y & \ref{f04} 	\\
23 &   IC~2501                     & O3-6           & A    & -    & A?   & -          & -       & A?          & E   & n & \ref{f08} 	\\
24 &   IC~4673                    & O(H)8        & -    &  -   &  N/D  & A      & A        & -          & N/D               & y & \ref{f09A} 	\\
25 &   IPHAS~190438                  & B[e]          & -    & -    & -   &  -         &  E        & -           &   -        & n & \ref{f09} 	\\
26 &   IPHAS~191716                     & O(H)~III-V(e)    & A   & A      & A?   & -    & A           & -           & -           & y & \ref{f13} 	\\
27 &   IPHAS~194751                     & B0~I         & -  & -     & -   & A           & A?       & -         &  -               & n & \ref{f10} 	\\
28 &   K~3-5                   & emission line? & -   & E?   & E?  & E?          & E?       & -          & E?          & n & \ref{f18} 	\\
29 &   K~3-14                     & O8-9 &  - & A & A & - & - & - & A & n & \ref{f009}       \\
30 &   K~3-21                     & cont. & - & - & - & - & - & - & - & n & \ref{f17}       \\
31 &   K~3-36                     & O? & - & - & - & - & - & A? & E  & n & \ref{f16B}       \\
32 &   K~3-66                     & O & A & - & - & - & - &  - & A &  n & \ref{f14A}       \\
33 &   K~4-10                     & O3-4 & A & - & A? & -   & - & A & E? & n & \ref{f04} 	\\
34 &   K~4-19                     & B1~I             & - & - & - & A & A & - & - & n & \ref{f11} 	\\
35 &   K~4-50                       & B? & - & - & - & A & - & - & -  & n & \ref{f09}    \\
36 &   K~6-34                     & O(H)4-6          & A    & -    &  A    & -         & A       & -         & N/D        & y    & \ref{f07} 	\\
37 &   Lo~8                     & O(H)3~Vz       & A   & A    & A    & -           & A        & A          & E  & y & \ref{f02} 	\\
38 &   Lo~16                     & O(H)3-4~Vz ((fc)) & A   & A    & A    &    -        & A       &   A        & E        & y & \ref{f03} 	\\
39 &   LoTr~8                     & [WR]            & -   &   E   &  -   &    -       &   -      &  -          & - & n & \ref{f18} 	\\
40 &   M~1-5                     & emission line? & - & E? & - & - & - & - & E? & n & \ref{f18}    \\
41 &   M~1-6                     & O6-9~I           & A  & -     & A?    & -         & -         & -  & A   & n & \ref{f08} 	\\
42 &   M~1-14                     & O(H)7-9             & A  & -   & A   & -         & -         & -     & A    & n & \ref{f12} 	\\
43 &   M~1-46                   & O(H)7~I(fc)      & A     & E   & A    & A           & -        & -          & A       & y & \ref{f07} 	\\
44 &   M~1-72                     & cont. & - & - & - & - & - & - & - & n & \ref{f17}       \\ 
45 &   M~2-7                     & O                & A?   & A    & A    & -           & -        & -          & -      & n & \ref{f14A} 	\\
46 &   M~2-19                   & O(H)6-8 III-V & A    & A    & A    &     -      & A        & -          & N/D           & n    & \ref{f05} \\
47 &   M~2-47                     & O3-6 & A & E? & A & - & - & A? & E  & n & \ref{f08}       \\
48 &   M~3-16                     & O(H)4-6 ~III-V   & A    & A    &  A    & A         & A       & -         &  N/D       & n    & \ref{f07} 	\\
49 &   NeVe~3-2                   & O(H)4~Vz       & A    & A    & N/D &   -          & A        & A?         & N/D    & y & \ref{f06} 	\\
50 &   NGC~6326                     & O(H)5-8 ((fc))  &   -  &  A   & A    &   A?        & A        &  -          & E?     &  y    & \ref{f09B} 	\\
51 &   Pe~1-9                    & O(H)          & A?   & A   & A    & A?      & A          & A?        & N/D              & y & \ref{f14} 	\\
52 &   PHR~J0745-3535                     & cont.         & - & - & - & - & - & - & - & y & \ref{f17} 	\\
53 &   PHR~J0809-1650                     & O(H)~III-V       &  A  & A  & A  & -          & A         & -     & -    & y & \ref{f09B} 	\\
54 &   PHR~J0834-2819                     &  DAO             & -   & A  & -  & -          & A         & -     & N/D  & y & \ref{f15} 	\\
55 &   PHR~J1244-6601                     & H-rich            & -   & -    & -    & -           & A       & -        & -   & n & \ref{f16} 	\\
56 &   PHR~J1424-5138                     & O(H)3~I          & A   & N/D   & A   &  -          & A        & A          & E & y & \ref{f02} 	\\
57 &   PHR~J1510-6754                     & O(H)e           & E    & E    & E   & E           & E        & -        & N/D  & y & \ref{f13} 	\\
58 &   PHR~J1736-3659                     & O(H)8~If         & A   & E    & N/D  & A           & A         &  -        & N/D     & y & \ref{f09A} 	\\
59 &   PHR~J1744-3355                     & H-rich           & -    & -    & -    &    -        &  A       & -          & N/D    & y & \ref{f16} 	\\
60 &   PHR~J1756-3342                     & O(H)3-4~Vz         & A    & A    & A    &  -           &  A      & A          & N/D    & y & \ref{f03} 	\\
61 &   PHR~J1757-2824                    & O(H)          &  -   &    A &  A   &     -      & A        &  -         & N/D           & y & \ref{f14} 	\\
62 &   PHR~J1804-2913                    & O(H)          & -    & A    & -    &    -       & A        & -          & N/D           & n & \ref{f14} 	\\
63 &   PM~1-29                     & B0.5~I & - & - & - & A & A & - & -  & n & \ref{f10}    \\
64 &   PM~1-295                   & O(H)9~I & - & - & -  & A & A & - & -  & n & \ref{f10} 	\\
65 &   PPA~J1747-3435                     & O(H)             & A    & A    & A    & -           & A        & -          & N/D    & n & \ref{f12} 	\\
66 &   PPA~J1855-2328                    & O(H)6 III-V  & A     & A   & A   & -          & A        &   -        &  -              & n & \ref{f07} 	\\
67 &   RCW~24                     & DA               & - & - & - & - & A & - & N/D  & y & \ref{f15} 	\\
68 &   Sa~1-8                   & O(H)4-8~III    & A   & A    & A   & -           & -        & -          & E           & n & \ref{f09B} 	\\
69 &   Sa~3-96                    & H-rich?  & -     & -   & -   & -          & A        &   -        &  -                  & y & \ref{f16A} 	\\
70 &   Sa~4-1                     & O(H)3~V     & A      & A     & A   & -        & A       & A         & E                 & y & \ref{f01} 	\\
71 &   SAB~41                     & H-rich           & -    & -    & -    & -            & A       & -          & -      & y & \ref{f16A} 	\\
72 &   SB~20                   & O(H)3-4       & A     & N/D & A    & -            & A        & A         & E           & y & \ref{f03} 	\\
73 &   Sp~1                     & O(H)3-5~(fc)     & A    & A+E?   &  A    & -           & A        & E?     & N/D    & y & \ref{f06} 	\\
74 &   Sp~4-1                     & O?        & A?     & -     & -   & -        & -       & -         & E                   & n & \ref{f16B} 	\\
75 &   TS~01                     & O(H) III-V  & A & A & N/D & - & - & - & A &  n & \ref{f13} 	\\
76 &   VBRC~7                     & H-rich?          & -    &  N/D  & -  & -           & A         & -          & -     & y & \ref{f16A} 	\\
77 &   Vy~2-3                     & O(H)3-4~I & A & E? & A &- & A? & A & E  & n & \ref{f02}    \\
78 &   WKK~136-337                     & O(H)3~III-V     & A    & A   &N/D&-&A& A         & N/D & y & \ref{f03} 	\\
 \hline 
\end{longtable}
%

%%%%%%%%%%%%%%%%%%%%%%%%%%%%%%%%%%%%%%%%%%%%%%%%%%%%%%%%%%%%%%%%%%%%%%%%%%%%%%%%%%%%%%%%%%%%%%%%%%%%%%%%%%%%%%%%%
%%%%%%%%%%%%%%%%%%%%%%%%%%%%%%%%%%%%%%%%%%%%%%%%%%%%%%%%%%%%%%%%%%%%%%%%%%%%%%%%%%%%%%%%%%%%%%%%%%%%%%%%%%%%%%%%%
%%%%%%%%%%%%%%%%%%%%%%%%%%%%%%%%%%%%%%%%%%%%%%%%%%%%%%%%%%%%%%%%%%%%%%%%%%%%%%%%%%%%%%%%%%%%%%%%%%%%%%%%%%%%%%%%%

\end{document}